\documentclass[prd,showpacs,nofootinbib,floatfix,preprintnumbers]{revtex4}%

\usepackage{amssymb,hyperref, amsmath}
\usepackage[dvips]{color}
\usepackage[dvips]{graphicx}
\usepackage{epsfig}
\preprint{JLAB-THY-08-896}
\preprint{TCDMATH-08-13}

\begin{document}
\newcommand{\beq}{\begin{equation}}
\newcommand{\eeq}{\end{equation}}
\newcommand{\tr}{\mbox{tr}\,}

\bibliographystyle{apsrev}

\title{First results from $2+1$ dynamical quark flavors on an anisotropic
  lattice: light-hadron spectroscopy and setting the strange-quark mass }

\author{Huey-Wen Lin}
\email{hwlin@jlab.org}
\author{Saul D. Cohen}
\author{Jozef Dudek}
\author{Robert G. Edwards}
\author{B\'alint Jo\'o}
\author{David G. Richards}
\affiliation{Thomas Jefferson National Accelerator Facility, Newport News, VA 23606, USA}

\author{John Bulava}
\author{Justin Foley}
\author{Colin Morningstar}
\affiliation{Department of Physics, Carnegie Mellon University, Pittsburgh, PA
  15213, USA}

\author{Eric Engelson}
\author{Stephen Wallace}
\affiliation{Department of Physics, University of Maryland, College Park, MD 20742, USA}

\author{K. Jimmy Juge}
\affiliation{Department of Physics, University of the Pacific, Stockton, CA 95211, USA}

\author{Nilmani Mathur}
\affiliation{Department of Theoretical Physics, Tata Institute of Fundamental Research, Mumbai 400005, India}

\author{Michael J. Peardon}
\author{Sin\'ead M. Ryan}
\affiliation{School of Mathematics, Trinity College, Dublin 2, Ireland}

\date{\today}
\begin{abstract}
\begin{center}
{(Hadron Spectrum Collaboration)}\\
\end{center}

We present the first light-hadron spectroscopy on a set of $N_f=2+1$
dynamical, anisotropic lattices.
A convenient set of coordinates that parameterize the
two-dimensional plane of light and strange-quark masses is introduced.
These coordinates are used to extrapolate data obtained at the simulated
values of the quark masses to the physical light and strange-quark point.
A measurement of the Sommer scale on these ensembles is made, and the
performance of the hybrid Monte Carlo algorithm used for generating the
ensembles is estimated.

\end{abstract}

\maketitle

\section{Introduction}

Understanding the internal structure of nucleons has been a central research
topic in nuclear and particle physics for many decades. As detailed experimental
data continue to emerge, improved theoretical understanding of the hadronic
spectrum will be needed to learn more about the complex, confining dynamics of quantum chromodynamics (QCD). Lattice calculations offer a means of linking experimental data to the Lagrangian of QCD, allowing access to the internal structure of any resonance.

At Jefferson Laboratory (JLab), an ambitious program of research into a range of
hadronic excitations is underway. To date, the Hall~B experiment has collected a large amount of data regarding the spectrum of excitations of the nucleons. The Excited Baryon Analysis Center (EBAC)\cite{Lee:2006xu,Matsuyama:2006rp} aims to review all observed nucleon excitations systematically and to extract reliable parameters describing transitions between resonances and the ground-state nucleons. The 12-GeV upgrade of JLab's CEBAF accelerator will make possible the GlueX
experiment, which will produce an unprecedented meson data-set through
photoproduction. A particular focus will be the spectrum of hadrons with exotic
quantum numbers which can arise when the gluonic field within a meson carries
non-vacuum quantum numbers. Such ``hybrid'' mesons offer a window into the
confinement mechanism and will be studied theoretically in some detail using
lattice methods. Lattice spectroscopy can determine the properties of
isoscalar mesons as well, including any possible candidate glueballs.

Accurate resolution of excited states using lattice QCD has proven difficult. In Euclidean space, excited-state correlation functions decay faster than the ground-state and at large times are swamped by the larger signals of lower states. To improve the chances of extracting excited states, better temporal resolution of correlation functions is extremely helpful. An anisotropic lattice, where the temporal domain is discretized with a finer grid spacing than its spatial counterpart, is one means of providing this resolution while avoiding the increase in computational cost that would come from reducing the spacing in all directions. This finer resolution must be combined with application of variational techniques to construct operators that overlap predominantly with excited states. 
In a series of
papers\cite{Basak:2005aq,Basak:2005ir,Basak:2006ww,Basak:2007kj,Dudek:2007wv},
  techniques to construct operators in irreducible representations of the cubic
  group to extract radially and orbitally excited states were presented.
  Application of these techniques to quenched anisotropic lattices shows clear
  signals as high as the eighth excited state. Also possible on the anisotropic
  lattice are studies of radiative transitions in meson systems\cite{Dudek:2006ej} and nucleon-$P_{11}$ transition form factors\cite{Lin:2008qv}.

Making the lattice discretization anisotropic comes with a price, however. Since
hypercubic symmetry is broken down to just the cubic group, relevant
(dimension-four) operators can mix in the lattice action. To ensure the
continuum limit of the lattice theory has full Lorentz invariance, a
nonperturbative determination of the lattice action parameters that enforce the
symmetry at finite lattice spacing in some low-energy observables has been
performed\cite{Edwards:2008ja}. In work to be reported elsewhere, a perturbative
determination of these action parameters is also being carried out by this collaboration\cite{Foley:2008xx}.

In this study, we perform three-flavor dynamical calculations with two degenerate light quarks and a strange quark. In a previous study, we tuned a three-flavor lattice action to ensure Lorentz symmetry is restored in appropriately chosen low-energy observables. We showed empirically that restoring the symmetry at
quark masses below 175~MeV requires only small changes to the action parameters, and no further determinations of these parameters is needed within the scope of this study. Our fermion action is a Sheikholeslami-Wohlert discretization, generalized to the anisotropic lattice\cite{Chen:2000ej}. The fermion fields interact with the gluons via 3-dimensionally stout-smeared\cite{Morningstar:2003gk} links, and the gluon action is Symanzik improved at the tree level of perturbation theory. To assess the cost of these dynamical calculations, we study the efficiency of the hybrid Monte Carlo (HMC) algorithm in large-scale production simulations.

Using current algorithms and computing resources, it remains impractical to run calculations at the physical value of the light-quark mass. An extrapolation of the light-quark dependence of simulation data is needed. While simulations straddling the correct strange-quark mass have been performed, determining the appropriate choice of mass in the Lagrangian is also problematic; a priori, this value is not known and changing the bare strange-quark mass affects all lattice observables in a delicate way. This work proposes a simple means of setting the lattice strange-quark mass by examining dimensionless ratios with mild behavior in the light-quark chiral limit. A new set of coordinates, parameterizing the space of theories with different light and strange-quark masses is introduced to help this process. We use the ratios $l_\Omega = 9m_\pi^2/4m_\Omega^2$ and $s_\Omega=9(2m_K^2 - m_\pi^2)/4m_\Omega^2$, inspired by expanding the pseudoscalar meson masses to leading order in chiral perturbation theory.

With this framework in place, the spectrum of some ground-state mesons and
baryons is determined and extrapolated to the physical quark masses using
leading-order chiral perturbation theory. The Sommer scale\cite{Sommer:1993ce} is determined (in units of the Omega-baryon mass) on a subset of our ensembles and extrapolated to the physical quark masses. Using this method, we intend to continue our exploration of excited-state hadrons, including isoscalar and hybrid mesons. A clear means of handling unstable states is needed, and our suite of measurement technology is currently under further development. A study of these techniques on $N_f=2$ dynamical lattices gives us confidence that more precise understanding of these states will be forthcoming.

The structure of this paper is as follows: In Sec.~\ref{Sec:Setup}, we will discuss the actions and algorithms used in this work, and the performance of the method used to generate Monte Carlo ensembles is examined. The details of the measurements we performed on these ensembles is presented in Sec.~\ref{Sec:Scale}. Sec.~\ref{Sec:Strange} presents the method we propose to set the strange-quark mass, including the dimensionless coordinates used for extrapolating quantities measured at unphysical quark masses to the physical theory. Our determination of a selection of states in the hadron spectrum and the Sommer scale is given in Sec.~\ref{Sec:Spec}. Some conclusions and future outlook are presented in Sec.~\ref{Sec:Conclusion}.

\section{Simulation Details}\label{Sec:Setup}

In this section details of the lattice action and the performance of the hybrid
Monte Carlo algorithm are presented.
Monte Carlo simulations were performed on lattices with grid spacings of $a_s$
and $a_t$ in the spatial and temporal directions respectively and with physical
volumes $L_s^3 \times L_t$ where $L_s = N_s a_s$ and $L_t = N_t a_t$.
Lattices with extents $N_s^3 \times N_t= 12^3 \times 96,16^3 \times
96, 16^3 \times 128 $ and $24^3 \times 128$ were employed.

\subsection{Action}\label{Subsec:Action}

The gauge and fermion actions used in this work are described in great detail in
our previous work\cite{Edwards:2008ja}. For completeness in this paper, we
briefly review the essential definitions. For more detailed definitions, see Ref.~\cite{Edwards:2008ja}.

For the gauge sector, we use a Symanzik-improved action with tree-level tadpole-improved coefficients:
\begin{eqnarray}\label{eq:aniso_syzG}
S_G^{\xi}[U] &=& \frac{\beta}{N_c\gamma_g} \left\{
\sum_{x,s\neq s^\prime} \left[  \frac{5}{6 u_s^4}\Omega_{{\cal P}_{ss^\prime}}(x)- \frac{1}{12 u_s^6}\Omega_{{\cal R}_{ss^\prime}}(x)\right]
+ \sum_{x,s}\gamma_g^2 \left[ \frac{4}{3 u_s^2  u_t^2}\Omega_{{\cal P}_{st}}(x) - \frac{1}{12 u_s^4 u_t^2}\Omega_{{\cal R}_{st}}(x)\right] \vphantom{\frac{1}{\xi}} \right\},
\end{eqnarray}
where $\Omega_W={\rm Re}{\rm Tr}(1-W)$ and $W={\cal P}$, the plaquette, or ${\cal R_{\mu\nu}}$, the $2\times1$ rectangular Wilson loop (length two in the $\mu$ direction and one in the $\nu$ direction) with $\{s,s^\prime\}\in\{x,y,z\}$.
The parameter $\gamma_g$ is the bare gauge anisotropy, $N_c=3$ indicates the
number of colors, $\beta$ is related to the coupling $g^2$ through $\beta=2 N_c/g^2$, and $u_s$ and $u_t$ are the spatial and temporal tadpole factors. This action has leading discretization error at $O(a_s^4,a_t^2,g^2 a_s^2)$ and possesses a positive-definite transfer matrix, since there is no length-two rectangle in time.

In the fermion sector, we adopt the anisotropic clover fermion action\cite{Chen:2000ej}:
\begin{eqnarray}
S_F^{\xi}[U, \overline{\psi},\psi ]
&=& \sum_{x} \overline{\hat \psi }(x) \frac{1}{ \tilde{u}_t} \left\{
 \tilde{u}_t \hat{m}_0
 +  \gamma_t \hat{W}_t +\frac{1}{\gamma_f}  \sum_s \gamma_s \hat{W}_s
 \right. \nonumber \\
 &-&\left.\frac{1}{2}
    \left[\frac{1}{2}\left(\frac{\gamma_g}{\gamma_f}+\frac{1}{\xi_R}\right)
        \frac{1}{\tilde{u}_t \tilde{u}_s^2} \sum_{s}\sigma_{ts}\hat{F}_{ts}
    +
      \frac{1}{\gamma_f}\frac{1}{\tilde{u}_s^3}
      \sum_{s<s^\prime} \sigma_{ss^\prime} \hat{F}_{ss^\prime}
    \right]
      \right\}\hat \psi (x), 
\label{eq:fermion-action}
\end{eqnarray}
where $\gamma_f$ is the bare fermion anisotropy and $\xi_R=a_s/a_t$ is the renormalized  anisotropy. $\gamma_{s,t}$, $\sigma_{st}$ and $\sigma_{ss^\prime}$ (with $\sigma_{\mu\nu}=\frac{1}{2}[\gamma_\mu,\gamma_\nu]$) are Dirac matrices.
Hats denote dimensionless variables which connect to dimensionful quantities as: quark field $\hat \psi = a_s^{3/2} \psi$, bare quark mass $\hat m_0 = m_0 a_t$, gauge field strength $\hat F_{\mu\nu} = a_\mu a_\nu F_{\mu\nu} = \frac{1}{4}{\rm Im}({\cal P}_{\mu\nu}(x))$ and ``Wilson operator'' $\hat{W}_\mu \equiv \hat \nabla_\mu - \frac{1}{2} \gamma_\mu \hat \Delta_\mu$ (with ${\hat \nabla}_\mu = a_\mu \nabla_\mu$, ${\hat \Delta}_\mu = a_\mu^2 \Delta_\mu$). The gauge links in the fermion action are 3-dimensionally stout-link smeared gauge fields with smearing weight $\rho=0.14$ and $n_\rho=2$ iterations. $\tilde{u}_s$ and $\tilde{u}_t$ are the spatial and temporal tadpole factors from smeared fields, respectively.

In our previous work\cite{Edwards:2008ja}, we found at $\beta=1.5$, that the tadpole factors are
\begin{equation}
u_s = 0.7336, \
u_t = 1, \qquad
\tilde{u}_s = 0.9267, \
\tilde{u}_t = 1.
\end{equation}
Tuning the anisotropy for all quark masses (even below the chiral limit) gives the desired $\gamma_{g,f}^*$
\begin{equation}
\gamma_g^* = 4.3, \quad
\gamma_f^* = 3.4.
\end{equation}

\subsection{Algorithm}

We use the rational HMC (RHMC) algorithm for gauge generation\cite{Clark:2006wq}. The theoretical aspects of our procedure were discussed in detail in Ref.~\cite{Edwards:2008ja}. Here we discuss only the aspects that are specific to the calculations presented in this work.

We use rational approximations for both the light-quark fields and for the strange quarks --- one field for each light-quark flavor and another one for the strange. We employ even-odd preconditioning for the Wilson clover operator, obtaining the Hamiltonian
\begin{equation}
H = {1 \over 2} \sum_{x,\mu} {\rm Tr}\ \pi^\dagger \pi - 2 \sum_{x} {\rm Tr} \ \log A_{ee}(m_l)  - \sum_{x} {\rm Tr}\ \log A_{ee}(m_s) + S_{F}(m_l) + S_{F}(m_l) + S_{F}(m_s) - S_{G}^{s} - S_{G}^{t},
\end{equation}
where $\pi$ are the momenta conjugate to the gauge fields; terms involving $A_{ee}$ contribute effects due to the parts of the preconditioned clover determinant coming from the submatrix connecting even sites; $S_G^{s}$ and $S_{G}^{t}$ are the parts of the gauge action involving loops in the spatial directions only and with loops including time direction respectively; and $S_{F}(m_l)$ and $S_{F}(m_s)$ are pseudofermion terms for the rational approximations to the fermion action corresponding to the light- and strange-quark fields respectively, which we discuss below.

The pseudofermion terms $S_{F}(m)$ employ a rational approximation to the fermion determinant of the even-odd preconditioned clover operator coming from the submatrix connecting the odd sites for a quark with mass $m$. This submatrix is
\begin{equation}
M(m; \tilde{U}) = A_{oo}(m;\tilde{U}) - D_{oe}(\tilde{U})
A^{-1}_{ee}(m;\tilde{U})D_{eo}(\tilde{U}),
\end{equation}
where $A_{oo}$ is the clover operator on the odd sites, $A^{-1}_{ee}$ is the
inverse clover operator on the even sites and $D_{oe} (D_{eo})$ is the Wilson
hopping term connecting odd sites with even (even sites with odd). In all the expressions, $\tilde{U}$ denotes stout-smeared gauge fields $U$, which were smeared as described in Subsection~\ref{Subsec:Action}.

To construct our pseudofermion actions, we use the rational approximation
$R^{{a \over b}}( M^\dagger M )$ in partial-fraction form:
\begin{equation}
R^{{a \over b}}(M^\dagger M)= \alpha \sum_i p_i \left(
M^\dagger M + q_i\right)^{-1} \approx \left(
M^\dagger M \right)^{{a \over b}},
\end{equation}
where we drop the quark-mass dependence of $M$ for clarity. The coefficients
$\alpha$, $p_i$ and $q_i$ define the approximation and are determined via the Remez algorithm\cite{Remez:1934,Remez:1962} applied over the spectral bounds of the operator $M^\dagger M$. In particular, we needed to compute approximations with $(a,b)=(-1,4)$ for evaluating the actions (see below), $(a,b)=(1,4)$ for pseudofermion refreshment and $(a,b)=(-1,2)$ for our molecular dynamics (MD). Our approximation bounds for the action are shown in Table~\ref{t:ActionApprox}. We solve the linear system resulting from applying $R^{a \over b}$ to pseudofermion fields using the multi-shift conjugate gradient algorithm\cite{Jegerlehner:1996pm}. We use a stopping relative residuum $r < 10^{-8}$ in our energy calculations where the residuum for pole $i$ is
\begin{equation}\label{e:residuum}
r_i = \frac{ \left|\left| \phi - \left( M^\dagger M + q_{i} \right) \psi_i \right|\right| }{  \left|\left| \phi \right|\right| },
\end{equation}
where $\phi$ is the pseudofermion field and $\psi_i$ is the solution corresponding to the $i^{\rm th}$ pole. However, since the multi-shift algorithm cannot be restarted, our stopping was based on estimates of $r_i$ accumulated with the short-term recurrence in the solver algorithm, which may be slightly different from the true residual as defined in Eq.~\ref{e:residuum} due to solver stagnation and rounding effects. To minimize rounding effects we accumulated sums and inner products using double precision.

\begin{table}
\begin{center}
\begin{tabular}{c|cccc|cccc}
\hline \hline
$V$ & \multicolumn{4}{c|}{light quark} & \multicolumn{4}{c}{strange quark} \\
    & $a_tm_l$ & bounds & no. poles & max. error & $a_tm_s$ & bounds & no. poles & max. error \\
\hline \hline
$24^3 \times 128$ & $-0.0840$ & $(5  \times 10^{-6}, 10)$ & 16 & $1.8 \times 10^{-8}$ & $-0.0743$ & $(10^{-4},10)$ & 12 & $8.8 \times 10^{-8}$ \\
\hline \hline
\end{tabular}
\end{center}
\caption{Details of the approximations used for the pseudofermionic action $R^{-{1 \over 4}}$. We show the bounds, the number of poles and the maximum error for the approximation to the light and strange pseudofermion terms.}\label{t:ActionApprox}
\end{table}

\begin{table}
\begin{center}
\begin{tabular}{c|cccc|cccc}
\hline \hline
$V$ & \multicolumn{4}{c|}{light quark} & \multicolumn{4}{c}{strange quark} \\
    & $a_tm_l$ & bounds & no. poles & max. error & $a_tm_s$ & bounds & no. poles & max. error \\
\hline \hline
$24^3 \times 128$ & $-0.0840$ & $(5  \times 10^{-6}, 10)$ & 12 & $2.5 \times 10^{-6}$ & $-0.0743$ & $(10^{-4},10)$ & 10 & $2.0 \times 10^{-6}$ \\
\hline \hline
\end{tabular}
\end{center}
\caption{Details of the approximations used for the force $R^{-{1 \over 2}}$. We show the bounds, the number of poles and the maximum error for the approximation to the light and strange pseudofermion terms.}\label{t:ForceApprox}
\end{table}

Our pseudofermion action terms are
\begin{equation}
 S_F = X^\dagger X, \quad X = R^{-{1 \over 4}}\left(M^\dagger M \right) \phi =
 \alpha \sum_i p_i \psi_i,
\end{equation}
individually for each flavor. We do not need to employ multiple pseudofermion fields per flavor in this study.

During our simulation, we adjusted our approximation range by measuring eigenvalue bounds every five trajectories during the process of thermalization. Thereafter we continued to measure the bounds to ensure we do not suffer from boundary violations.

Our molecular dynamics process employs a rational force
\begin{equation}
F = - \alpha \sum_i p_i \psi^\dagger_i \left( \frac{d M^\dagger}{dU} M + M^\dagger \frac{dM}{dU} \right) \psi_i,
\end{equation}
where $i$ runs over the number of poles in the approximation $R^{-{1\over 2}}$. Time derivatives are evaluated over the stouted gauge field $\tilde{U}$, and only the final sum is recursed down to compute the force for the thin links $U$.

We employ a multiple-timescale integration scheme for the molecular dynamics
evolution\cite{Sexton:1992nu} by nesting a second-order
Omelyan\cite{Omelyan:2003,Takaishi:2005tz} integration step at each timescale.
Our largest forces come from the temporal directions: the gauge force from
$S_{G}^{t}$ and the temporal forces generated by the pseudofermions. To mitigate
the numerical effort needed\cite{Morrin:2006tf}, we place the $S_{G}^{t}$ term in the action on a finer timescale than the other terms, and to deal with the temporal forces from the pseudofermion terms, employ an anisotropic timestep with temporal timestep $dt$ of length
\begin{equation}
dt_t  = dt_s / \xi_{\rm MD},
\end{equation}
where $\xi_{\rm MD} = 3.5$.
Apart from the above, we find the forces from $S_{G}^{s}$ and the spatial forces from the $S_{F}$ terms to be within a factor of 2 of each other, so we place them on the same timescale. The forces from the ${\rm Tr} \log A_{ee}$ terms are very small in comparison but have small numerical cost, so we place them on the same timescale as $S_{G}^s$ and $S_{F}$.

We note that the Hamiltonian for the MD does not need to be known as accurately
as the one for the energy calculations; all that is required is for the MD to be
reversible, area-preserving, and (as a practical matter) for the acceptance rate
to be reasonable. To save on numerical effort we solved our systems of linear
equations only to a residuum $r_{\rm MD}$ of at most $r_{\rm MD} < 10^{-6}$.
Correspondingly, we never required the rational approximation to $R^{-{1 \over
  2}}$ to have a maximum error better than $10^{-6}$, resulting in a smaller
  number of poles in the approximation than we need for the energy calculations.
  Further, to make the MD even less numerically intensive, we follow
  Refs.~\cite{Clark:2004cq,Clark:2006fx} by relaxing the requirements on the
  residua for individual poles in $R^{-{1\over 2}}$. We use a range of $r_{\rm
    MD} < 10^{-4}$ for the smallest shifts and $r_{\rm MD} < 10^{-6}$ for the
    larger shifts. We tune our molecular dynamics to attain an overall
    acceptance rate close to 70\%. We show in Table~\ref{t:ForceApprox} the
    bounds of the MD rational approximation used for the run with $V=24^3 \times
    128$, $(a_tm_l, a_tm_s)=(-0.0840,-0.0743)$. We show the residua requested in the MD evolution  in Table~\ref{t:ForceResidua}, and the timesteps and the resulting acceptance rate in Table \ref{t:StepSizes}.

\begin{table}
\leavevmode
\begin{center}
\begin{tabular}{c|c|c|c|c|c}
\hline \hline
$V$ & $(a_tm_l, a_tm_s)$& poles for $a_tm_l$ & residua & poles for $a_tm_s$ & residua \\
\hline
$24^3 \times 128$ & $(-0.0840,-0.0743)$ & 12
& \begin{tabular}{c}
$10^{-4}, 10^{-4}, 5 \times 10^{-5},$\\
$ 5 \times 10^{-5}, 5 \times 10^{-5}, 10^{-5},$\\
$ 10^{-5}, 5 \times 10^{-6}, 5 \times 10^{-6},$ \\
$ 5 \times 10^{-6}, 3 \times 10^{-6}, 10^{-6}$
\end{tabular} & 10 &
\begin{tabular}{c}
$ 10^{-4}, 10^{-4}, 5 \times 10^{-5},$ \\
$ 10^{-5}, 10^{-5}, 10^{-5},$\\
$ 5 \times 10^{-6}, 5 \times 10^{-6}, 3 \times 10^{-6},$\\
$ 10^{-6} $
\end{tabular} \\
\hline
\end{tabular}
\end{center}
\caption{Requested residua for the poles in the MD force approximation from the
  smallest shifts (leftmost) to larger shifts (rightmost).}\label{t:ForceResidua}
\end{table}

\begin{table}
\begin{center}
\begin{tabular}{cccccc}
\hline \hline
$V$ & $(a_tm_l, a_tm_s)$ & $dt^{1}_s$ & $dt^{2}_s/dt^{1}_s$ & $\xi_{\rm MD} $ & Accept Rate \\
\hline
$24^3 \times 128$ & $(-0.0840, -0.0743)$ & $\frac{1}{16}$ & $\frac{1}{4}$ & $3.5$ & $0.71$ \\
\hline
\end{tabular}
\end{center}
\caption{The two timescales used in the molecular dynamics integration. The spatial timestep for the coarse scale is $dt^1_s$, and for the finer scale it is $dt^{2}_s$, which we display as a fraction of $dt^{1}_s$ here. We also show our MD timestep anisotropy. On each scale, $dt^{i}_t = dt^{i}_s / \xi_{\rm MD}$. Finally, we show the average acceptance rate for the molecular dynamics with these step sizes.}\label{t:StepSizes}
\end{table}

\subsection{Thermalization and Autocorrelation}

During the first segment of each gauge ensemble generation, some special conditions apply. We do not apply the acceptance test during the first $O(10)$ trajectories in each series, which allows a fast initial approach to the vicinity of the equilibrium. Such a scheme is particularly important in the case of simulations starting from totally ordered or disordered configurations. Wherever possible, however, we begin the algorithm with an equilibrated configuration from a simulation at nearby parameters. Also, during this phase (as mentioned above), the minimum and maximum eigenvalue bounds are updated every 5 trajectories.

\begin{figure}
\includegraphics[width=0.45\textwidth]{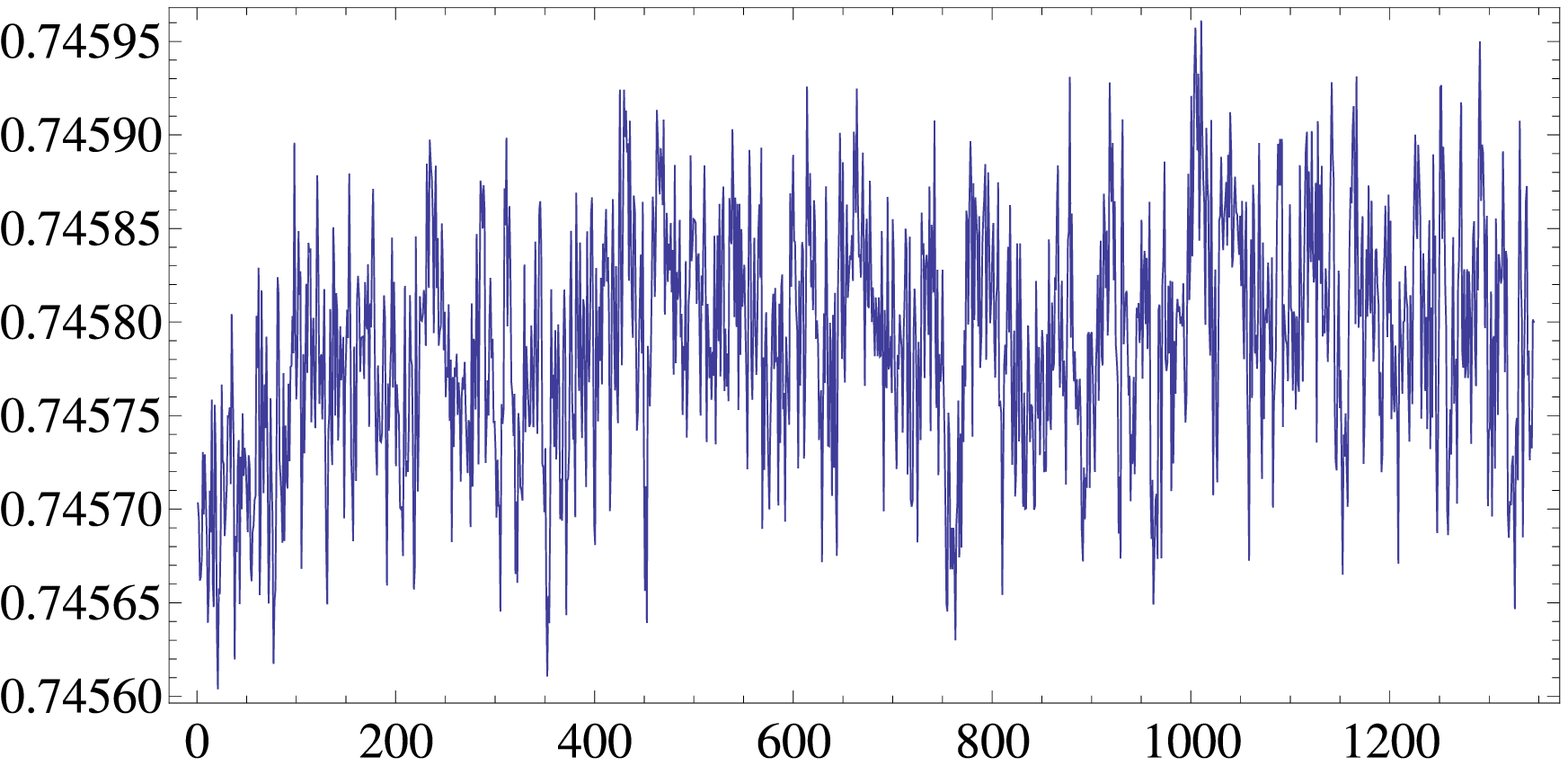}
\includegraphics[width=0.45\textwidth]{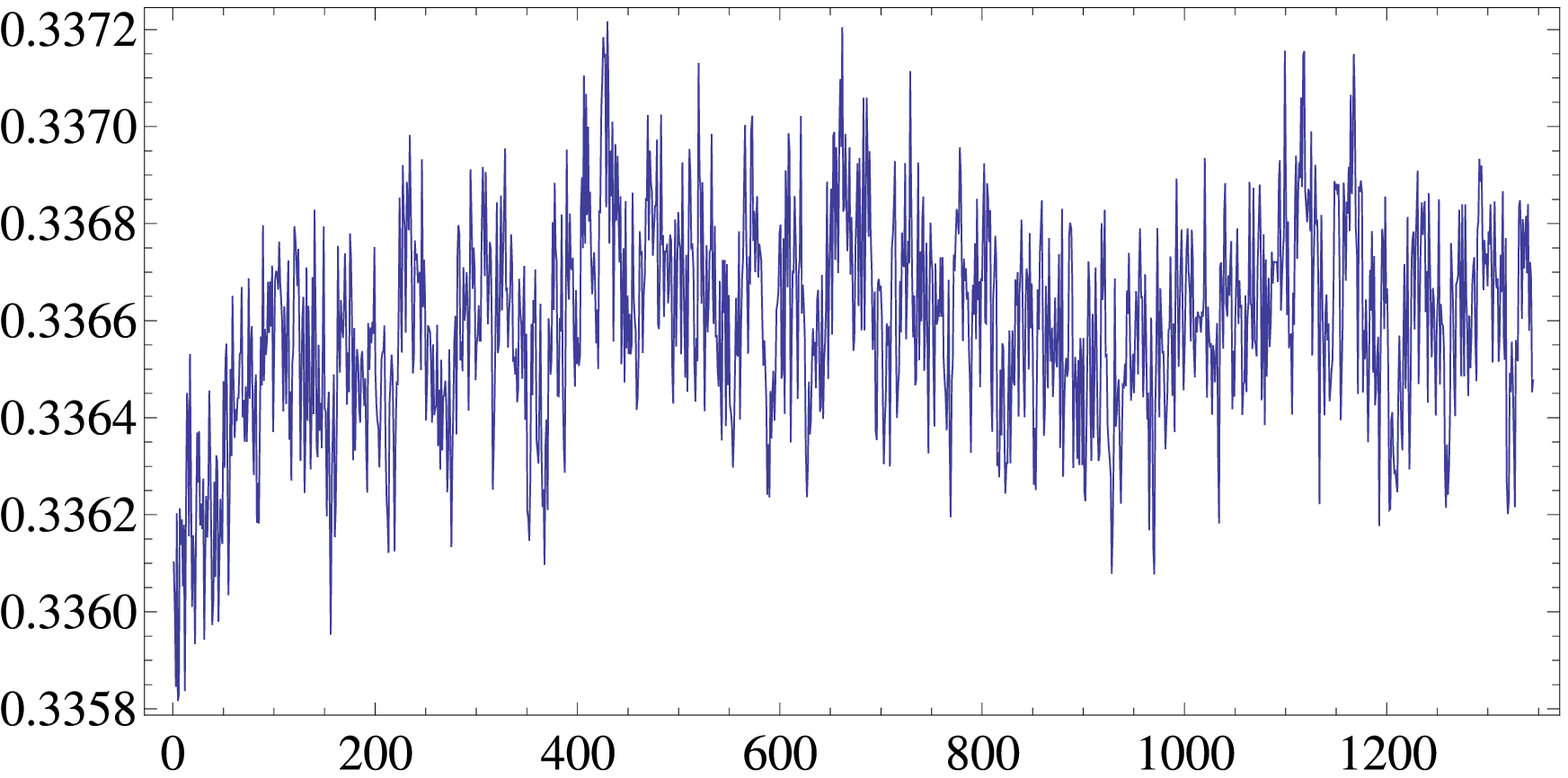}
\caption{Temporal (left column) and spatial (right column) plaquette history
  from the ensemble with $24^3\times 128$ volume and $a_tm_l=-0.0840$. The x-axis is in units of trajectories.}
\label{fig:plaq}
\end{figure}

Figure~\ref{fig:plaq} shows its plaquette history for $24^3\times 128$ volume
and $a_tm_l=-0.0840$. Both plaquette histories (one excluding temporal links and the other including only plaquettes with temporal links) show that equilibrium is reached long before 1000 RHMC trajectories. Therefore, to allow for thermalization of our gauge ensembles during the RHMC, we discard the initial 1000 trajectories from each set.

\begin{figure}
\includegraphics[width=0.65\textwidth]{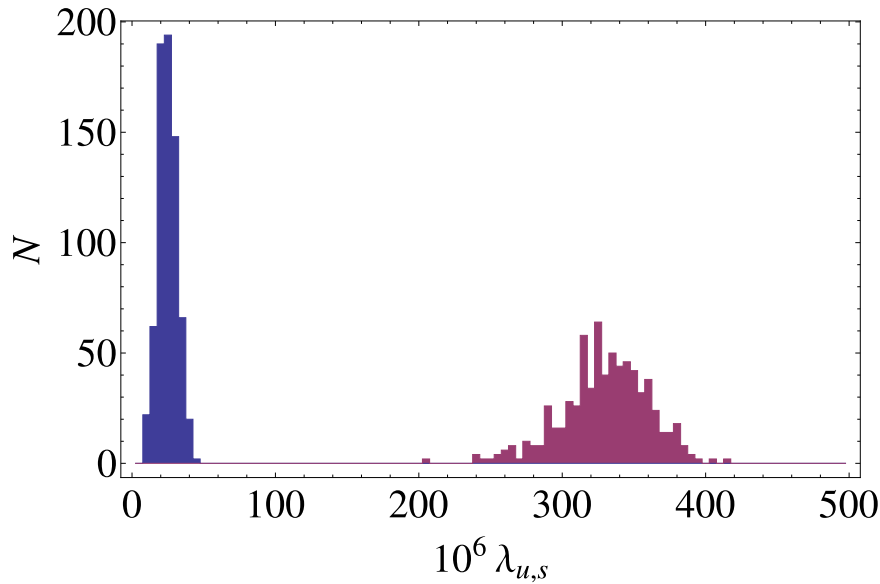}
\caption{Histogram of the lowest eigenvalues of the Dirac operator for the light
  and strange quarks from the ensemble with $24^3\times 128$ volume and $a_tm_l=-0.0840$.}
\label{fig:histus}
\end{figure}

Figure~\ref{fig:histus} shows a histogram of the lowest eigenvalues of the Dirac
operator $M^\dagger M$ for the light and strange quarks from the ensemble with
$24^3\times 128$ volume and $a_tm_l=-0.0840$. The lowest eigenvalues remain safely above the minimum eigenvalue bounds in which our rational approximation is valid. In addition, they show a clear gap away from zero, where the stability of the algorithm might be compromised.

\begin{figure}
\includegraphics[width=0.45\textwidth]{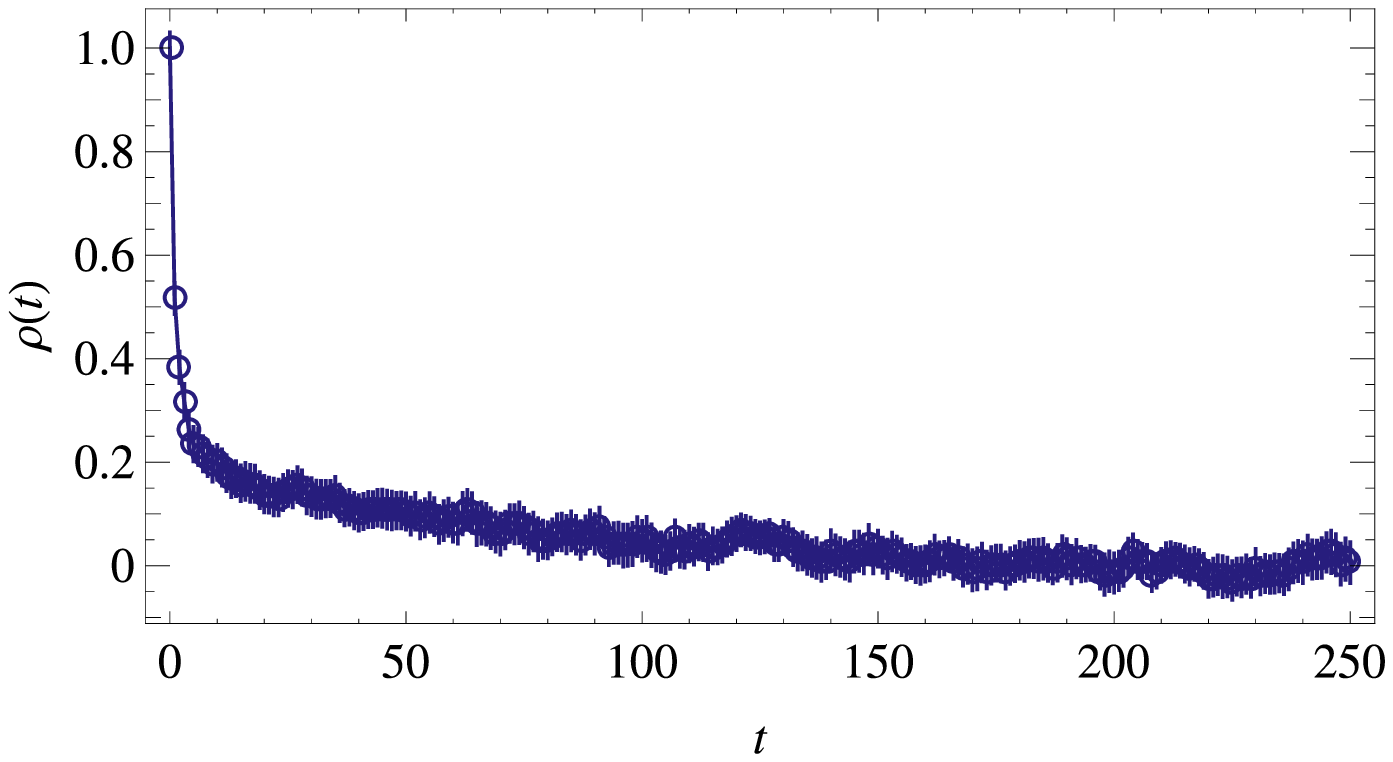}
\includegraphics[width=0.45\textwidth]{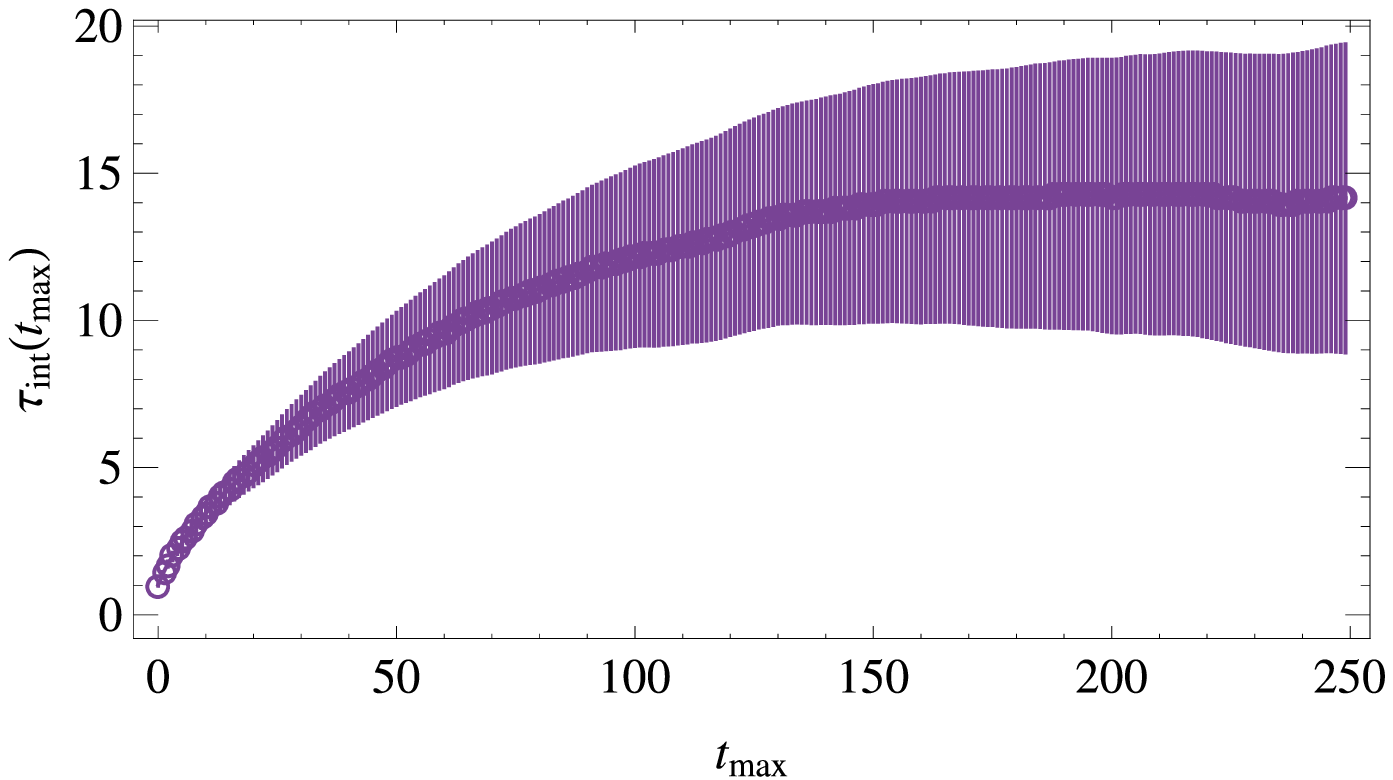}
\includegraphics[width=0.45\textwidth]{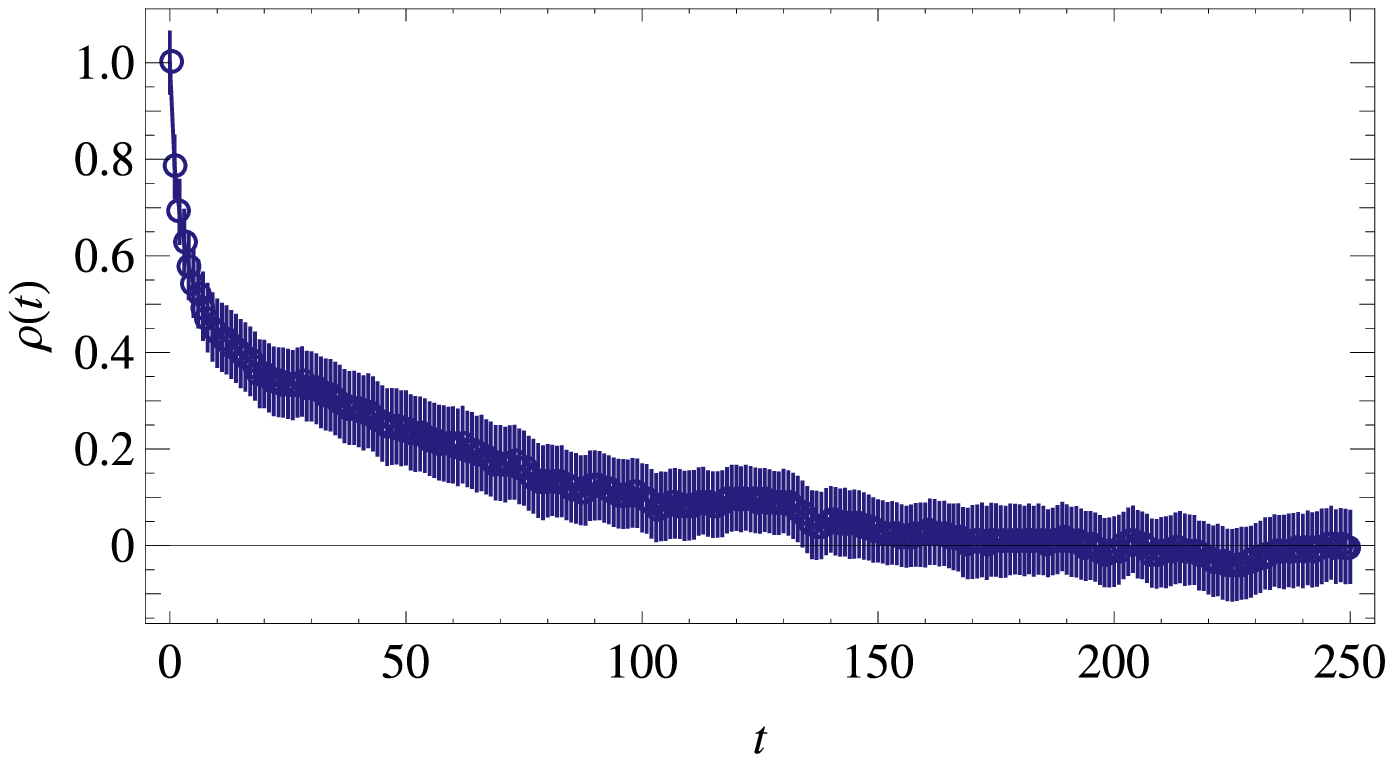}
\includegraphics[width=0.45\textwidth]{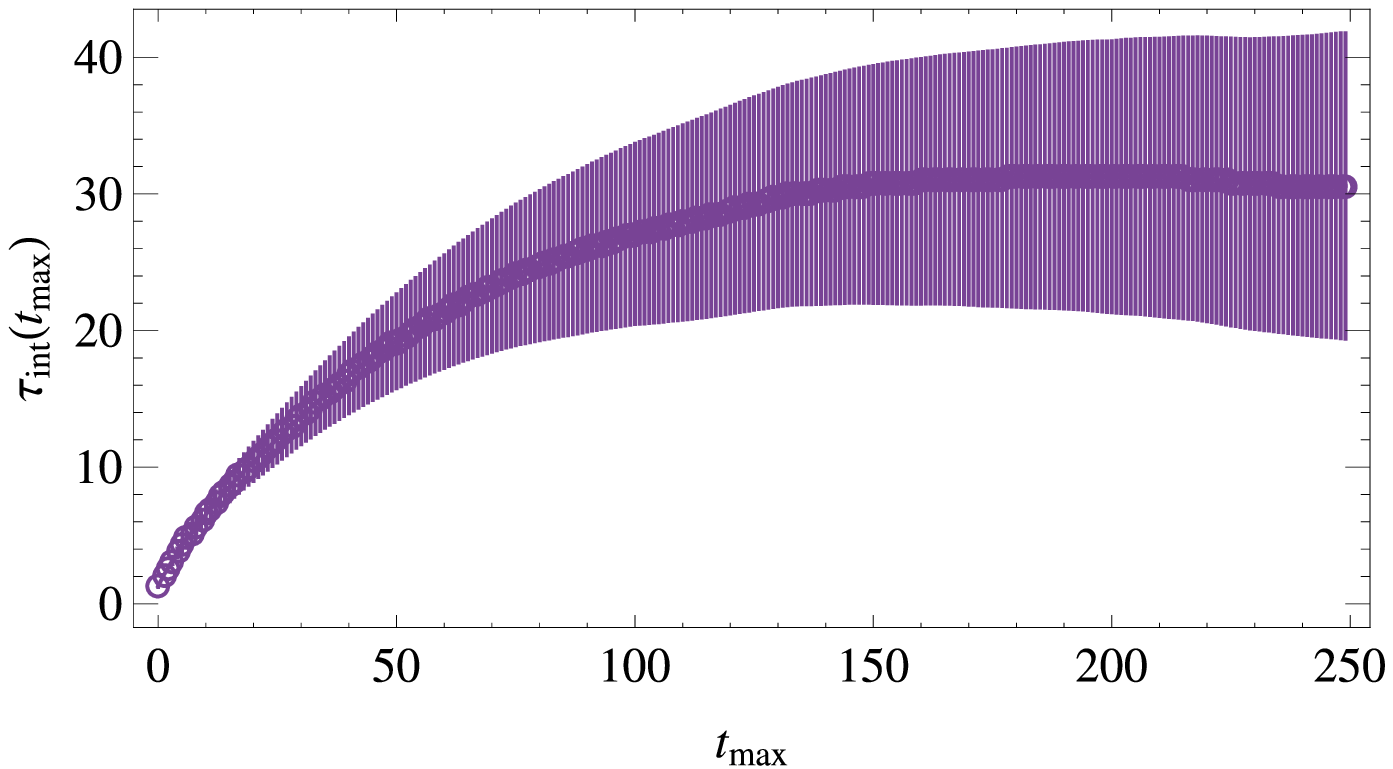}
\caption{Autocorrelation $\rho(t)$ and integrated autocorrelation length
  $\tau_{\rm int}$ (in trajectories) for the unsmeared (above) and smeared (below) plaquette
    involving only spatial links from the ensemble with $16^3\times 128$ volume
    and $a_tm_l=-0.0808$.}
\label{fig:plaq_auto}
\end{figure}

\begin{figure}
\includegraphics[width=0.45\textwidth]{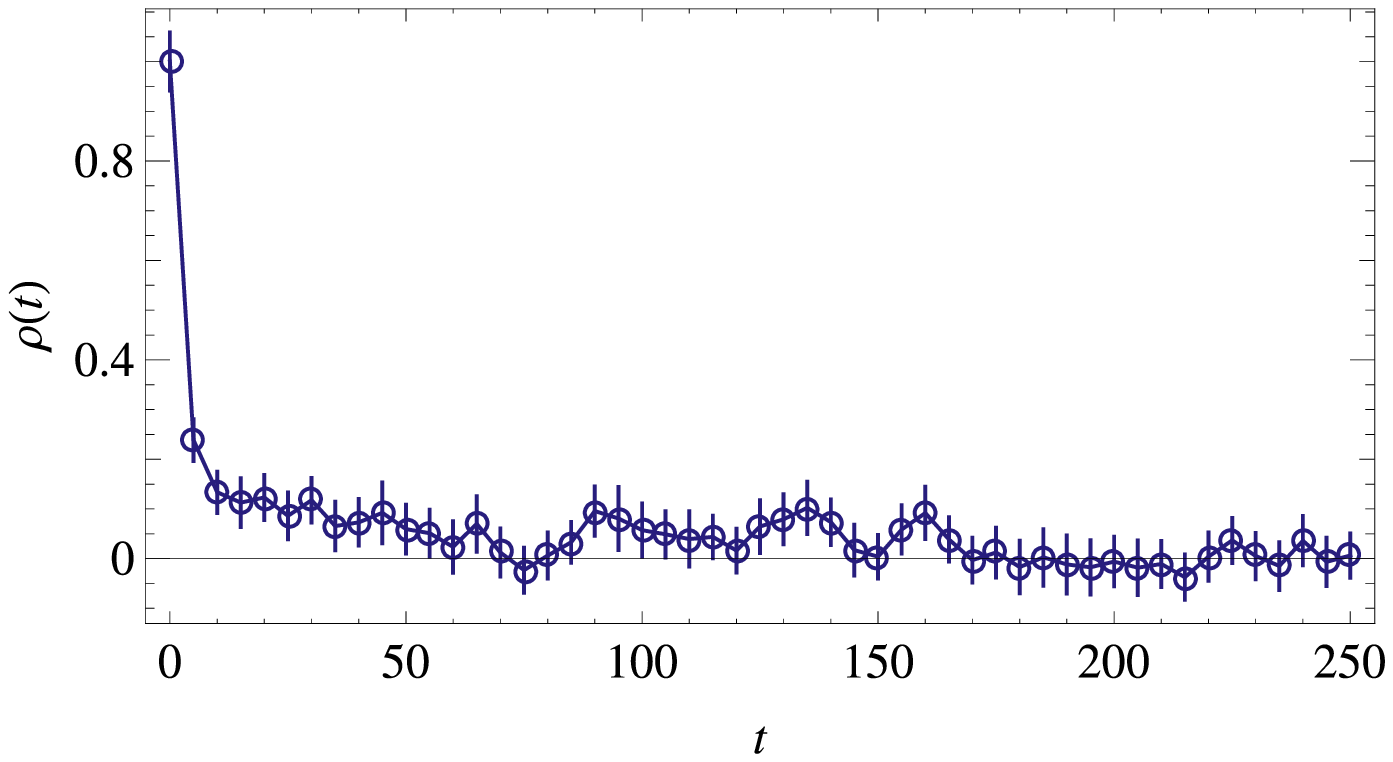}
\includegraphics[width=0.45\textwidth]{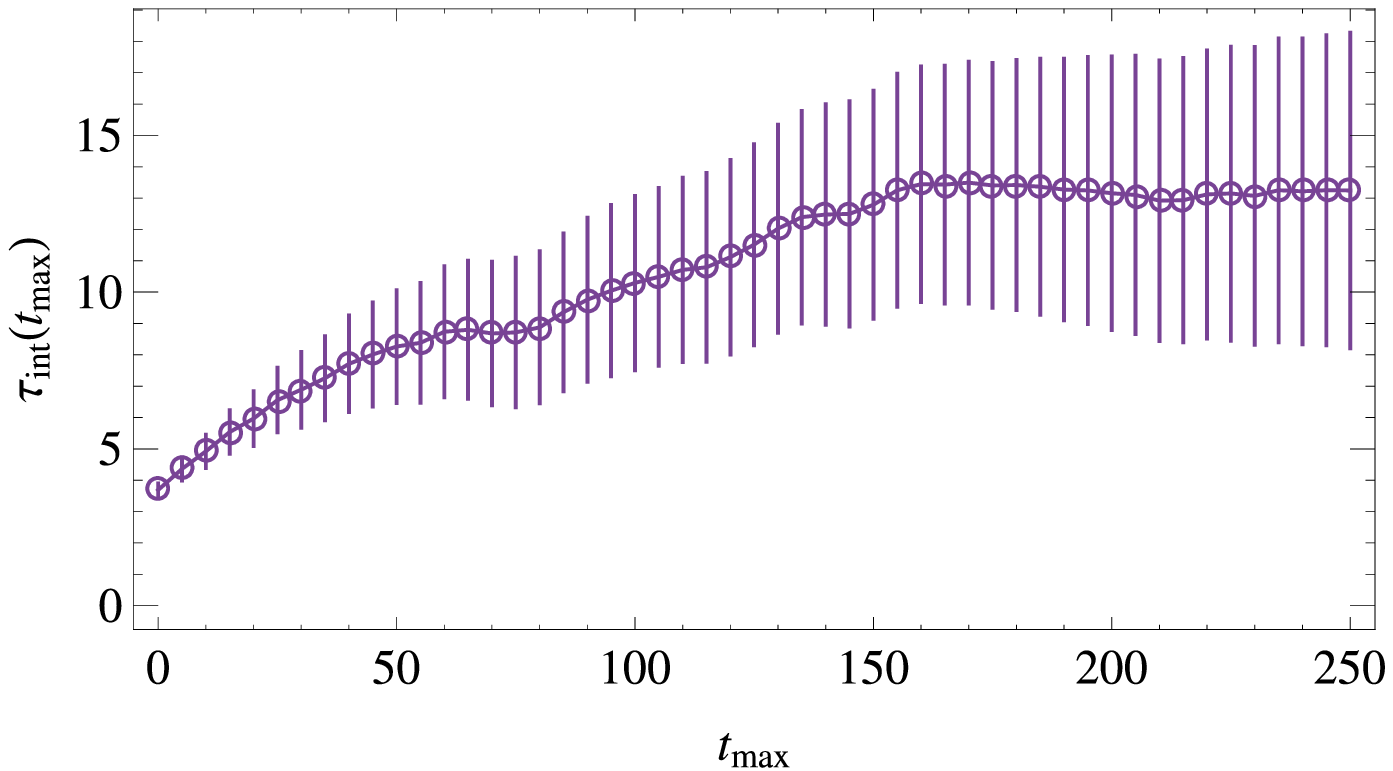}
\includegraphics[width=0.45\textwidth]{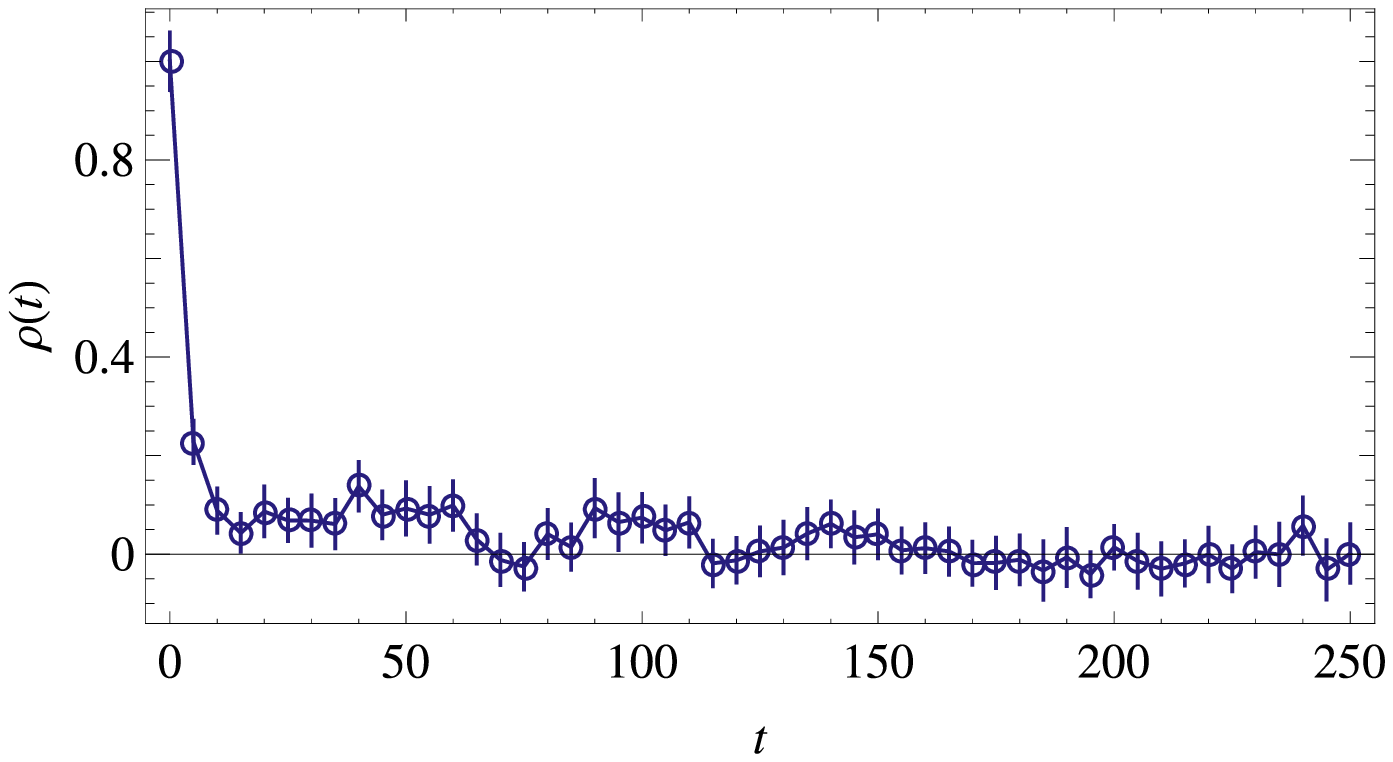}
\includegraphics[width=0.45\textwidth]{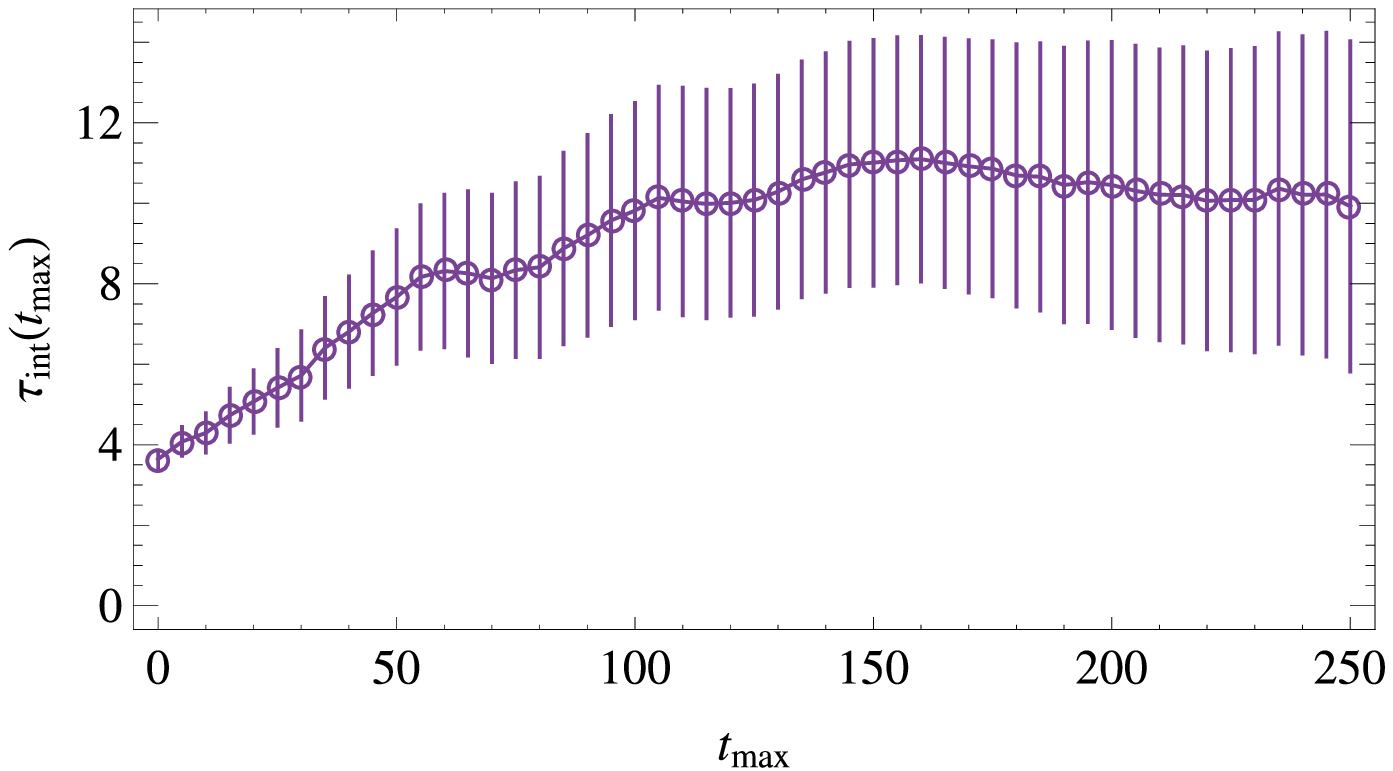}
\caption{Autocorrelation $\rho(t)$ and integrated autocorrelation length
  $\tau_{\rm int}$ (in trajectories) for the up/down (above) and strange (below) quark eigenvalues
    from the ensemble with $16^3\times 128$ volume and $a_tm_l=-0.0808$.}
\label{fig:mdagm_auto}
\end{figure}

\begin{figure}
\includegraphics[width=0.45\textwidth]{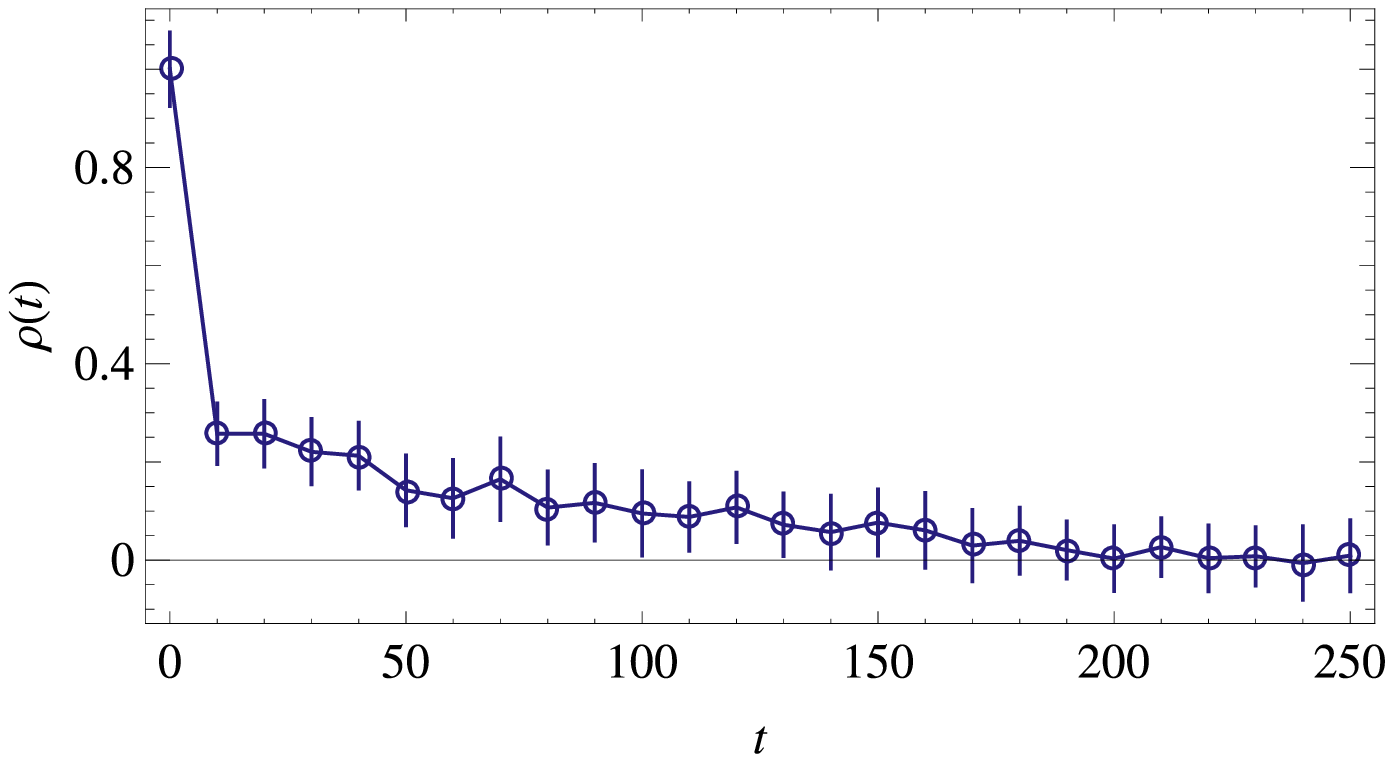}
\includegraphics[width=0.45\textwidth]{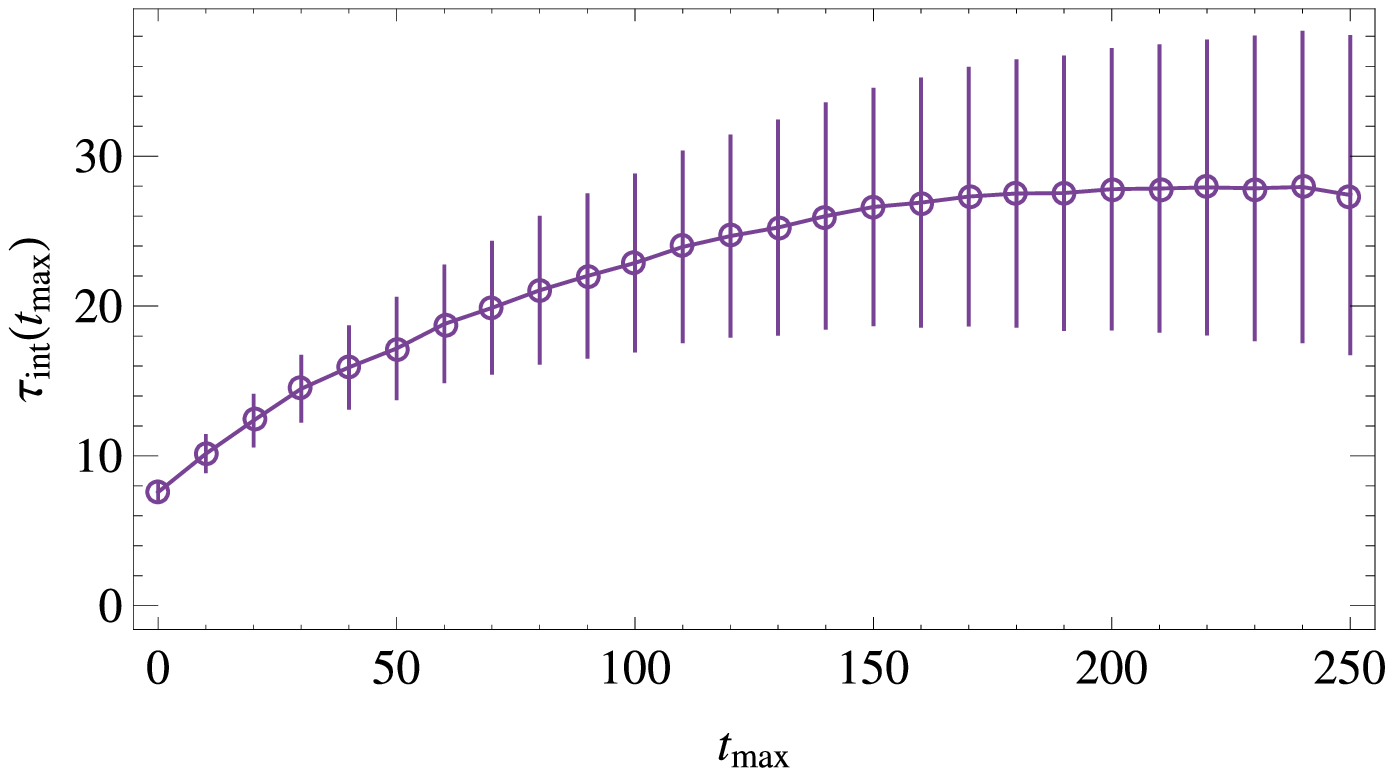}
\includegraphics[width=0.45\textwidth]{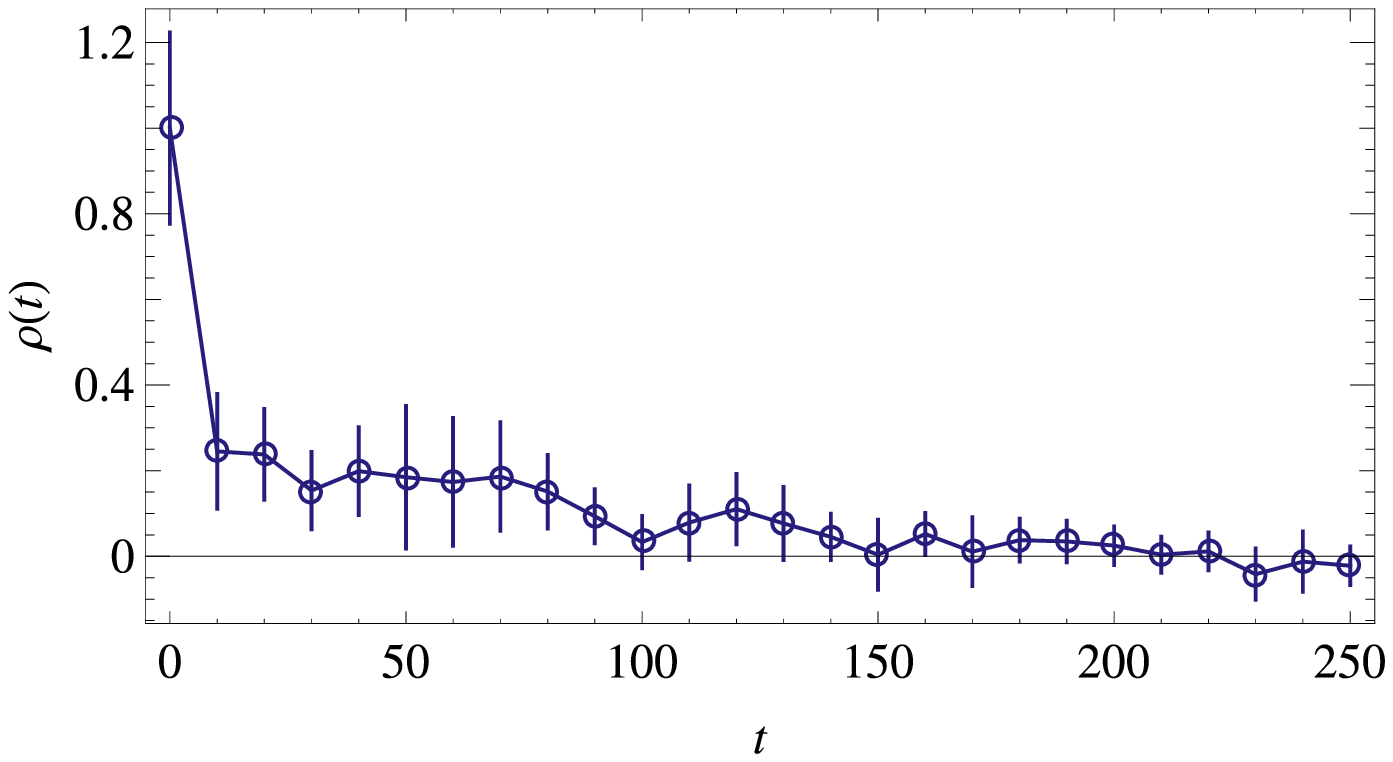}
\includegraphics[width=0.45\textwidth]{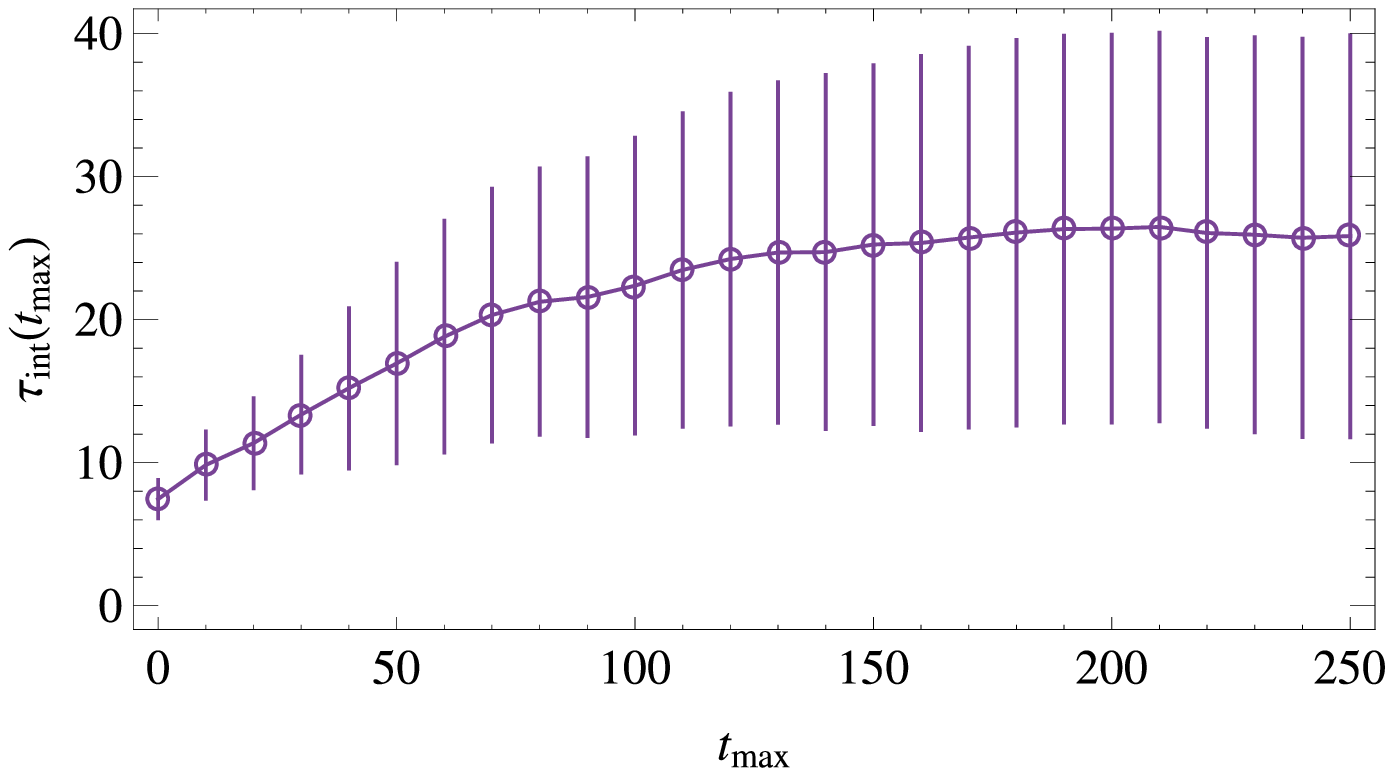}
\caption{Autocorrelation $\rho(t)$ and integrated autocorrelation length
  $\tau_{\rm int}$ (in trajectories) for the pion (above) and proton (below) correlator at $t=30$
    from the ensemble with $16^3\times 128$ volume and $a_tm_l=-0.0808$.}
\label{fig:correlator_auto}
\end{figure}

The autocorrelation function is defined as
\begin{eqnarray}\label{eq:auto-corr}
\rho(t) &=& \langle ({\cal O}(t^\prime)-\langle {\cal O} \rangle)({\cal
    O}(t^\prime+t)-\langle {\cal O} \rangle) \rangle,
\end{eqnarray}
where $\langle ... \rangle$ means taking an average over the samples, $t$ is the trajectory difference in the autocorrelation (from 1 to $N$ total trajectories), and different $t^\prime$ (also indexing trajectory number) are averaged. To calculate the integrated autocorrelation length $\tau_{\rm int}$ with jackknife-estimated errorbar, we first divide the configurations into blocks of size $N_b$; we calculate $\rho_j(t)$ for jackknife index $j$ by ignoring contributions when either $t$ or $t^\prime+t$ is located within the $j^{\rm th}$ block and replacing $\langle {\cal O} \rangle$ by $\langle {\cal O} \rangle_j$, the mean value without the $j^{\rm th}$ block. With a jackknife data set of length $N/N_b$, we calculate integrated autocorrelation length,
\begin{eqnarray}\label{eq:int-auto-corr}
\tau_{\rm int}(t_{\rm max}) &=& \frac{1}{2} +\frac{1}{\rho(t=0)}\sum_{t=1}^{t_{\rm max}}
\rho (t),
\end{eqnarray}
using standard jackknife procedure. The autocorrelations of the spatial
plaquette from gauge ensemble $a_tm_l=-0.0808$, $16^3\times 128$ are shown in
Figure~\ref{fig:plaq_auto}; the integrated autocorrelation length for the
stout-smeared plaquette is about 30 trajectories, which is around twice as large as the un-smeared ones. The integrated autocorrelation length for lowest light and strange eigenvalues are around 13 and 10 trajectories respectively; shown in Figure~\ref{fig:mdagm_auto}. Figure~\ref{fig:correlator_auto} shows the case of pion and proton correlators
at $t=30$ on our largest spectrum measurement (518 configurations) ensemble,
   $a_tm_l=-0.0808$, $16^3\times 128$. The integrated autocorrelation length is about 30 trajectories.

\section{Measurement Techniques}\label{Sec:Scale}
In this section, the methods used to determine relevant spectroscopy data
on the Monte Carlo ensembles are described.
We have used well-established lattice spectroscopy technology throughout this
calculation.

To better access ground-state correlation functions, we use the variational
method\cite{Michael:1985ne,Luscher:1990ck}. Consider the generalized
eigenvalue problem
\begin{eqnarray}
C(t)v=\lambda(t,t_0)C(t_0)v,
\end{eqnarray}
where $t_0$ is chosen as the earliest time at which our model (given below) well describes the correlator $C$. $C_{ij}(t)$ is a two-point correlation function, composed from the operators ${\cal O}_i$ and ${\cal O}_j$. The correlation matrix can be approximated by a sum over the lowest $N$ states:
\begin{eqnarray}
C_{ij}(t) &=& \sum_{n=1}^\infty {z_{ni}}^* z_{nj} e^{-E_n(t-t_0)} \\
    &\approx& \sum_{n=1}^N {u_{ni}}^* u_{nj} e^{-E_n(t-t_0)},
\end{eqnarray}
where $E_n$ is the energy of the $n^{\rm th}$ state, and $u_m \cdot z_n = \delta_{mn}$. We extract the energies from the eigenvalues
\begin{eqnarray}
\lambda_n(t,t_0)=e^{-E_n(t-t_0)},
\end{eqnarray}
which are obtained by solving
\begin{eqnarray}
C(t_0)^{-1/2}C(t)C(t_0)^{-1/2}v_n = \lambda_n(t,t_0)v_n.
\end{eqnarray}

\subsection{Hadron Correlation Functions}\label{sec:Measurements}

We perform measurements starting from trajectory 1000 on every $10^{\rm th}$ trajectory, using the EigCG inverter (developed by A.~Stathopoulos et al. in Ref.~\cite{Stathopoulos:2007zi}) to calculate quark propagators (with CG residual set to $10^{-8}$). We use 4 sources on each configuration, where a random source location is selected for the first source, and the remaining three are uniformly shifted by $N_{x,y,z}/2$ and $N_t/4$; this arrangement should reduce potential autocorrelations between configurations.  We bin the data over spans of 5 measurements.

\begin{figure}
\includegraphics[width=0.85\textwidth]{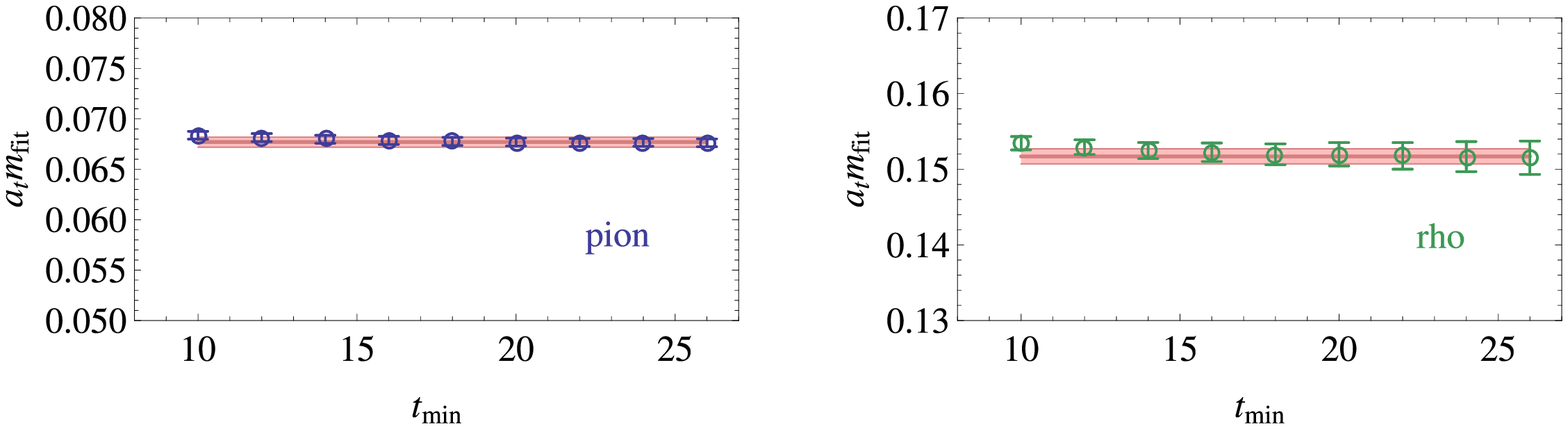}
\includegraphics[width=0.85\textwidth]{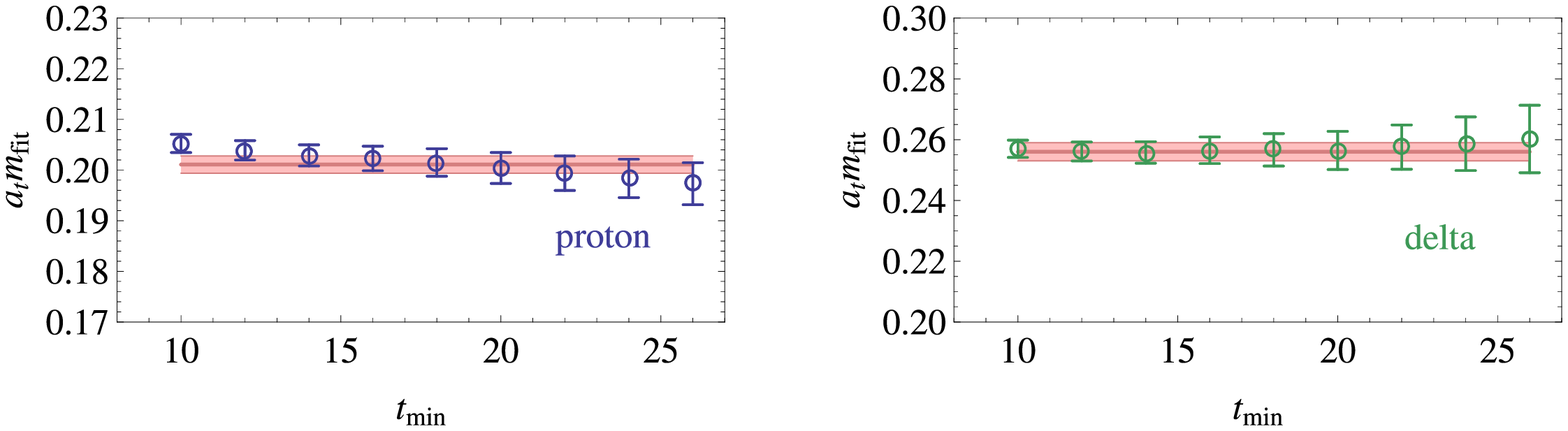}
\caption{Pion and rho (above) and proton and Delta (below) fitted masses as
  functions of $t_{\rm min}$ where $t_{\rm source}$ has been shifted to $t=0$.
  The bands indicated the final fitted masses summarized in
  Tables~\ref{tab:mixed_mesons} and \ref{tab:mixed_baryons}.
}
\label{fig:tmin}
\end{figure}

In this work, we construct a $3\times3$  correlator matrix $C_{ij}$ by using 3 different Gaussian smearing widths ($\sigma\in\{3.0,5.0,6.5\}$) on the hadron operators. We extract the ground-state principal correlator and fit the ground-state mass using a cosh form. (We also try an exponential form on the principal correlator, and the fit results are consistent.) The $t_{\rm min}$ dependences (with $t_{\rm max} \approx 50$) of the fitted masses are shown in Figure~\ref{fig:tmin} for the pion, rho, nucleon and Delta. The fitted masses are very consistent between various choices of starting time in the fits.

We use meson interpolating fields of the form $\bar{q}\Gamma q$, which overlap with the physical states listed in Table~\ref{tab:meson_operator}; charge conjugation $C$ applies only to particles with zero net flavor. The estimated $\eta$ mass is $\sqrt{\frac{1}{3}m_\pi^2+\frac{2}{3}m_{s\bar{s}}^2}$. The ground-state masses are summarized in Table~\ref{tab:mixed_mesons}.
We have two volumes ($12^3$ and $16^3$) of the lightest ensemble, $m_s=-0.0540$,
   and two ($16^3$ and $24^3$) on $a_tm_s=-0.0743$: no major finite-volume
   effects are observed, except for the $a_0$ mass from the $a_tm_s=-0.0743$ ensemble and baryon states from $a_tm_s=-0.0540$.
\begin{table}
\begin{center}
\begin{tabular}{c|c|cccc}
\hline\hline
$J^{PC}$ & $\Gamma$                 & $I=1$  & $I=0$      & $S=1$     \\
\hline
$0^{-+}$ & $\gamma_5$               & $\pi$  & $s \bar{s}$ & $K$     \\
$1^{--}$ & $\gamma_\mu$             & $\rho$ & $\phi$     & $K^*$   \\
$0^{++}$ & 1                        & $a_0$ & & \\ 
$1^{++}$ & $\gamma_\mu\gamma_5$     & $a_1$  & & \\  
$1^{+-}$ & $\gamma_\mu\gamma_\nu$   & $b_1$  & & \\  
\hline\hline
\end{tabular}
\end{center}
\caption{\label{tab:meson_operator} Meson interpolating operators. States
  are sorted into columns according to the degree of strangeness from 0 (left two columns) to 1 (right column) and then according to total isospin.}
\end{table}
\begin{table}
{\footnotesize	
\begin{tabular}{cccc|ccccccccccc}
\hline\hline
\hline
$N_s$ & $N_t$ & $a_tm_l$ & $a_tm_s$ & $a_tm_\pi$ & $a_tm_K$ & $a_tm_\eta$ &
$a_tm_\rho$ & $a_tm_{K^*}
$ & $a_tm_\phi$ & $a_tm_{a_0} $ & $a_tm_{a_1} $ & $a_tm_{b_1} $ & $m_\rho/m_\pi$ & $N_{\rm cfg}$ \\
\hline
12 & 96 & $-$0.0540 & $-$0.0540 &  0.2781(9) &  0.2781(9) &  0.2781(9) &  0.334(3) &  0.334(3) &  0.334(3) &  0.44(4) &  0.474(15) &  0.480(18) &  0.833(7) & 92 \\
12 & 96 & $-$0.0699 & $-$0.0540 &  0.1992(17) &  0.2227(15) &  0.2450(13) &  0.268(2) &  0.2860(21) &  0.3031(18) &  0.37(2) &  0.389(15) &  0.377(13) &  0.742(9) & 110 \\
12 & 96 & $-$0.0794 & $-$0.0540 &  0.1393(17) &  0.1841(13) &  0.2231(11) &  0.201(7) &  0.236(5) &  0.268(3) &  0.33(7) &  0.317(14) &  0.330(16) &  0.69(2) & 95 \\
12 & 96 & $-$0.0826 & $-$0.0540 &  0.1144(19) &  0.1691(17) &  0.2142(15) &  0.194(7) &  0.232(4) &  0.266(3) &  0.22(4) &  0.306(15) &  0.266(19) &  0.59(2) & 84 \\
16 & 96 & $-$0.0826 & $-$0.0540 &  0.113(3) &  0.1669(15) &  0.2112(15) &  0.185(5) &  0.222(4) &  0.258(3) &  0.28(4) &  0.28(3) &  0.28(2) &  0.61(2) & 25 \\
12 & 96 & $-$0.0618 & $-$0.0618 &  0.2322(15) &  0.2322(15) &  0.2322(15) &  0.286(5) &  0.286(5) &  0.286(5) &  0.415(20) &  0.436(15) &  0.459(18) &  0.812(12) & 50 \\
16 & 128 & $-$0.0743 & $-$0.0743 &  0.1483(2) &  0.1483(2) &  0.1483(2) &
 0.2159(6) &  0.2159(6) &  0.2159(6) &  0.287(6) &  0.317(5) &  0.325(5) &
 0.6867(17) & 79 \\
16 & 128 & $-$0.0808 & $-$0.0743 &  0.0996(6) &  0.1149(6) &  0.1196(5) &
 0.173(2) &  0.1819(21) &  0.1901(18) &  0.222(11) &  0.252(6) &  0.269(5) &
 0.574(6) & 518 \\
16 & 128 & $-$0.0830 & $-$0.0743 &  0.0797(6) &  0.1032(5) &  0.1100(4) &
 0.1623(16) &  0.1733(10) &  0.1845(11) &  0.196(18) &  0.236(8) &  0.263(8) &
 0.491(6) & 266 \\
16 & 128 & $-$0.0840 & $-$0.0743 &  0.0691(6) &  0.0970(5) &  0.1047(5) &
 0.154(3) &  0.1663(16) &  0.1788(13) &  0.159(15) &  0.222(7) &  0.238(8) &
 0.448(7) & 224 \\
24 & 128 & $-$0.0840 & $-$0.0743 &  0.0681(4) &  0.0966(3) &  0.1045(3) &
 0.1529(10) &  0.1660(6) &  0.1788(6) &  0.194(14) &  0.233(4) &  0.242(6) &
 0.446(3) & 287 \\
\hline\hline
\end{tabular}
}
\caption{\label{tab:mixed_mesons}Meson masses for $N_f=3$ and $N_f=2+1$ (in
    temporal lattice units).}
\end{table}

\begin{figure}
\includegraphics[width=0.32\textwidth]{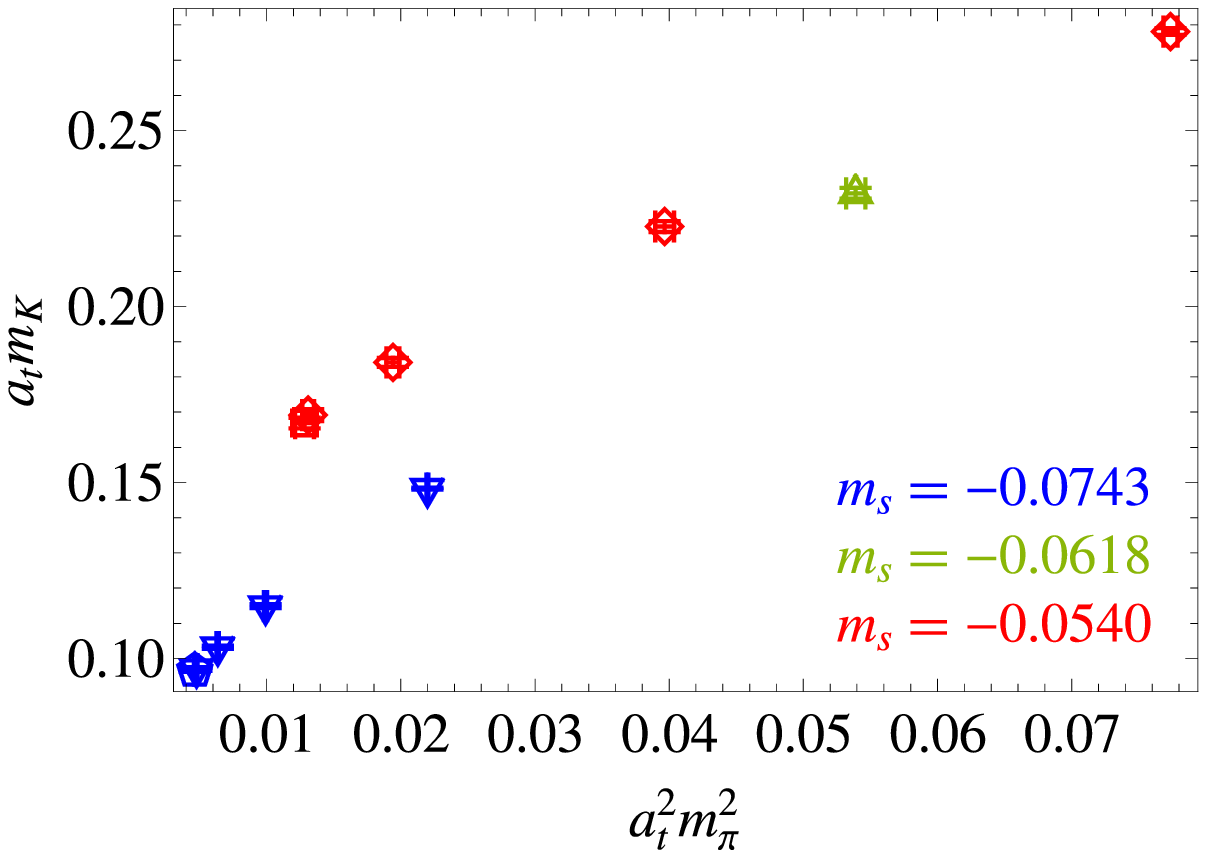}
\includegraphics[width=0.32\textwidth]{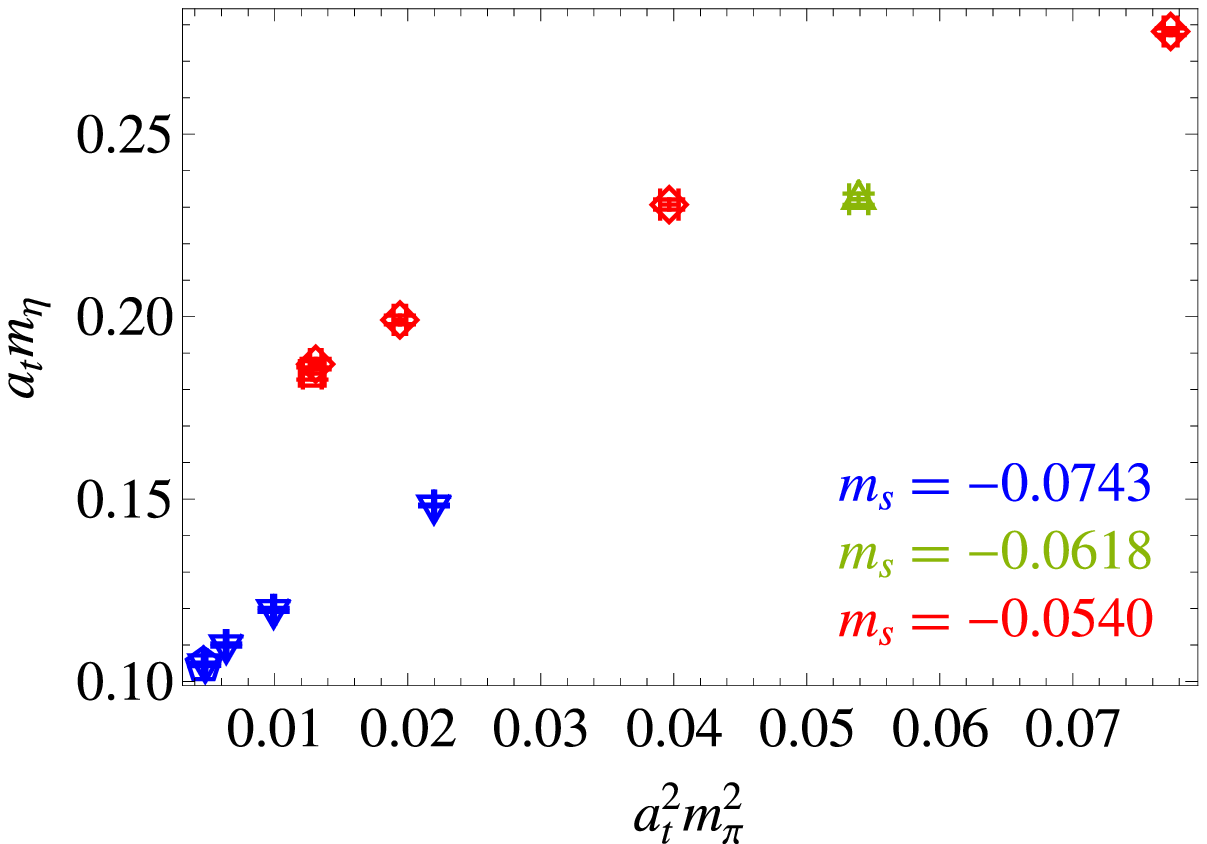}
\includegraphics[width=0.32\textwidth]{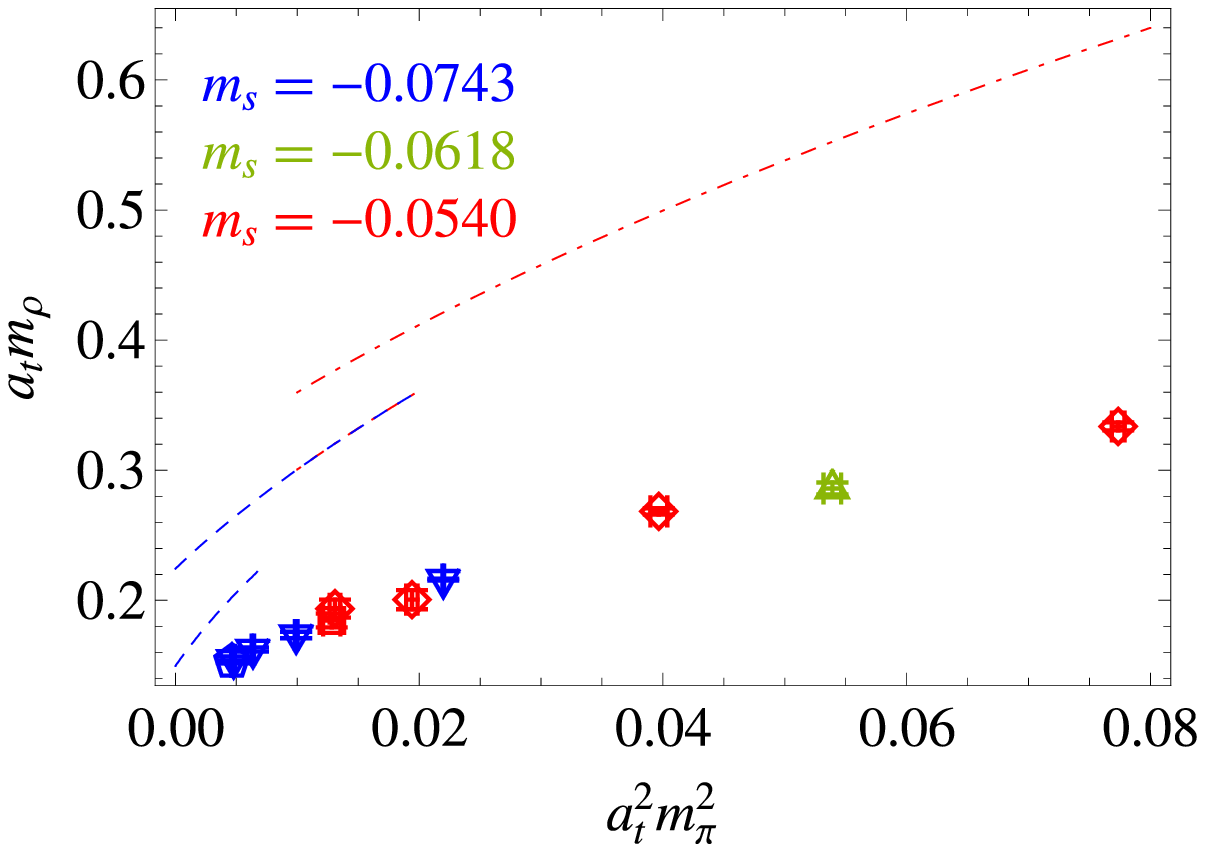}
\includegraphics[width=0.32\textwidth]{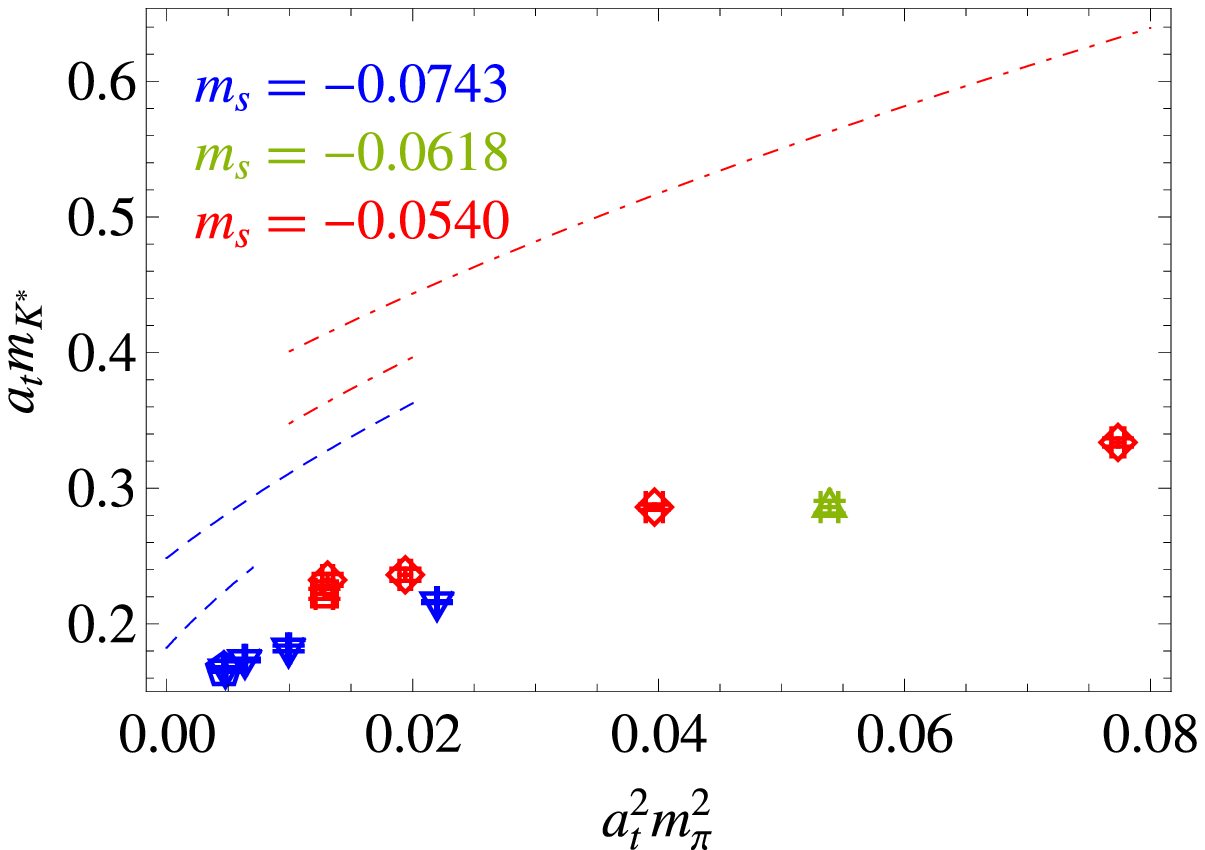}
\includegraphics[width=0.32\textwidth]{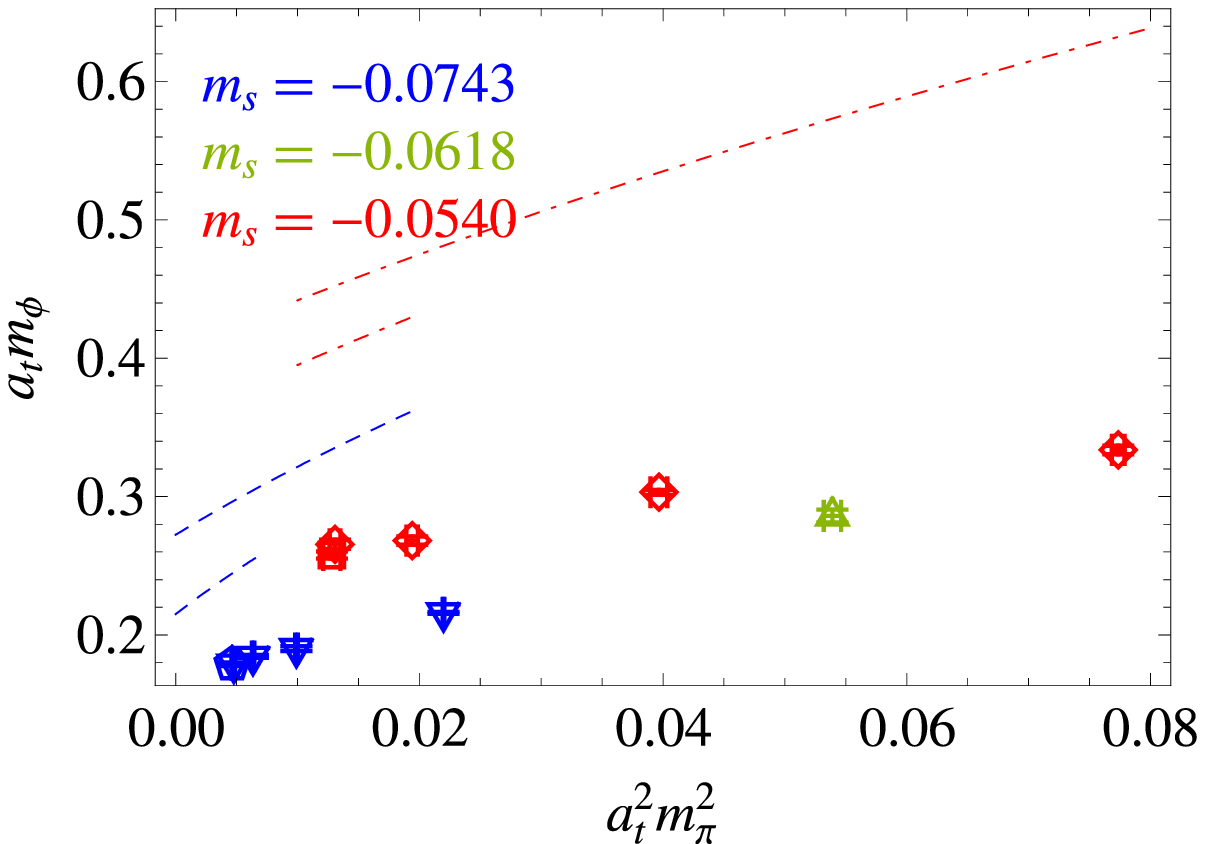}
\includegraphics[width=0.32\textwidth]{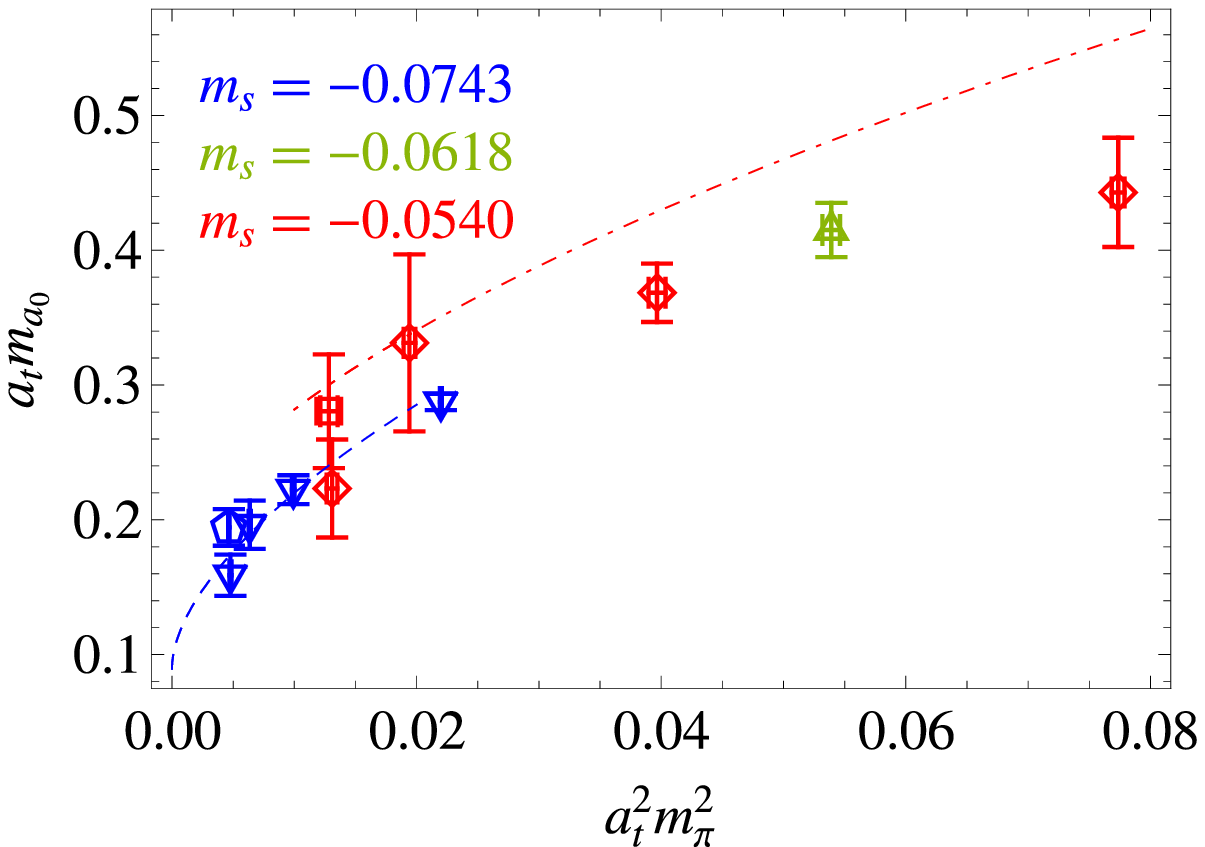}
\includegraphics[width=0.32\textwidth]{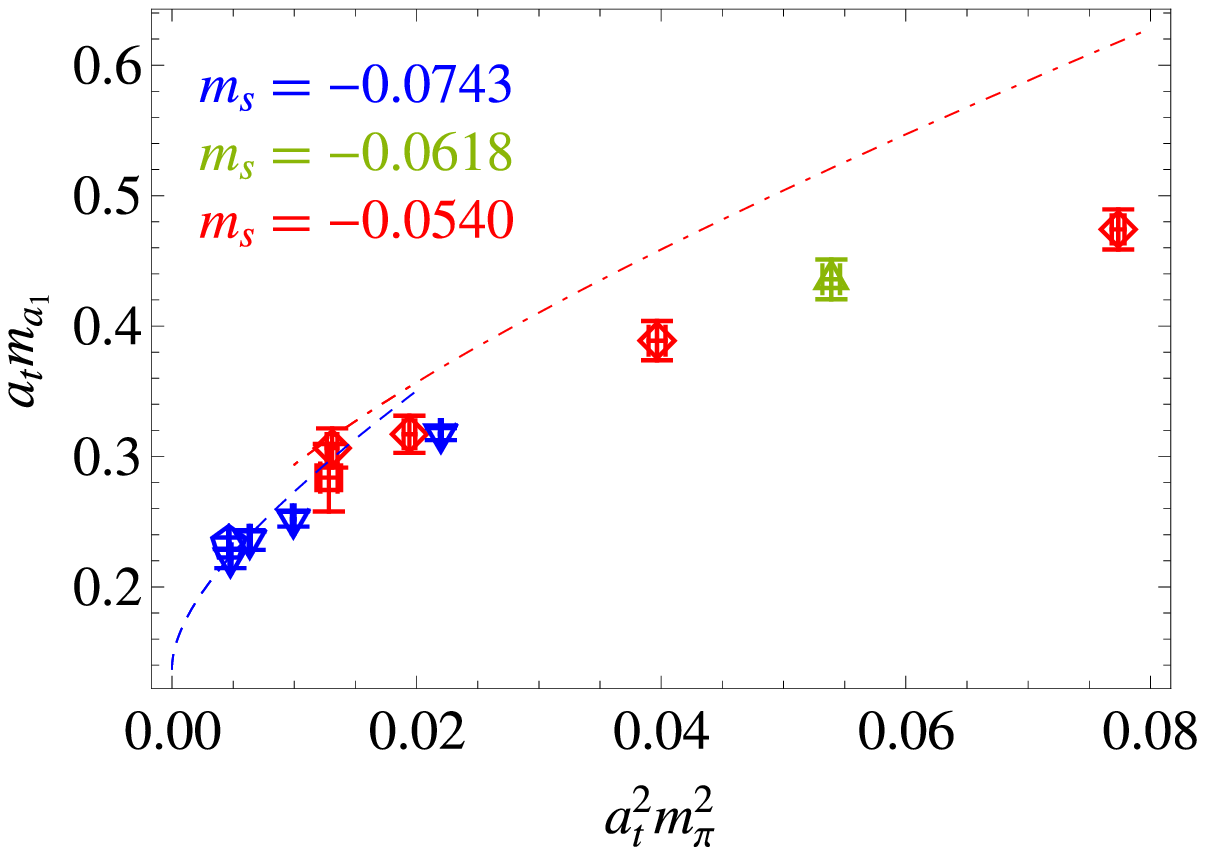}
\includegraphics[width=0.32\textwidth]{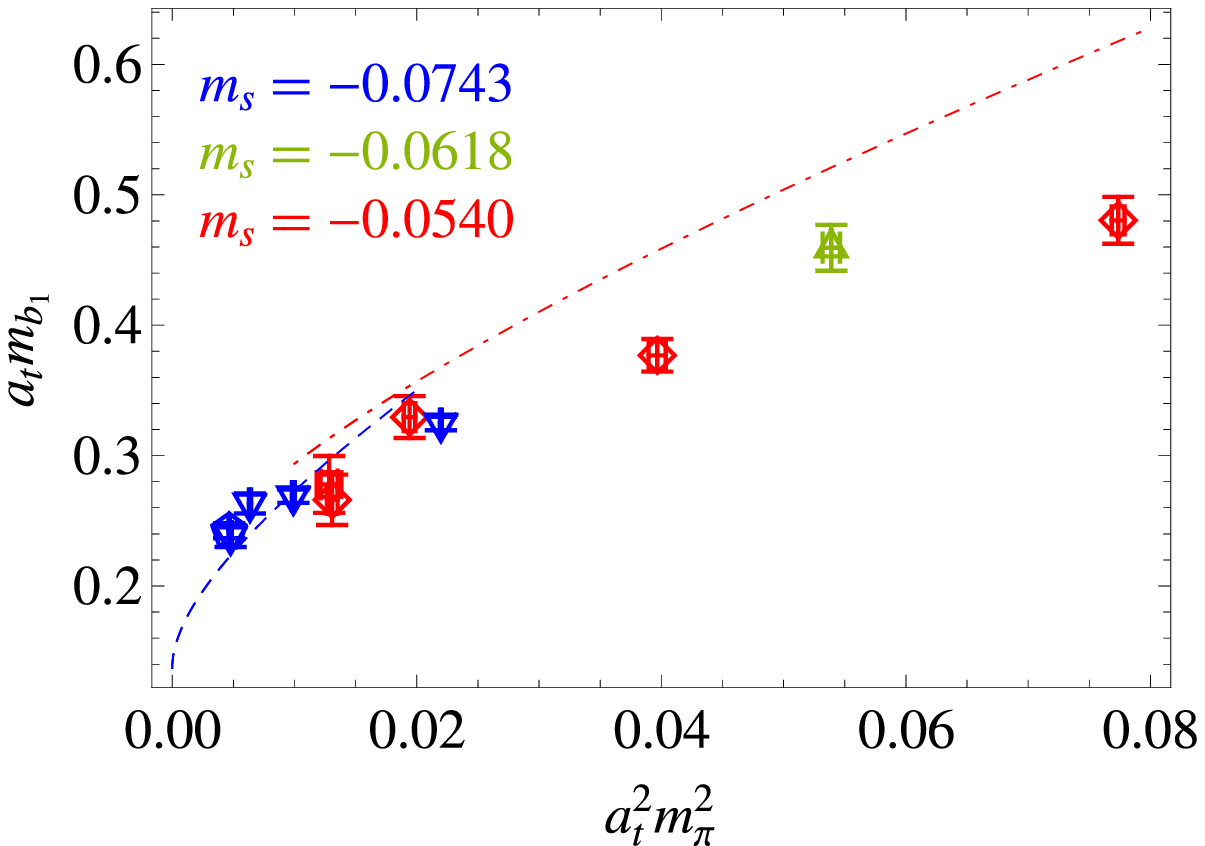}
\includegraphics[width=0.32\textwidth]{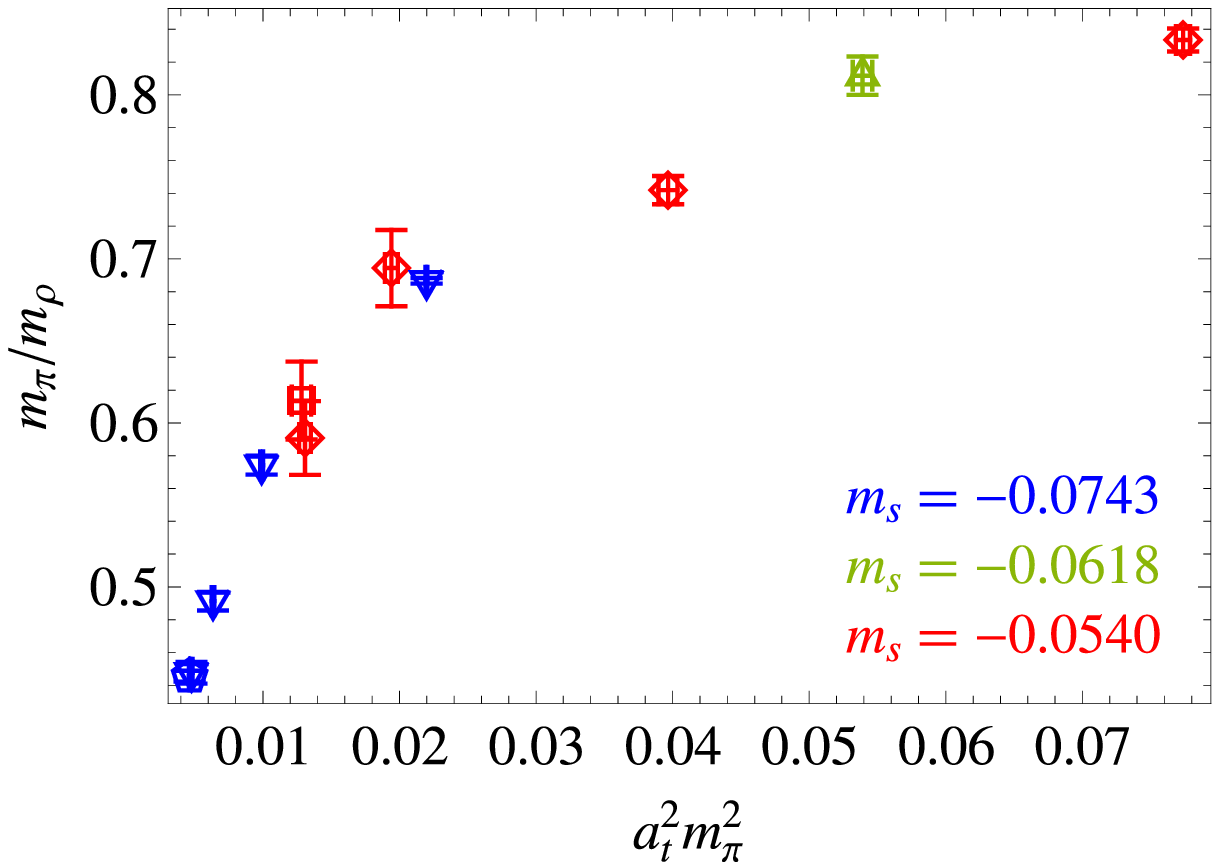}
\caption{All measured meson masses as functions of the squared pseudoscalar
  masses. The diamonds and squares are measured with $m_s=-0.0540$ but with two
    different volumes, $12^3\times96$ and $16^3\times96$; the upward-pointing
    triangles are those with $m_s=-0.0618$ and $12^3\times96$ volume; the
    downward triangles and pentagons are measured with $m_s=-0.0743$ and two
    different volumes, $16^3\times128$ and $24^3\times128$. The (red) dot-dashed
    lines indicate the decay thresholds for the $12^3$ (upper) and $16^3$
    (lower) $m_s=-0.0540$ ensembles, while the (blue) dashed lines are for the
    $16^3$ (upper) and $24^3$ (lower) $m_s=-0.0743$. The lowest decay thresholds
    are: $\rho \rightarrow \pi (p) + \pi (-p)$, ${K^*} \rightarrow K (p) + \pi
    (-p)$, $\phi \rightarrow K (p) + \overline{K}(-p)$, $a_0 \rightarrow \pi (0) + \eta (0)$, $a_1 \rightarrow \pi (0) + \rho (0)$ and $b_1 \rightarrow \pi (0) + \omega (0)$ (with $\omega$ approximated by $\rho$) where the minimum allowed momentum $p$ on the lattice is $\frac{2\pi}{L_s}$.
}
\label{fig:all-meson}
\end{figure}

\begin{figure}
\includegraphics[width=0.32\textwidth]{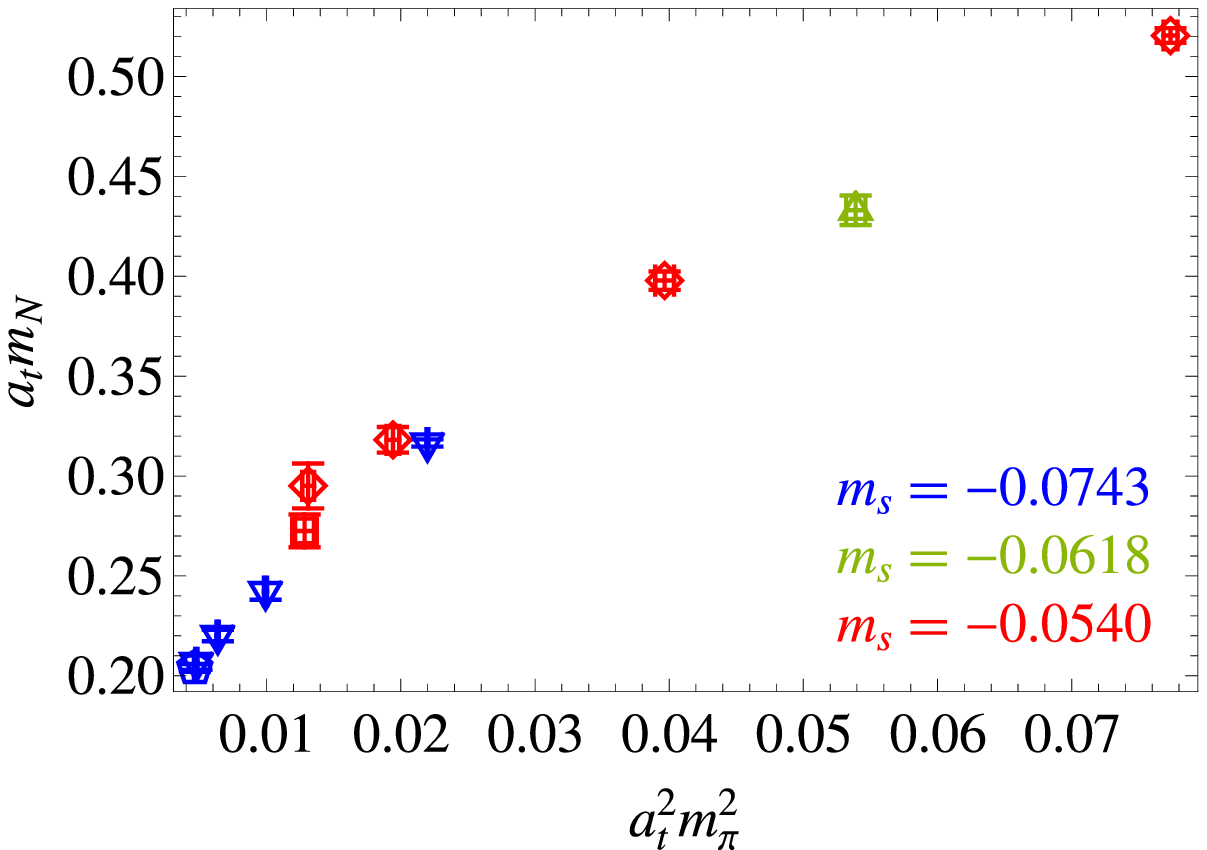}
\includegraphics[width=0.32\textwidth]{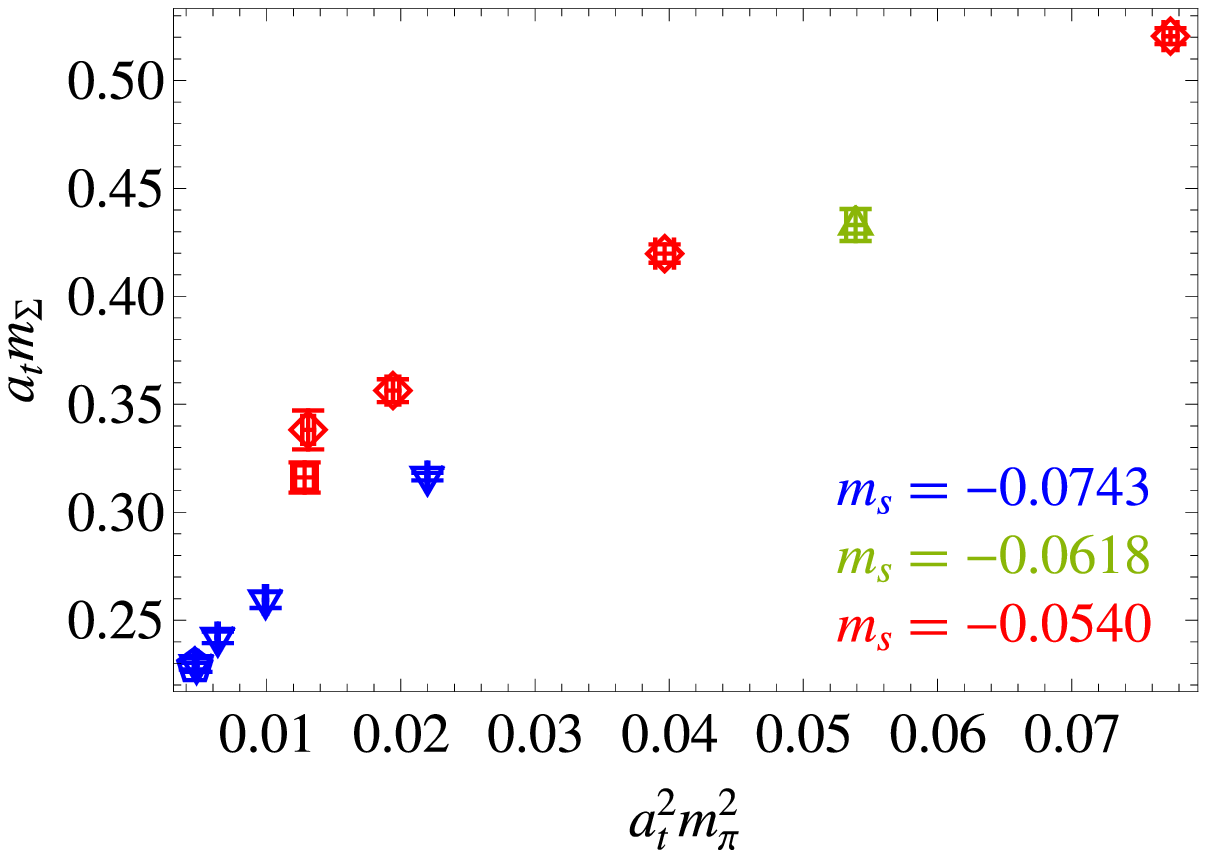}
\includegraphics[width=0.32\textwidth]{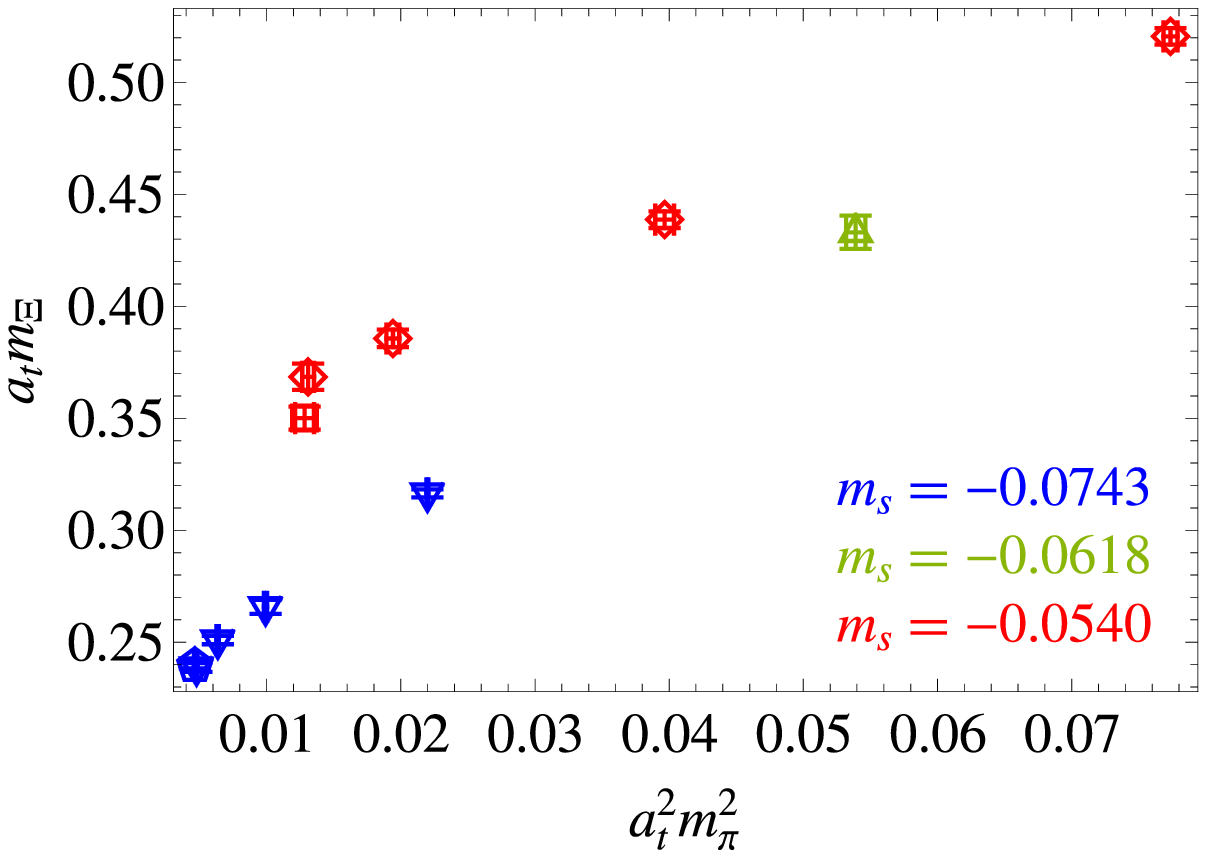}
\includegraphics[width=0.32\textwidth]{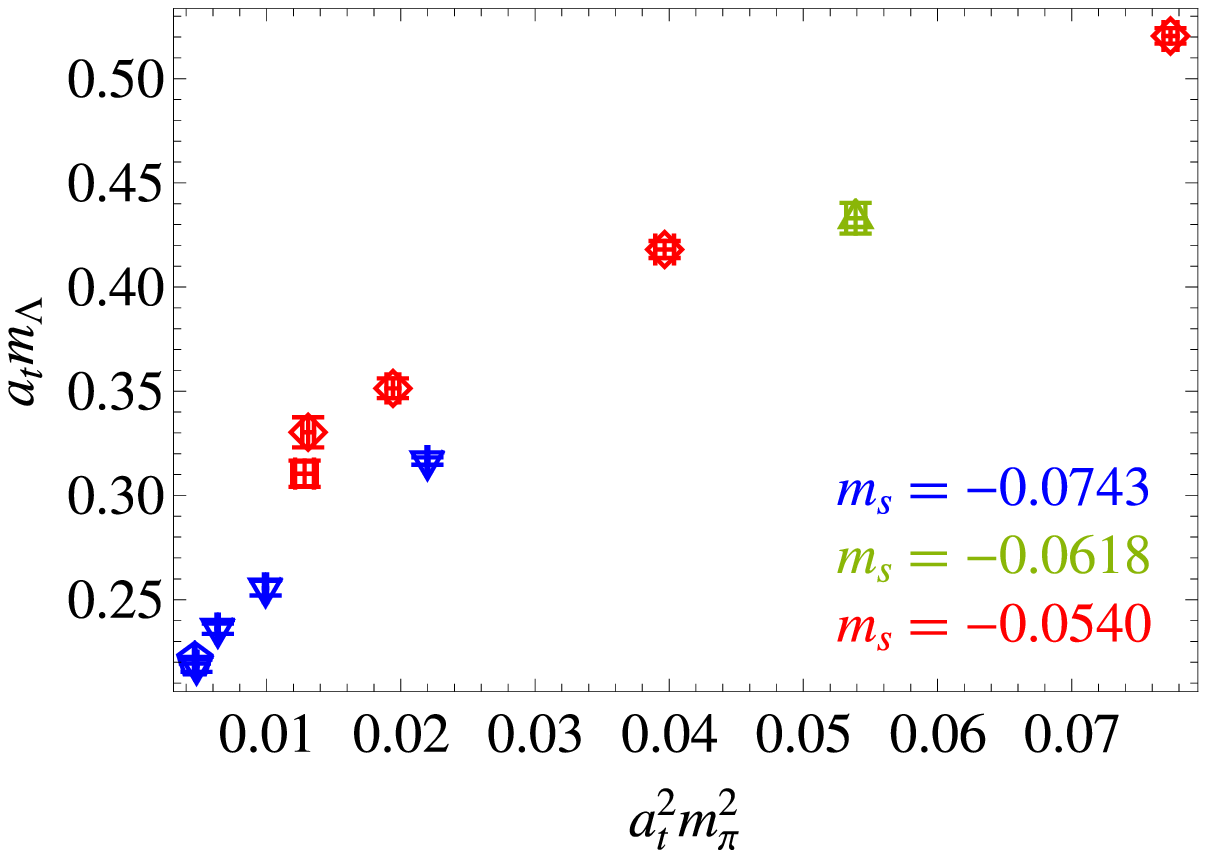}
\includegraphics[width=0.32\textwidth]{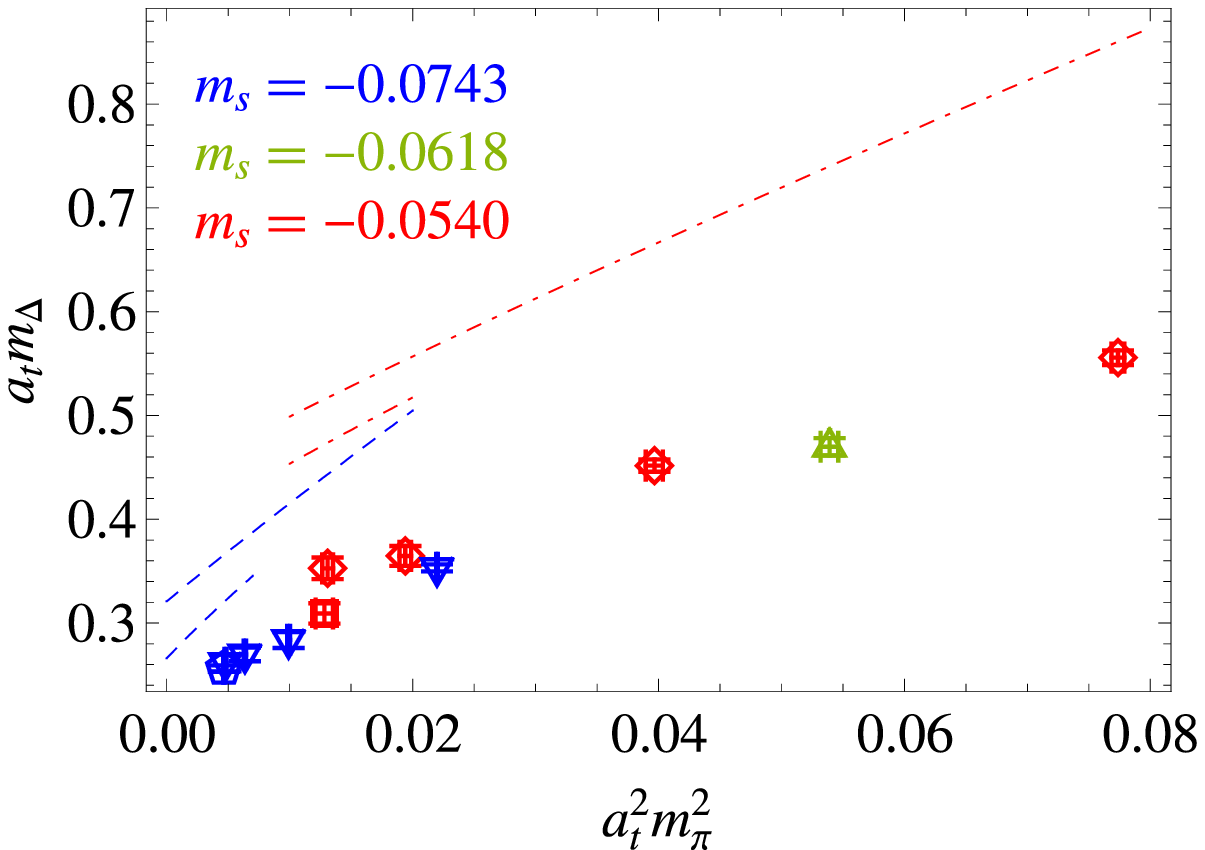}
\includegraphics[width=0.32\textwidth]{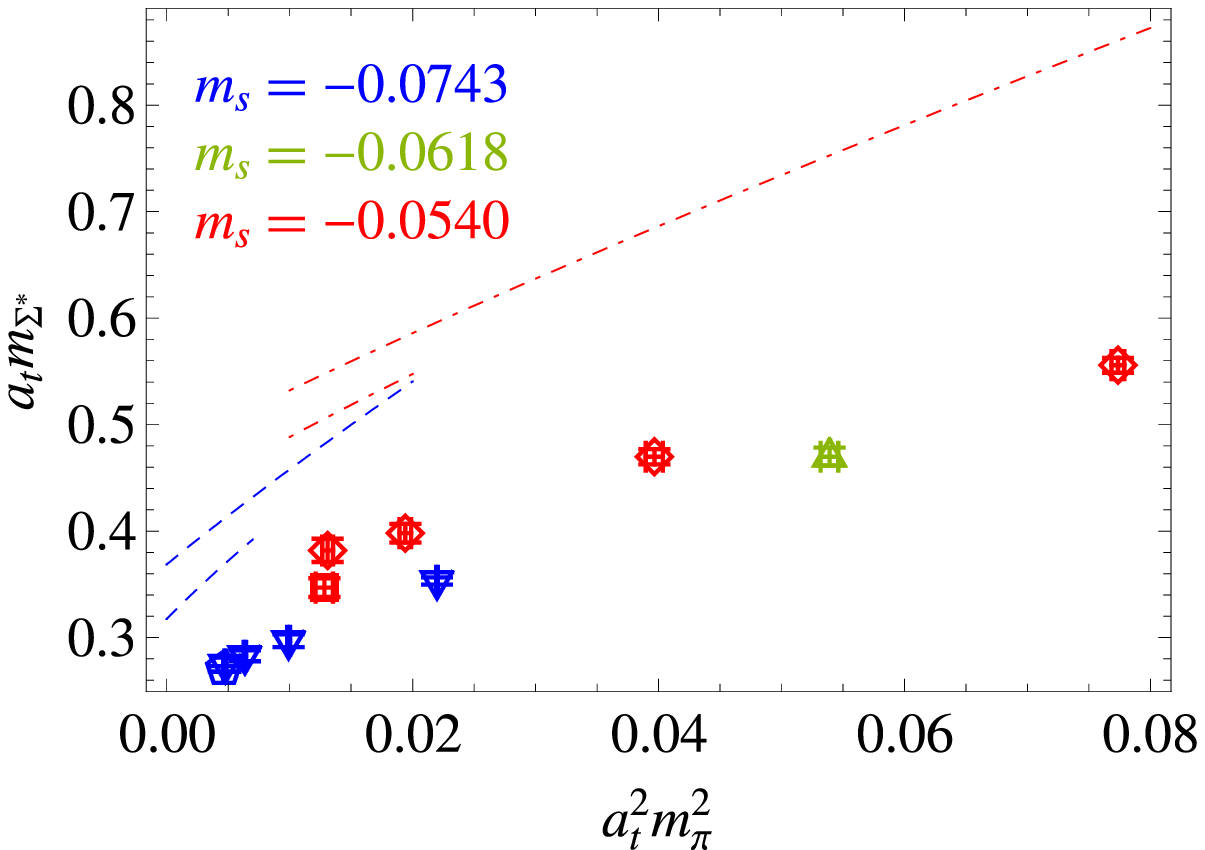}
\includegraphics[width=0.32\textwidth]{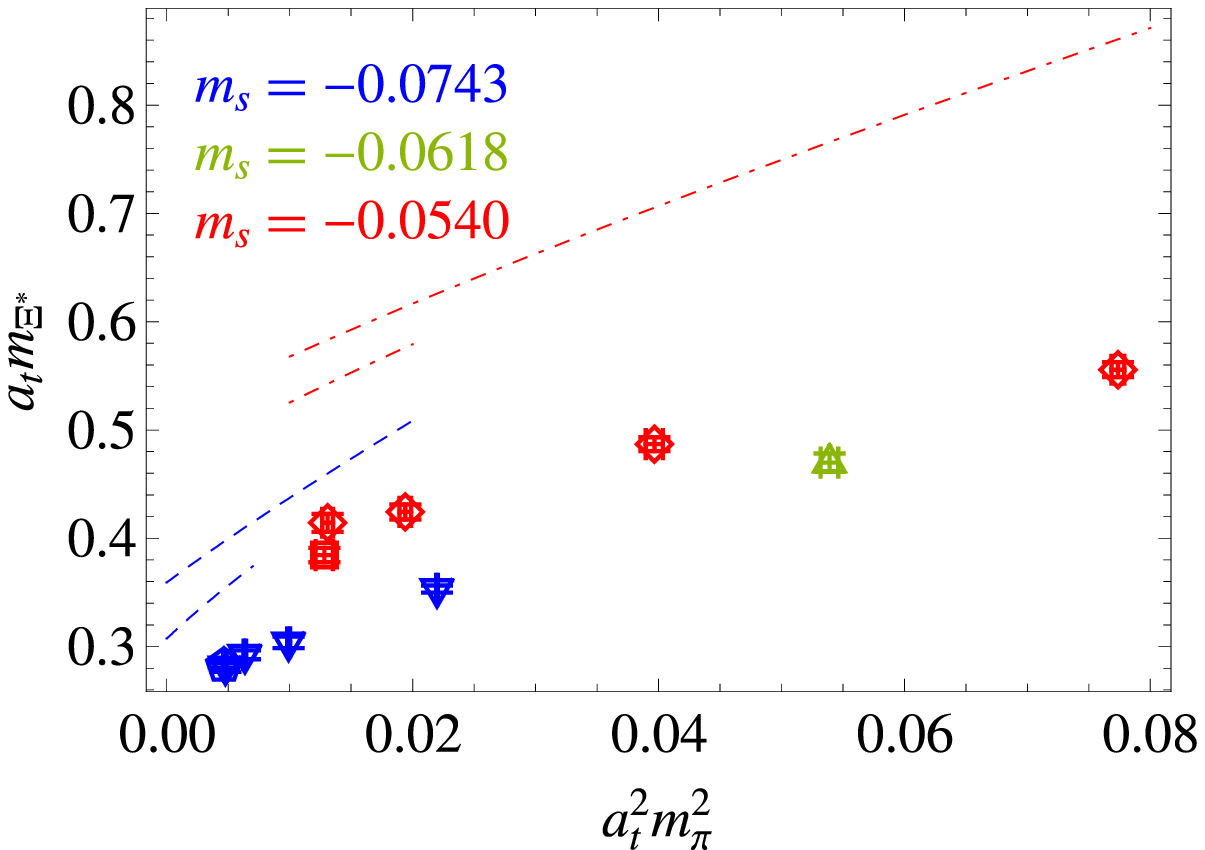}
\includegraphics[width=0.32\textwidth]{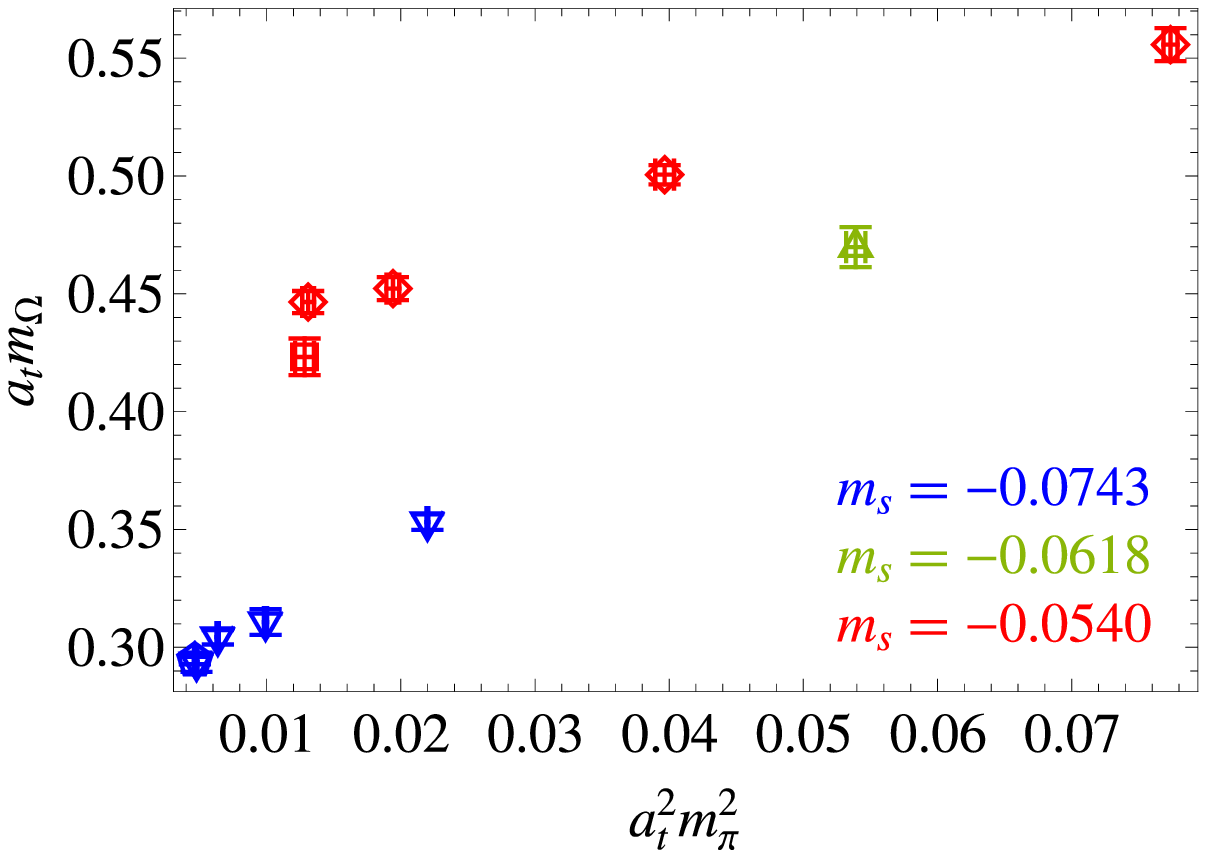}
\caption{All measured baryon masses as functions of the squared pseudoscalar
  masses. The diamonds and squares are measured with $m_s=-0.0540$ but with two
    different volumes, $12^3\times96$ and $16^3\times96$; the upward-pointing
    triangles are those with $m_s=-0.0618$ and $12^3\times96$ volume; the
    downward triangles and pentagons are measured with $m_s=-0.0743$ and two
    different volumes, $16^3\times128$ and $24^3\times128$. The (red) dot-dashed
    lines indicate the decay thresholds for the $12^3$ (upper) and $16^3$
    (lower) $m_s=-0.0540$ ensembles, while the (blue) dashed lines are for the
    $16^3$ (upper) and $24^3$ (lower) $m_s=-0.0743$. The lowest decay threshold
    are: $\Delta \rightarrow N (p) + \pi (-p)$,  $\Sigma^* \rightarrow \Lambda
    (p) + \pi (-p)$, $\Xi^* \rightarrow \Xi (p) + \pi (-p)$ where the minimum allowed momentum $p$ on the lattice is $\frac{2\pi}{L_s}$.
}
\label{fig:all-baryon}
\end{figure}

The octet baryons are calculated using the interpolating field $(q_1 C \gamma_4 \gamma_5 q_2) q_1$ (with $q_i=u/d$ or $s$ quark); the $\Lambda$ uses $2(u C \gamma_5 d) s + (s C \gamma_5 d) u + (u C \gamma_5 s) d$; and the decuplet uses $2(q_2 C (1/2)(1+\gamma_4) \gamma_- q_1) q_1 + (q_1 C (1/2)(1+\gamma_4) \gamma_- q_1) q_2$ (with $\gamma_-=\gamma_x-\gamma_y$). The calculated octet and decuplet ground-state masses are summarized in Table~\ref{tab:mixed_baryons}.
We observe a finite-volume discrepancy in the baryon sector on the lightest
ensemble, $a_tm_s=-0.0540$. When we extrapolate the hadron masses to the
physical limit, we will exclude the small volume sets: $12^3$ with
$a_tm_s=-0.0540$ and $16^3$ with $a_tm_s=-0.0743$.
\begin{table}
\begin{center}
\begin{tabular}{cccc|ccccccccccc}
\hline\hline
$N_s$ & $N_t$ & $a_tm_l$ & $a_tm_s$ & $a_tm_\pi$ & $a_tm_N$ & $a_tm_\Sigma$ &
$a_tm_\Xi$ &
$a_tm_\Lambda$ & $a_tm_\Delta$ & $a_tm_{\Sigma^*}$ & $a_tm_{\Xi^*}$ & $a_tm_\Omega$ & $N_{\rm
  cfg}$ \\
\hline
 12 & 96 & $-$0.0540 & $-$0.0540 &  0.2781(9) &  0.521(4) &  0.521(4) &  0.521(4) &  0.521(4) &  0.556(7) &  0.556(7) &  0.556(7) &  0.556(7) & 92 \\
 12 & 96 & $-$0.0699 & $-$0.0540 &  0.1992(17) &  0.398(5) &  0.420(4) &  0.439(4) &  0.418(4) &  0.452(6) &  0.470(7) &  0.487(6) &  0.501(4) & 110 \\
 12 & 96 & $-$0.0794 & $-$0.0540 &  0.1393(17) &  0.318(6) &  0.356(5) &  0.386(4) &  0.351(5) &  0.365(10) &  0.398(9) &  0.424(7) &  0.452(5) & 95 \\
 12 & 96 & $-$0.0826 & $-$0.0540 &  0.1144(19) &  0.295(11) &  0.338(9) &  0.369(6) &  0.330(7) &  0.353(10) &  0.382(11) &  0.414(8) &  0.447(5) & 84 \\
 16 & 96 & $-$0.0826 & $-$0.0540 &  0.113(3) &  0.273(8) &  0.316(7) &  0.350(5) &  0.310(6) &  0.309(10) &  0.347(9) &  0.385(7) &  0.423(8) & 25 \\
 12 & 96 & $-$0.0618 & $-$0.0618 &  0.2322(15) &  0.433(7) &  0.433(7) &  0.433(7) &  0.433(7) &  0.470(8) &  0.470(8) &  0.470(8) &  0.470(8) & 50 \\
 16 & 128 & $-$0.0743 & $-$0.0743 &  0.1483(2) &  0.3165(18) &  0.3165(18) &
 0.3165(18) &  0.3165(18) &  0.353(3) &  0.353(3) &  0.353(3) &  0.353(3) & 79 \\
 16 & 128 & $-$0.0808 & $-$0.0743 &  0.0996(6) &  0.242(4) &  0.259(4) &
 0.266(3) &  0.256(3) &  0.284(8) &  0.297(6) &  0.304(5) &  0.311(6) & 518 \\
 16 & 128 & $-$0.0830 & $-$0.0743 &  0.0797(6) &  0.220(3) &  0.242(3) &
 0.2510(19) &  0.236(2) &  0.270(7) &  0.283(5) &  0.292(4) &  0.304(3) & 266\\
 16 & 128 & $-$0.0840 & $-$0.0743 &  0.0691(6) &  0.207(4) &  0.229(3) &
 0.239(2) &  0.218(3) &  0.262(8) &  0.275(7) &  0.282(5) &  0.294(4) & 224 \\
 24 & 128 & $-$0.0840 & $-$0.0743 &  0.0681(4) &  0.2039(19) &  0.2287(15) &
 0.2395(12) &  0.2209(15) &  0.256(3) &  0.271(3) &  0.282(2) &  0.2945(16) &
287 \\
\hline\hline
\end{tabular}
\end{center}
\caption{\label{tab:mixed_baryons}Baryon masses for $N_f=3$ and $N_f=2+1$ (in
    temporal lattice units).}
\end{table}
Figures~\ref{fig:all-meson} and \ref{fig:all-baryon} 
show the squared--pion-mass dependence of these quantities. We note that for non-strange hadrons, the sea strange-quark dependences are relatively mild.

At low pion masses, not all the states we calculate on the lattice are safe from
decays. To check which particles may decay, we compare the particle masses to the
threshold two-particle energies in each channel. The vector mesons could decay
to two pseudoscalars in a $P$-wave: $\rho \rightarrow \pi (p) + \pi (-p)$, ${K^*}
\rightarrow K (p) + \pi (-p)$ and $\phi \rightarrow K (p) + \overline{K} (-p)$,
  where the minimum allowed momentum $p$ on the lattice is $\frac{2\pi}{L_s}$.
  In Figure~\ref{fig:all-meson}, we plot the lowest two-particle energy
  threshold for the $m_s=-0.0540$ data with our two lattice extents
  (dot-dashed line) and $a_tm_s=-0.0743$ (dashed). All vector mesons in our calculation are well below threshold. The scalar mesons could decay to $\pi\eta$ in an $S$-wave, which puts the states slightly below the threshold. Similarly for the $a_1$ and $b_1$ mesons: $a_1 \rightarrow \pi (0) + \rho (0)$ and $b_1 \rightarrow \pi (0) + \omega (0)$. In this case, we approximate the $\omega$ by the $\rho$, since their masses are similar.
The $a_0$, $a_1$ and $b_1$ (especially for the $m_s=-0.0743$ ensemble) are slightly below the decay threshold. Fortunately, we have two volumes on the lightest $m_l$ set; if these states are not single-particle, the ratios of their overlap factors between the two volumes would be of order 2 or higher\cite{Mathur:2004jr}. We find the ratios (using point-point correlators) to be $0.87(12)$, $0.97(6)$ and $1.01(6)$, which indicates that our measurements are of single-particle states.
The decuplet baryons are free from decays into an octet baryon. The lowest decay
modes are: $\Delta \rightarrow N (p) + \pi (-p)$, $\Sigma^* \rightarrow \Lambda
(p) + \pi (-p)$, $\Xi^* \rightarrow \Xi (p) + \pi (-p)$, as shown in Figure~\ref{fig:all-baryon}. Overall, most of the particles are stable.

\begin{figure}
\includegraphics[width=0.45\textwidth]{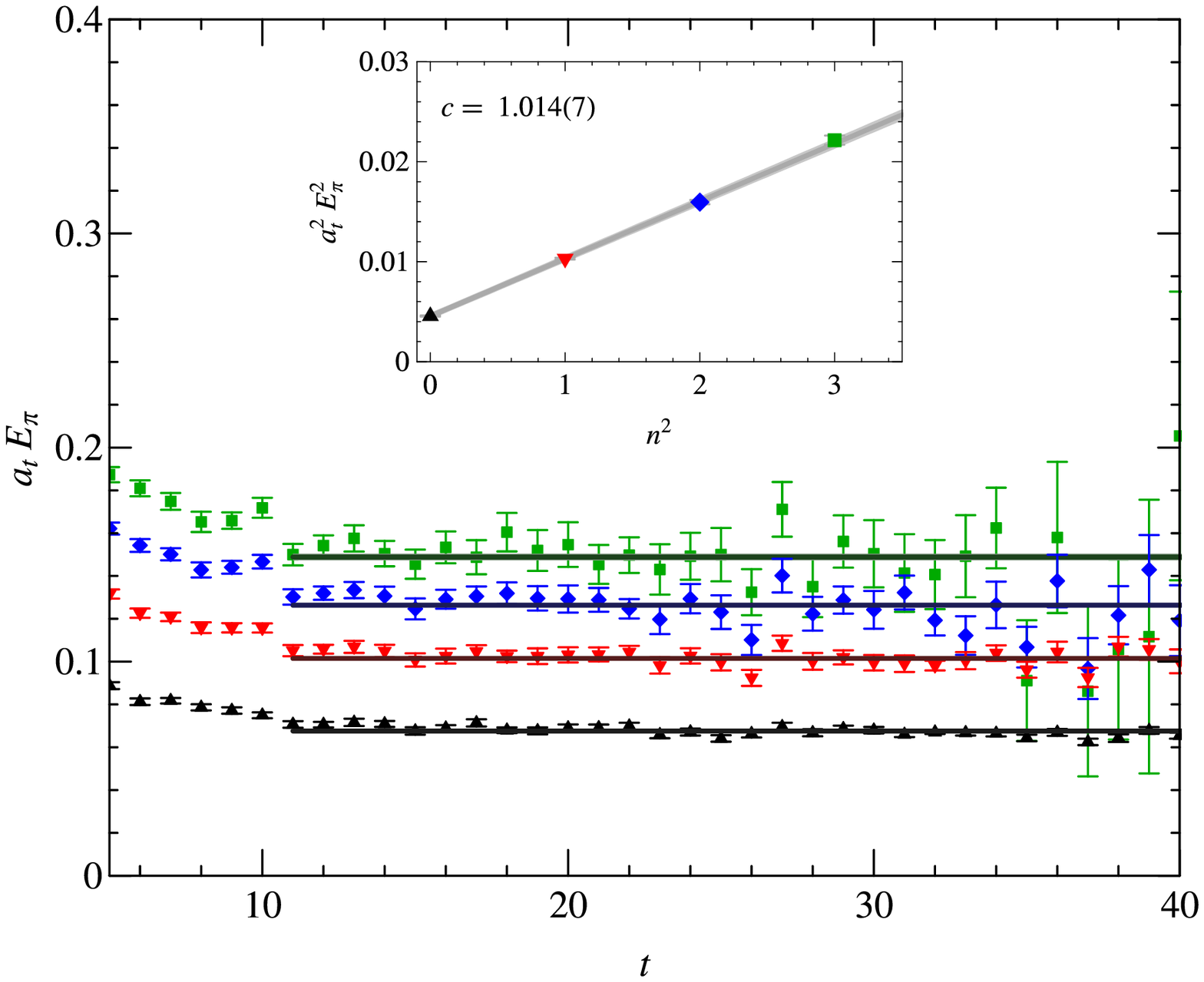}
\includegraphics[width=0.45\textwidth]{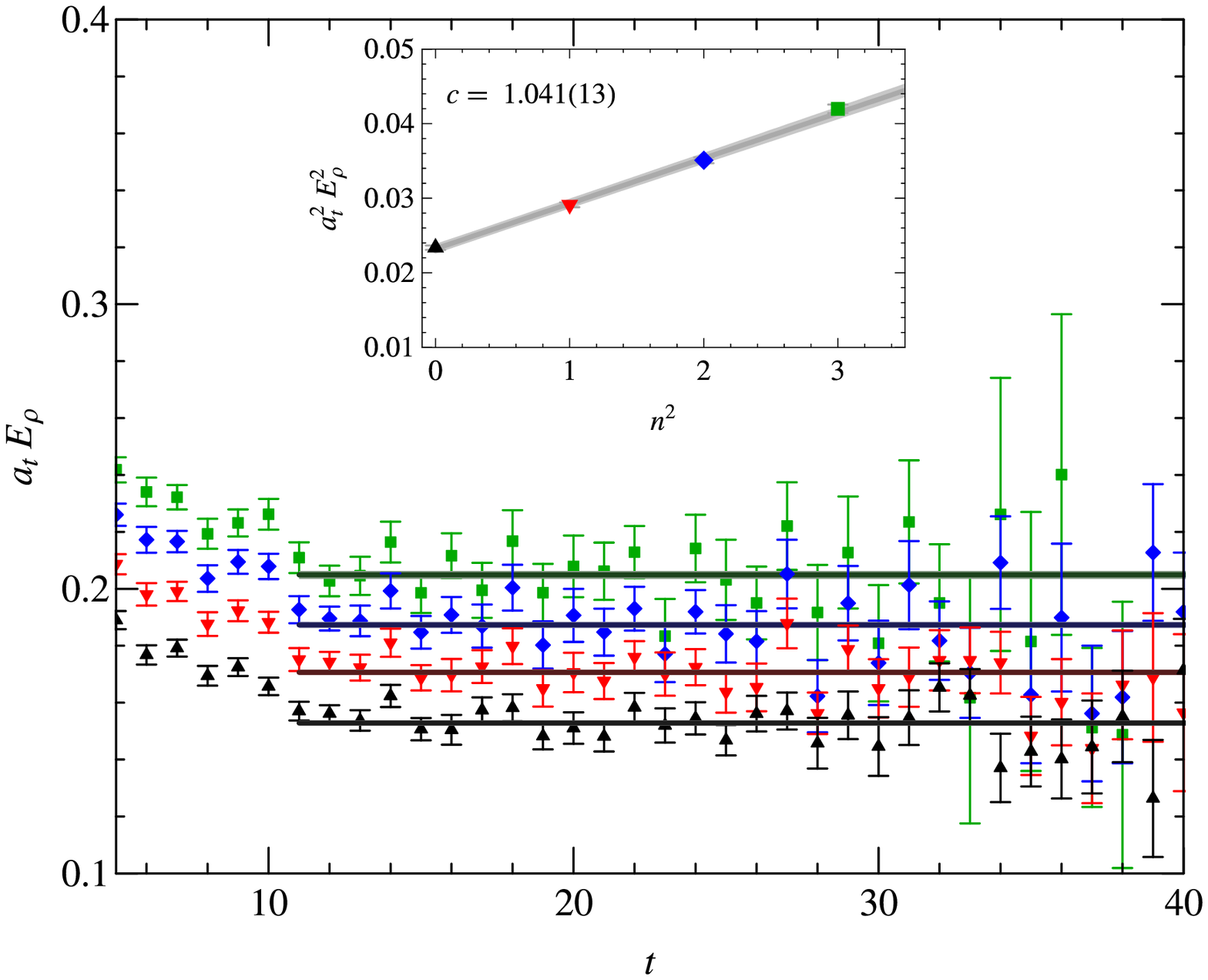}
\caption{Effective-mass plots using the ground-state principal correlators from
  the pion and rho-meson masses at 4 different momenta on the lightest ensemble
    ($m_s=-0.0743$) with volume $24^3\times 128$. The insets show the energy
    squared in temporal lattice units versus $n^2$, which is related to the
    momentum by $p^2=\frac{4\pi^2}{L_s^2}n^2$.
}
\label{fig:disp}
\end{figure}

Finally, we measure the renormalized fermion anisotropy on the $N_f=2+1$
lattices. We tuned the fermion anisotropy in the three-flavor calculation in
Ref.~\cite{Edwards:2008ja}, where we found that the fermion action coefficients
are consistent for bare PCAC quark masses up to about 175~MeV.
Figure~\ref{fig:disp} shows the meson dispersion on the $24^3\times 128$,
  $a_tm_l=-0.0840$ and $a_tm_s=-0.0743$ ensembles. The effective mass plots are shown for the ground-state principal correlators at momenta $p=\frac{2\pi}{L_s}n$ with $n \in \{0,1,2,3\}$; the fitted range and extracted energies are shown as straight lines across the effective mass plots. The inset shows the fitted renormalized fermion anisotropy at each $n^2$.
The speed of light $c$ is measured from the energy of the boosted hadron using
$a_t^2 E_H(p)^2=a_t^2 E_H(0)^2
+\frac{c^2}{\xi_R^2}\frac{4\pi^2}{N_s^2}n^2$.
The values of $c$ from the pion and
rho mesons are about two and three standard deviations away from unity. Such
a small deviation is also expected on isotropic lattices.
For example, $c$ for the pion and rho
are about two standard deviations away from unity on the MILC coarse asqtad
lattice ensembles\cite{Bernard:2001av}.

\subsection{Static-Quark Potential}\label{sec:Potential}

$V(r)$, the energy of two static color sources separated by distance $r$
provides a useful reference scale
for spectrum calculations. This is most usefully described by the Sommer
parameter $r_0$, defined by the condition
\begin{eqnarray}\label{eq:SQPr0}
- r^2 \frac{\partial V(r)}{\partial r} |_{r=r_0} = 1.65.
\end{eqnarray}

The potential is computed by measuring correlations between operators creating a
static color source in the fundamental representation of SU(3), connected
via a gauge covariant parallel tranporter to a source in the $\bar{3}$
representation. The
gauge connector can be formed by any sum of path-ordered products of link
variables that respects the symmetry of rotations about the inter-source axis.
Better ground-state operators are formed by using stout-smeared link variables
in the path-ordered connections and by using an operator optimized using the
variational method.

A basis of five operators is constructed from the set of straight connectors
and staples linking the mid-point between the two color sources. In forming the
temporal correlators, straight, unsmeared temporal links are used for the
propagator of the static source. A five-by-five correlation matrix,
$G_{ij}(r,t)$ is then computed for a range of time separations $t$ and
values of $r \in \{1,N_x/2\}$ along a lattice axis.
As outlined in the previous sub-section, this correlation matrix can be analysed
using the variational method to make a more reliable ground-state energy
extraction.

\begin{figure}
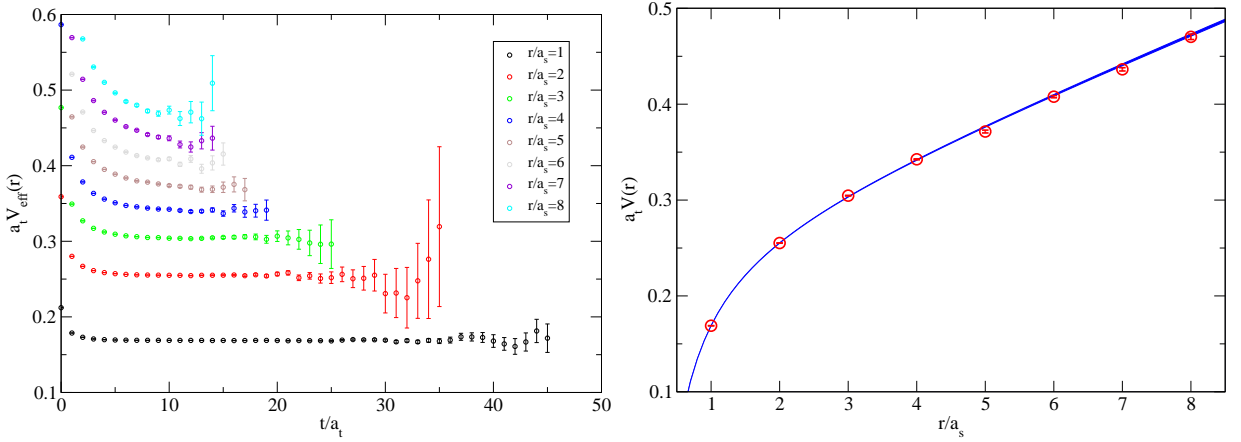

\includegraphics[width=0.45\textwidth]{plots/Veff.eps}
\includegraphics[width=0.45\textwidth]{plots/V.eps}
\caption{Results for the static-quark potential for the $a_tm_l=-0.0808$,
  $a_tm_s=-0.0743$ mass set. The left panel shows the effective energies,
  $a_tV_{\rm eff}(r)$ for each $r$.
    The right panel shows the resulting fit to the potential using Eq.~\protect\ref{eq:static_pot_th}.
}\label{fig:qpot}
\end{figure}

Once the potential energy for a range of values of $r$ has been determined,
the data are compared with the Cornell model,
\begin{eqnarray}
V(r) = V_0 + \frac{\alpha}{r} + \sigma r,
\label{eq:static_pot_th}
\end{eqnarray}
and best-fit values for the parameters $\alpha, \sigma$ and $V_0$ are
determined. See an example from one of our ensembles ($a_tm_l=-0.0808$ and
  $a_tm_s=-0.0743$) in Figure~\ref{fig:qpot}. In all cases, a small range of $r$ values that span $r_0$ are used.
Once values of these parameters are computed, a value of $r_0$ was derived
from Eq.~\ref{eq:SQPr0}.
The QCD flux-tube is expected to break in the presence of dynamical quarks and
the ground-state of the system should not be modelled by the Cornell potential
at large separations. We fit the data to Eq.~\ref{eq:SQPr0} in a
sufficiently small range of $r$ such that this issue does not arise.
No good evidence of this ``string-breaking'' effect was
observed in our data at larger separations. This observation fits with
previous investigations\cite{Bali:2005fu}, which
established the need to include appropriate operators that construct two
disconnected static-source--light-quark systems to measure the
full spectrum.

\begin{table}[h]
\begin{center}
\begin{tabular}{cccc|cc}
\hline\hline
$N_s$ & $N_t$ & $a_tm_l$ & $a_tm_s$ & $r_0/a_s$  \\
\hline
16 &  96 & $-$0.0826 & $-$0.0540 & 3.221(25) \\
12 &  96 & $-$0.0794 & $-$0.0540 & 3.110(31) \\
12 &  96 & $-$0.0699 & $-$0.0540 & 2.752(77) \\
12 &  96 & $-$0.0540 & $-$0.0540 & 2.511(14) \\
12 &  96 & $-$0.0618 & $-$0.0618 & 2.749(37) \\
16 & 128 & $-$0.0840 & $-$0.0743 & 3.646(10) \\
16 & 128 & $-$0.0830 & $-$0.0743 & 3.647(14) \\
16 & 128 & $-$0.0808 & $-$0.0743 & 3.511(12) \\
16 & 128 & $-$0.0743 & $-$0.0743 & 3.214(10) \\
\hline\hline
\end{tabular}
\end{center}
\caption{\label{tab:r0}Sommer scale $r_0/a_s$.}
\end{table}

\section{Choosing the bare strange-quark mass}\label{Sec:Strange}

The appropriate value for the strange-quark mass in the lattice action is
not known a priori. The Wilson formulation makes the task of choosing a
sensible value for this parameter more difficult, as the breaking of chiral
symmetry at the action level induces an additive mass renormalization. In
dynamical simulations, changes to the strange-quark mass parameter in the
action cause all observables to change. We suggest a helpful starting point for
solving this issue is to determine where reference simulations lie in a
parameterized two-dimensional coordinate system.  Note that to
leading order in chiral perturbation theory, the pseudoscalar masses are related
to the quark masses via
\begin{eqnarray}
m_P^2 = 2 B (m_{q_1}+m_{q_2}),
\end{eqnarray}
where $B$ is a low-energy constant and $m_{q_i}$ are the quark masses that
compose the meson. The light-quark dependence can be eliminated using the linear
combination $(2 m_K^2 - m_\pi^2)$.
A useful property of a new coordinate system would be to remove all explicit
  dependence on the lattice cut-off.
Such a dependence can be suppressed (if not completely removed) by taking
ratios of hadron masses.
One good candidate is the $\Omega$ baryon mass, which is stable against QCD
decays and which has a simplified chiral extrapolation due to its lack of
light valence quarks.
An alternative is the $\Xi(1/2)$, which also decays only weakly and is
statistically clean to measure.
Appendix~A shows a comparison between these two choices.
Therefore, we suggest two dimensionless coordinates, $l_\Omega$ and $s_\Omega$:
\begin{eqnarray}\label{eq:ls-Omega}
l_\Omega  &=& \frac{9m_\pi^2}{4 m_\Omega^2},\\
s_\Omega  &=& \frac{9 (2 m_K^2 - m_\pi^2)}{4 m_\Omega^2},
\end{eqnarray}
where the factor of $9/4$ is a convenient normalization which makes $l_\Omega =
s_\Omega = 1$ in the static-quark limit. Note that three-flavor-degenerate
theories lie on the diagonal line across the unit square.
Table~\ref{tab:l_s} summarizes all the $\{l,s\}_\Omega$ values calculated in
this work. Hadron masses are taken from Tables~\ref{tab:mixed_mesons}
and \ref{tab:mixed_baryons} in Sec.~\ref{Sec:Spec}.

\begin{figure}
\includegraphics[width=0.5\textwidth]{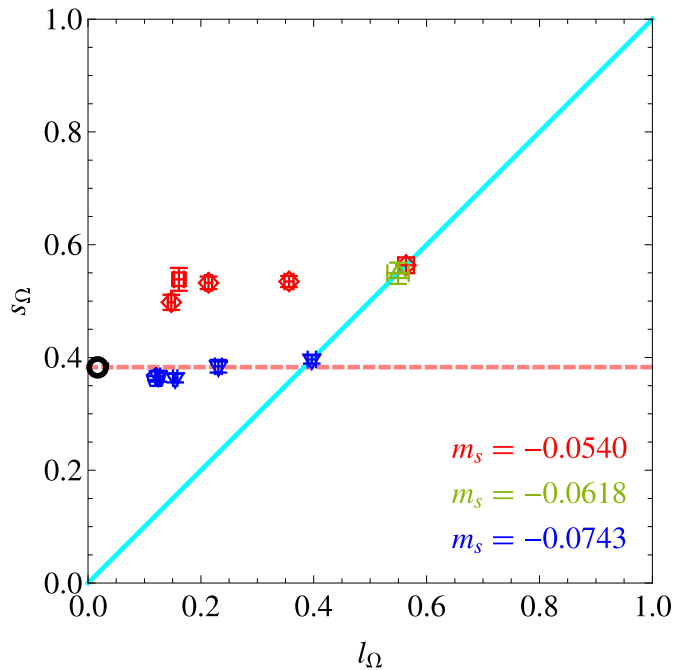}
\caption{
The location of the dynamical ensembles used in this work in the
$s_\Omega$-$l_\Omega$ plane. The circle (black) indicates
the physical point \{$l_\Omega^{\rm phys}$, $s_\Omega^{\rm phys}$\}.
The (red) diamonds and squares are generated on $12^3\times 96$ and
$16^3\times96$ lattices with $a_tm_s=-0.0540$; the (green) upper triangle
is the ensemble on a $12^3\times 96$ lattice with $a_tm_s=-0.0618$ and
(blue) upside-down triangles and pentagons represent the ensembles on
$16^3\times128$ and $24^3\times128$ lattices with $a_tm_s=-0.0743$. Detailed
parameters can be found in in Table~\ref{tab:l_s}.
The horizontal dashed (pink) line indicates constant $s_\Omega$ from the
physical point and the diagonal line indicates three-flavor degenerate
theories.
}
\label{fig:jlab2}
\end{figure}

In Figure~\ref{fig:jlab2}, we locate all the simulations performed in this
work using their $l_\Omega$-$s_\Omega$ coordinates.
The dashed line runs horizontally from the physical point.
We add two more strange-mass candidates: $-0.0618$, $-0.0743$ which are the
points on the diagonal line. The choice of $-0.0743$ seems to anchor the correct
$m_s$ value for $N_f=3$ within one standard deviation of physical. Since we
expect only a few percent deviation coming from the next-to-leading effects on
$s_\Omega$, we settle on $a_tm_s=-0.0743$ for our final choice of strange bare mass; the points to the left of the $N_f=3$ points are $N_f=2+1$ points with fixed strange input parameters. At the lightest simulation point, $s_\Omega$ only differs from $N_f=3$ by less than $2\sigma$. The running of the quantity $s_\Omega$ is indeed small and is thus a good means for tuning the strange-quark mass for fixed-$\beta$ simulation.
We note however, that the trajectory followed by simulations as bare lattice
parameters are changed is dependent on the details of the lattice action and is
not universal; different actions may follow different paths as their bare parameters
change.

\begin{table}
\begin{center}
\begin{tabular}{cccc|ccccccccc}
\hline\hline
$N_s$ & $N_t$ & $a_tm_l$ & $a_tm_s$ & $l_\Omega$ & $s_\Omega$ & $a_tm_\pi$ &
$a_tm_K$ & $a_tm_\Omega$ & $a_tm_\phi$ \\
\hline
 12 & 96 & $-$0.0540 & $-$0.0540 &  0.564(14) &  0.564(14) &  0.2781(9) &  0.2781(9) &  0.556(7) &  0.334(3) \\
 12 & 96 & $-$0.0699 & $-$0.0540 &  0.356(8) &  0.535(10) &  0.1992(17) &  0.2227(15) &  0.501(4) &  0.3031(18) \\
 12 & 96 & $-$0.0794 & $-$0.0540 &  0.214(6) &  0.532(11) &  0.1393(17) &  0.1841(13) &  0.452(5) &  0.268(3) \\
 12 & 96 & $-$0.0826 & $-$0.0540 &  0.148(6) &  0.498(13) &  0.1144(19) &  0.1691(17) &  0.447(5) &  0.266(3) \\
 16 & 96 & $-$0.0826 & $-$0.0540 &  0.161(9) &  0.539(20) &  0.113(3) &  0.1669(15) &  0.423(8) &  0.258(3) \\
 12 & 96 & $-$0.0618 & $-$0.0618 &  0.549(19) &  0.549(19) &  0.2322(15) &  0.2322(15) &  0.470(8) &  0.286(5) \\
 16 & 128 & $-$0.0743 & $-$0.0743 &  0.397(7) &  0.397(7) &  0.1483(2) &  0.1483(2) &  0.353(3) &  0.2159(6) \\
 16 & 128 & $-$0.0808 & $-$0.0743 &  0.231(6) &  0.384(11) &  0.0996(6) &  0.1149(6) &  0.311(6) &  0.1901(18) \\
 16 & 128 & $-$0.083 & $-$0.0743 &  0.154(4) &  0.363(8) &  0.0797(6) &  0.1032(5) &  0.304(3) &  0.1845(11) \\
 16 & 128 & $-$0.0840 & $-$0.0743 &  0.124(4) &  0.367(10) &  0.0691(6) &  0.0970(5) &  0.294(4) &  0.1788(13) \\
 24 & 128 & $-$0.0840 & $-$0.0743 &  0.1205(15) &  0.363(4) &  0.0681(4) &  0.0966(3) &  0.2945(16) &  0.1788(6) \\
\hline\hline
\end{tabular}
\end{center}
\caption{\label{tab:l_s}Values of $l_\Omega$ and $s_\Omega$.}
\end{table}

\section{Extrapolation to the physical quark masses}\label{Sec:Spec}
\label{sec:Extrapolations}

Following the discussion in Sec.~\ref{Sec:Strange}, we adopt the
coordinates $l_\Omega$ and $s_\Omega$ to perform extrapolation of the meson
and baryon masses. To avoid the ambiguity in the lattice-spacing
determination, we extrapolate mass ratios $\frac{a_t m_H}{a_t m_\Omega}$ using
the simplest ansatz consistent with leading order chiral effective theory,
\beq\label{eq:m-ratio-fit}
\left(\frac{m_H}{m_\Omega}\right)^n = c_0 + c_l l_\Omega + c_s s_\Omega
\eeq
with $n=2$ for pseudoscalar mesons and $n=1$ for all other hadrons.
With such a parameterization, care is needed to take account of the statistical
errors of $l_\Omega$ and $s_\Omega$ in the fit. Consider a general fit of the
form $f=a+b x+c y$ where $f$, $x$ and $y$ are all quantities with statistical
error. We wish to find the combination of $a$, $b$, $c$ which minimizes
\begin{eqnarray}
\label{eq:f-xyvar}
\sum_i \frac{\left(f(a,b,c;x_i,y_i) - \langle f_i \rangle
    \right)^2}{\sigma_{f_i}^2+ b^2 \sigma_{x_i}^2+ c^2 \sigma_{y_i}^2},
  \end{eqnarray}
where $i$ indexes different data points $\{x,y,f\}$, $\langle \dots
\rangle$ indicates a mean over all configurations and $\sigma$ is the
statistical error of each quantity.
The extrapolation (minimizing a quantity as in Eq.~\ref{eq:f-xyvar} with
    $f=\frac{m_H}{m_\Omega}$, $x=l_\Omega$ and $y=s_\Omega$) is taken to
physical $\{l,s\}_\Omega$ and we then take $m_\Omega$ as experimental input
to make physical predictions.

\subsection{Hadrons}\label{sec:hadrons}
\begin{figure}
\includegraphics[width=0.75\textwidth]{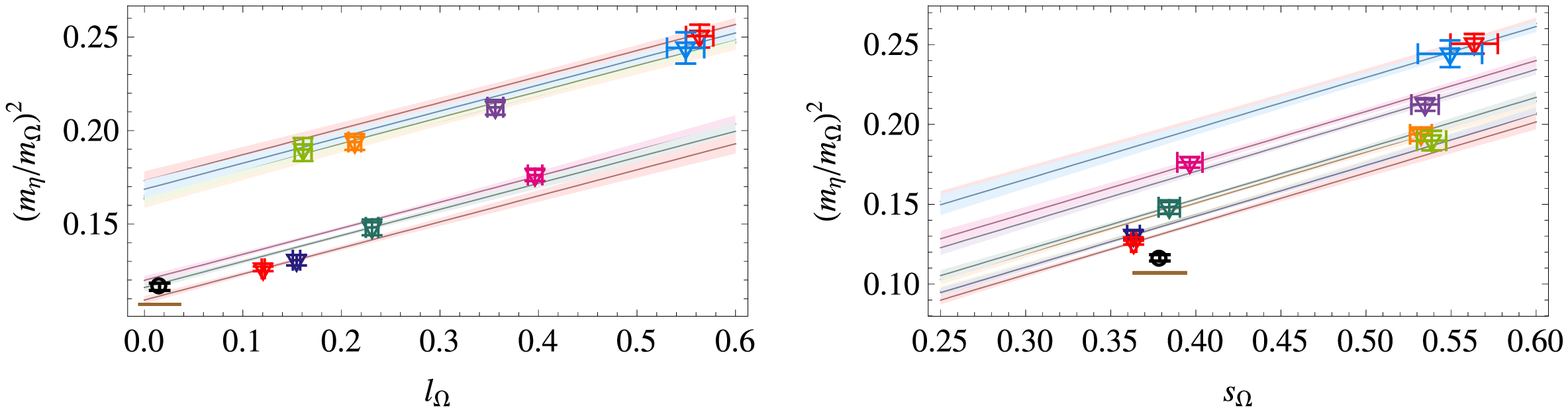}
\includegraphics[width=0.75\textwidth]{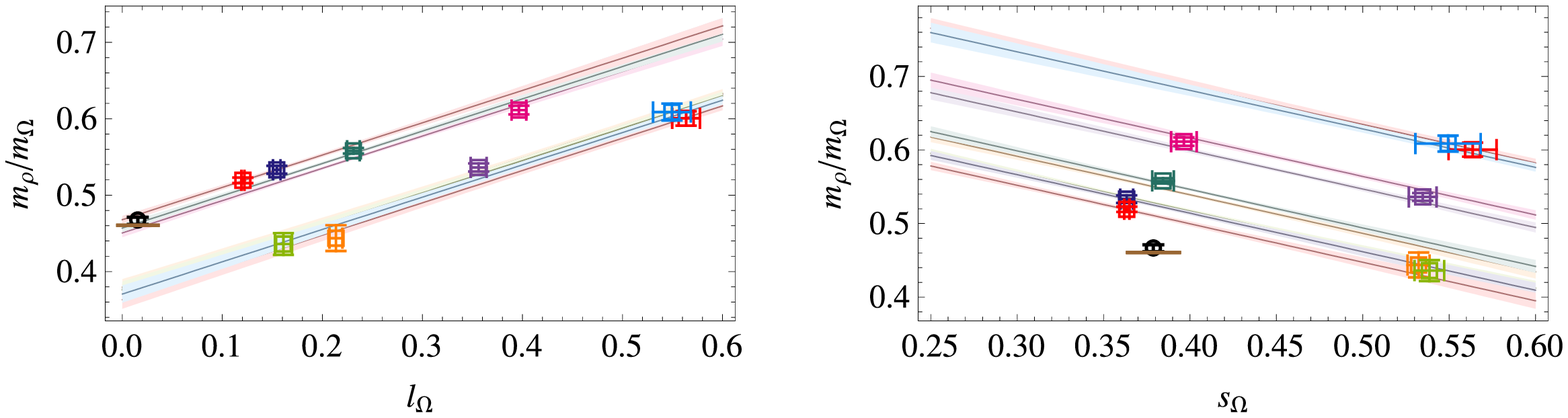}
\includegraphics[width=0.75\textwidth]{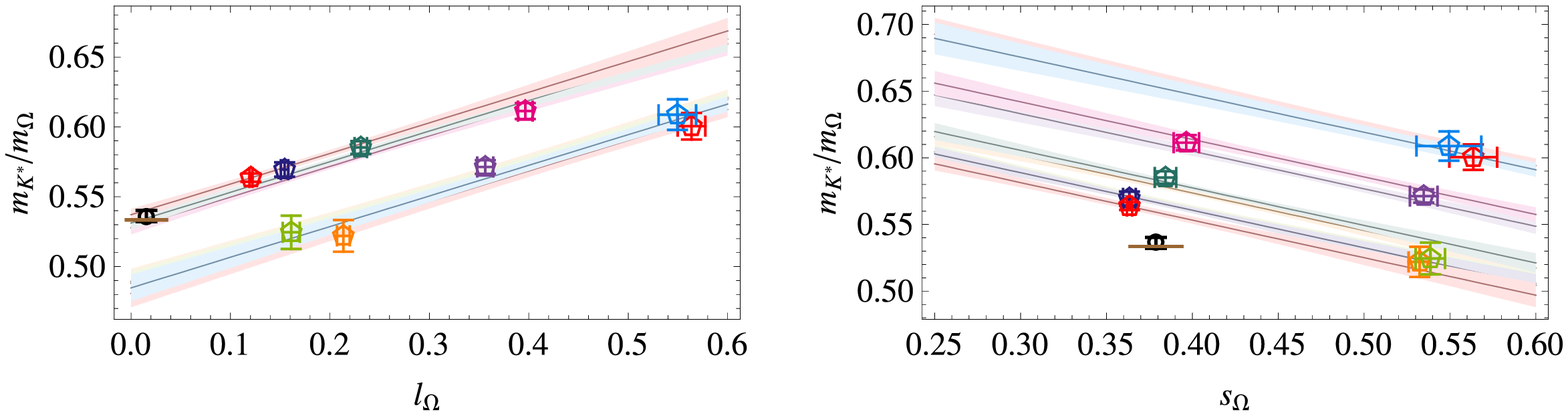}
\includegraphics[width=0.75\textwidth]{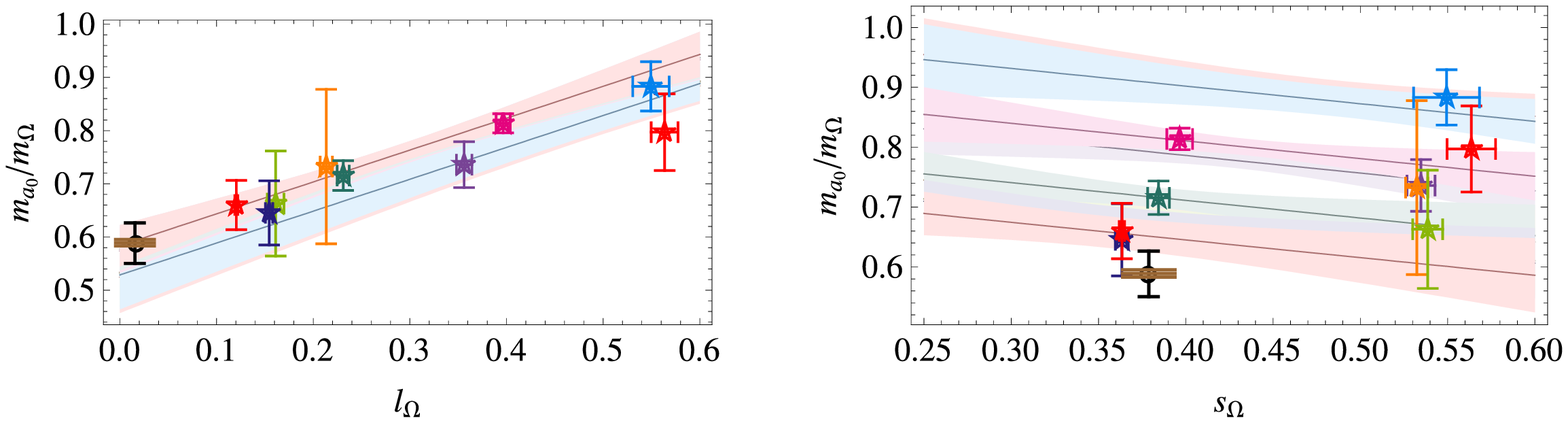}
\includegraphics[width=0.75\textwidth]{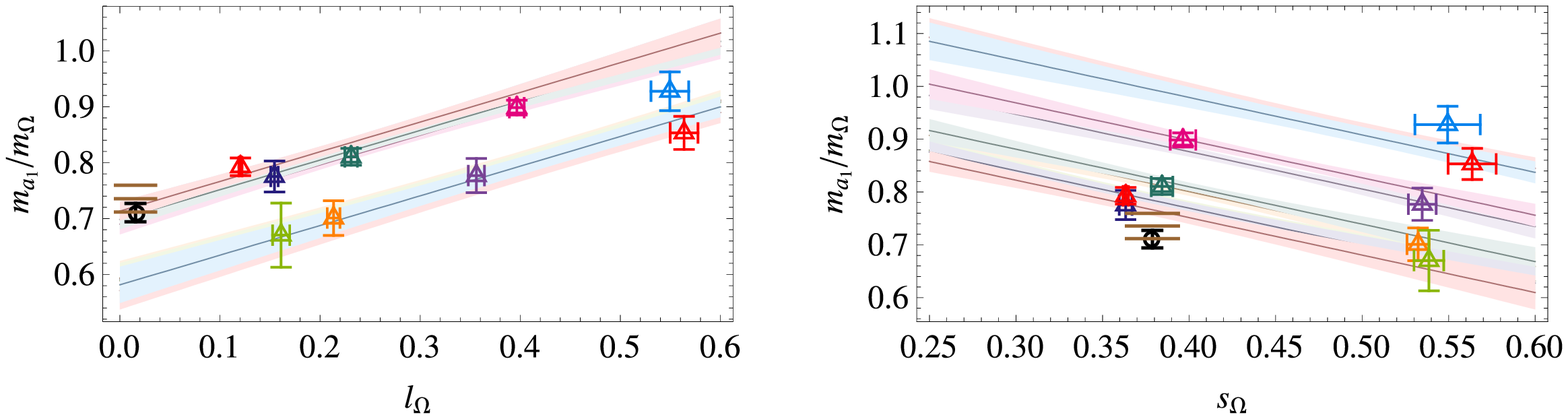}
\includegraphics[width=0.75\textwidth]{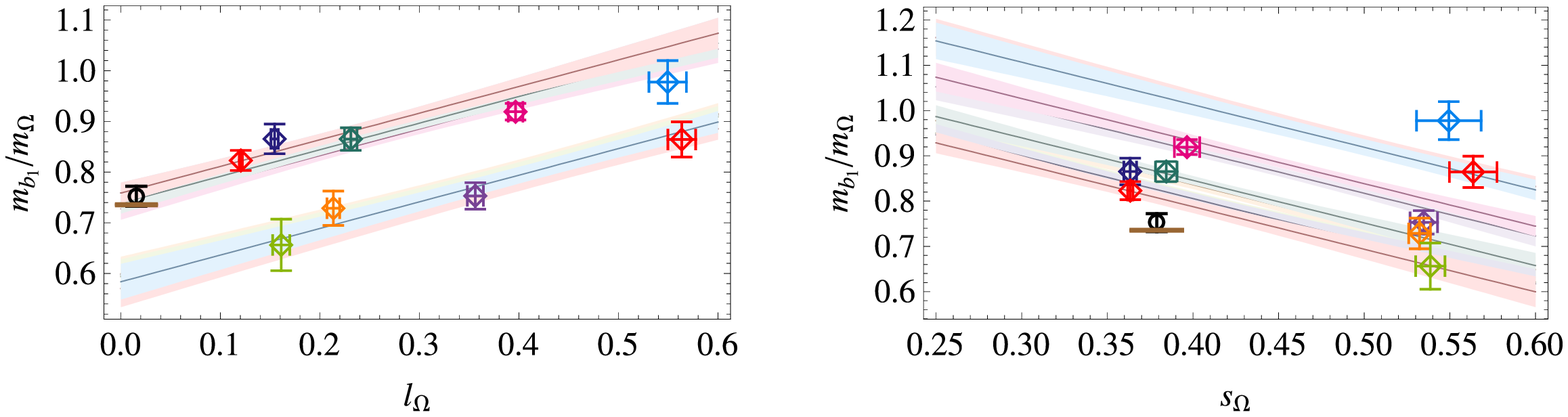}
\caption{Selected meson mass ratios as functions of $l_\Omega$ and $s_\Omega$.
 Differently shaded (or colored) points correspond to the $a_tm_{l,s}$
    combinations in Figure~\ref{fig:color}; detailed numbers can be
    found in Table~\ref{tab:mixed_mesons}. The smaller-volume ensembles
    $\{a_tm_l,a_tm_s\}=\{-0.0826,-0.0540\}$ and $\{-0.0840,-0.0743\}$ are excluded from the fits. The lines indicate the ``projected'' leading chiral extrapolation fit in $l_\Omega$ and $s_\Omega$ while keeping the other one fixed. The black (circular) point is the extrapolated point at physical $l_\Omega$ and $s_\Omega$.
}
\label{fig:all-lsO-meson}
\end{figure}

\begin{figure}
\includegraphics[width=0.75\textwidth]{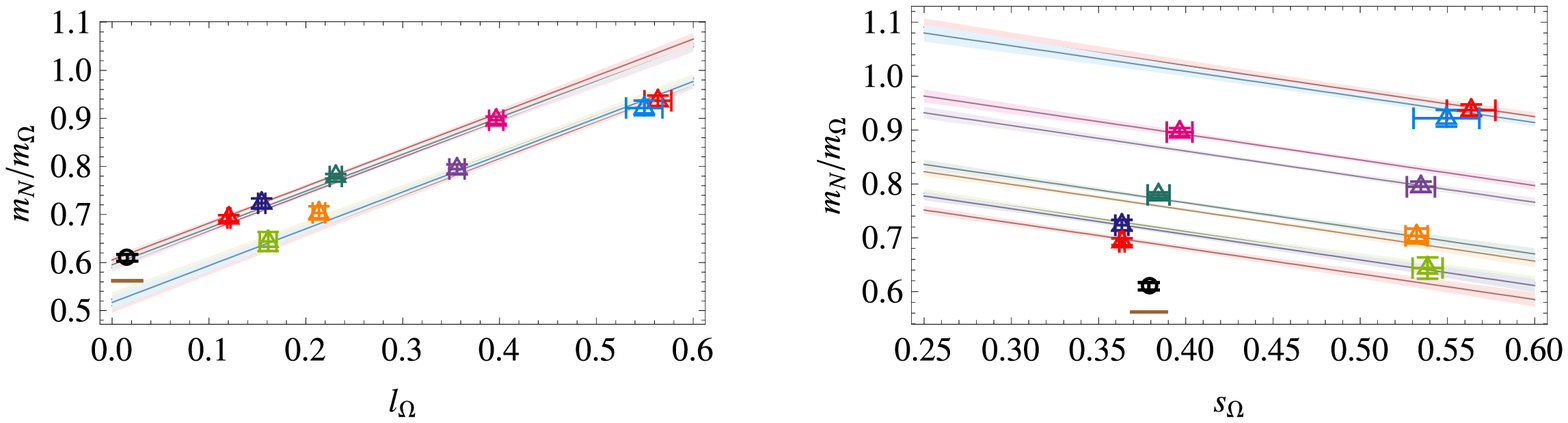}
\includegraphics[width=0.75\textwidth]{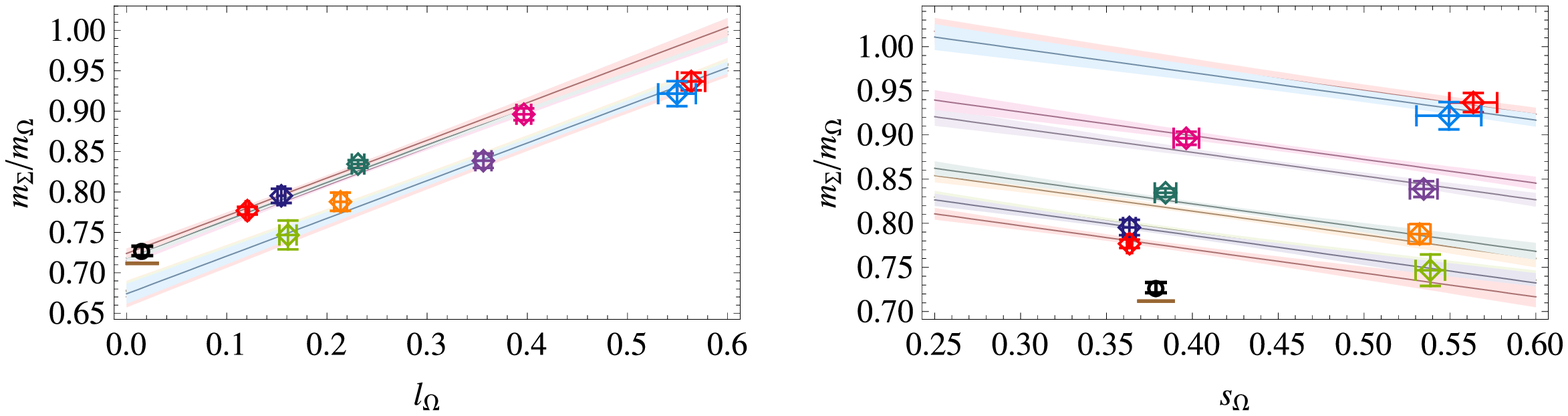}
\includegraphics[width=0.75\textwidth]{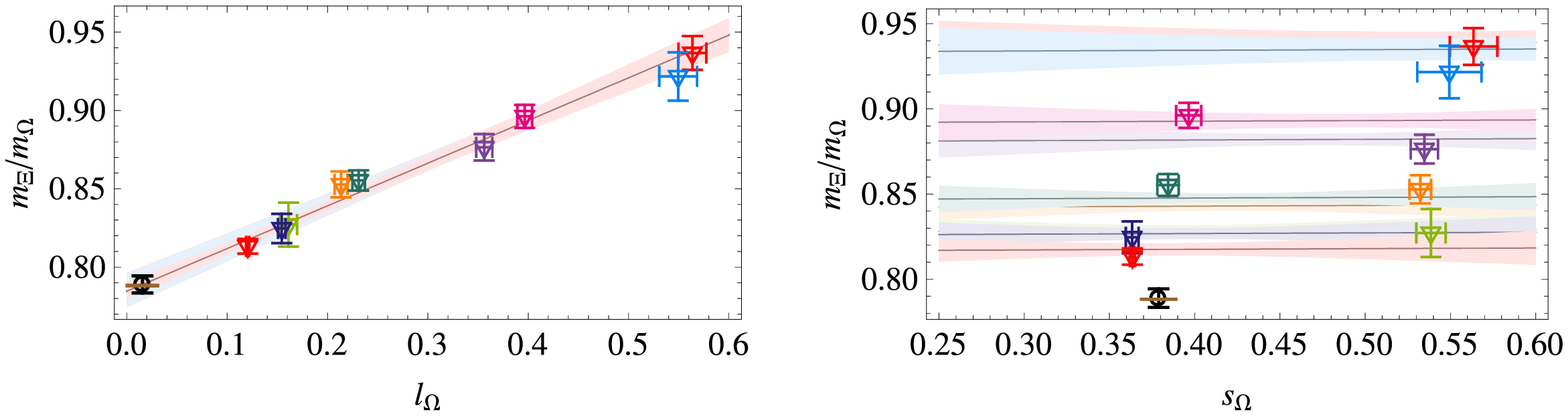}
\includegraphics[width=0.75\textwidth]{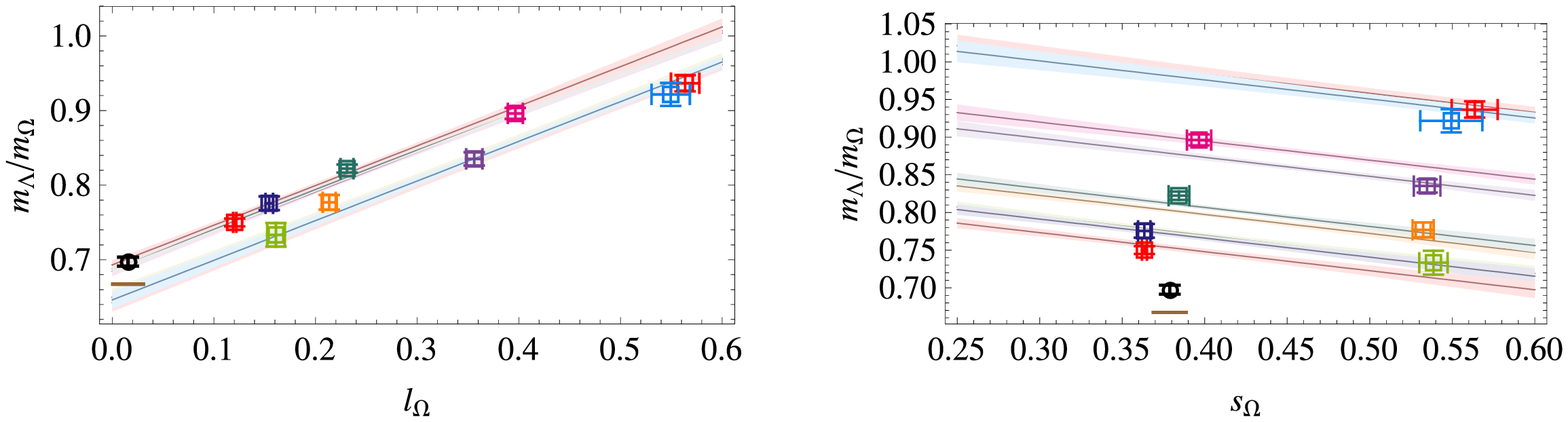}
\includegraphics[width=0.75\textwidth]{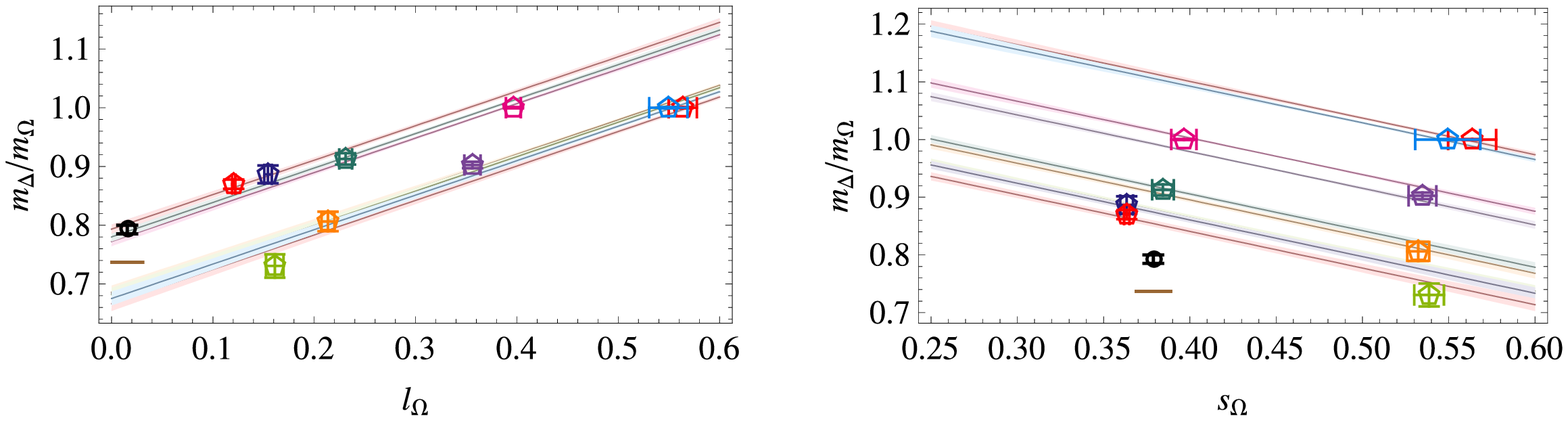}
\includegraphics[width=0.75\textwidth]{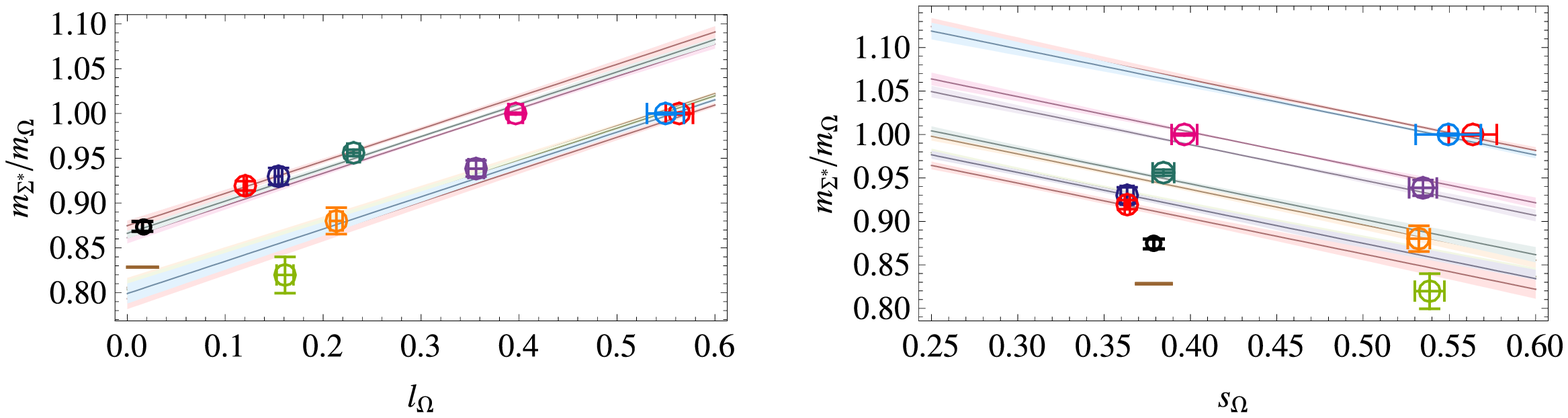}
\caption{Selected baryon mass ratios as functions of the $l_\Omega$ and
  $s_\Omega$. Differently shaded (or colored) points correspond to the $a_tm_{l,s}$
    combinations in Figure~\ref{fig:color};
  detailed numbers can be found in Table~\ref{tab:mixed_baryons}. The
    smaller-volume ensembles $\{a_tm_l,a_tm_s\}=\{-0.0826,-0.0540\}$ and $\{-0.0840,-0.0743\}$ are excluded from the fits. The lines indicate the ``projected'' leading chiral extrapolation fit in $l_\Omega$ and $s_\Omega$ while keeping the other one fixed. The black (circular) point is the extrapolated point at physical $l_\Omega$ and $s_\Omega$.
}
\label{fig:all-lsO-baryon}
\end{figure}

The $\chi^2/{\rm dof}$ for the fits of hadronic data are all around or smaller
than 1 for both meson and baryon masses.
Figures~\ref{fig:all-lsO-meson} and \ref{fig:all-lsO-baryon} show the
``sliced'' plots of selected mass ratio with fixed $l_\Omega$ (or $s_\Omega$).
The $s_\Omega$ are almost a constant for the same sea $a_tm_s$; this is why we see almost a single extrapolated line in the left column of the figures. The hadron masses linearly increase with $l_\Omega$ and decrease with $s_\Omega$.
The ratio of $m_\Xi/m_\Omega$ is almost constant with respect to $s_\Omega$, indicating its insensitivity to the sea strange mass.
The strange-mass dependence is almost completely canceled out in such a
combination.
Later in the appendix, we see the quantity $(2m_K^2-m_\pi^2)/m_\Xi^2$ is
relatively constant with respect to changes in $m_\pi^2$.

\begin{figure}
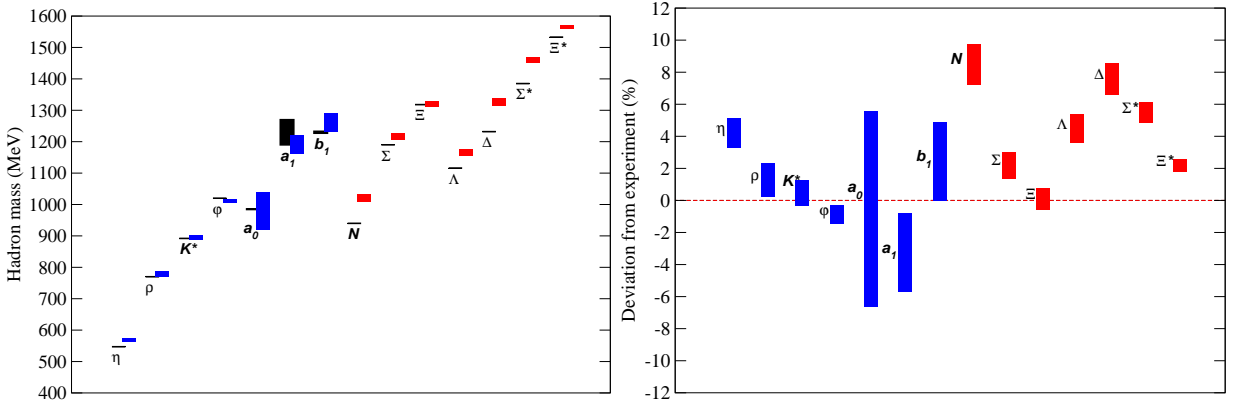

\includegraphics[width=0.45\textwidth]{plots/hadron_spec.ps}
\includegraphics[width=0.45\textwidth]{plots/hadron_spec_per.ps}
\caption{Summary of the extrapolated hadron masses compared with their experimental values.
}
\label{fig:summary}
\end{figure}

Figure~\ref{fig:summary} and the first column in Table~\ref{tab:baryon-final} summarize all of our extrapolated masses along with the experimental values. The second half of the plots shows the relative discrepancy in percent between our calculation and the experimental numbers.
The meson sector appears to have good agreement with experiment; overall, 0.1--4.3\% discrepancy from experimental values. The biggest discrepancy comes in the $\eta$, which we estimate using a combination of light and strange pesudoscalar mesons.
All the vector mesons are in good consistency with experiment; the $\rho$ meson is only $1.2 \sigma$ away. These vector mesons are below the decay threshold on our ensemble; no decays are observed. The extrapolated scalar meson $a_0$ is consistent with the resonance at 980~MeV. The $b_1$ meson is slightly higher than $a_1$, and both of them are 2--$3\sigma$ away from experiment.

The baryon sector, on the other hand, does not work as well as the meson extrapolation. Non-strange baryons, such as nucleon and Delta, have the biggest discrepancy, by as much as $8.5\sigma$. This is likely due to contributions from next-to-leading-order chiral perturbation theory (or pion-loop contributions), which are not as negligible as the meson ones. This becomes evident as we increase the number of strange quarks in the baryon: the discrepancy is smaller in the Sigma and cascade. To have better control of the chiral extrapolation to higher order, we must have better statistics on these measurements; this will be a task for the near future once we complete all of our gauge generation.

Finally, we compare the extrapolation results using all $a_tm_s$ ensembles and
using a single ensemble of either $a_tm_s=-0.0540$ or $a_tm_s=-0.0743$ alone;
results are summarized in Table~\ref{tab:baryon-final}. In both cases, the
$\phi$ measurements are in good agreement with experiment; this is expected once
$\Omega$ is fixed to 1.672~GeV. However, the kaon masses from $a_tm_s=-0.0540$
ensemble are almost 17\% away from experimental ones. This is also not
surprising since the $a_tm_s=-0.0540$ ensemble was selected using the $J$
parameter strange-quark mass setting, where MILC had seen 14--25\% discrepancy
in the strange-quark mass tuning. The kaon mass from the $a_tm_s=-0.0743$
ensemble, on the other hand, is only $3\sigma$ away from the physical one, which
is relatively close for a tuning using degenerate light and strange masses. The
extrapolations using $a_tm_s=-0.0743$ alone versus all $a_tm_s$ ensembles are in
rough agreement within a few $\sigma$, indicating that $a_tm_s=-0.0743$ is a good candidate for gauge generation.

\begin{table}
\begin{center}
\begin{tabular}{c|cccccccccc}
\hline\hline
$$ & all & $a_tm_s=-0.0743$ & $a_tm_s=-0.0540$ \\
\hline
 $m_K$ &   n/a &  0.476(6)[0.14](3.74) &  0.578(13)[0.17](16.7) \\
 $m_\eta$ &  0.570(5)[0.04](4.29) &  0.546(5)[0.25](0.2) &  0.677(12)[0.62](23.79) \\
 $m_\rho$ &  0.780(8)[0.72](1.27) &  0.812(8)[0.12](5.4) &  0.64(3)[2.96](16.52) \\
 $m_{K^*}$ &  0.896(7)[0.49](0.5) &  0.912(7)[0.1](2.27) &  0.828(20)[1.7](7.19) \\
 $m_\phi$ &  1.011(6)[0.42](0.84) &  1.012(7)[0.15](0.75) &  1.007(18)[0.97](1.3) \\
 $m_{a_0}$ &  0.98(6)[0.3](0.13) &  0.98(7)[0.11](0.49) &  1.06(17)[0.05](7.73) \\
 $m_{a_1}$ &  1.19(3)[0.7](3.39) &  1.23(3)[1.1](0.1) &  1.03(7)[0.09](16.17) \\
 $m_{b_1}$ &  1.26(3)[1.09](2.28) &  1.33(4)[0.41](8.47) &  1.03(7)[0.53](16.18) \\
\hline
$m_p$ &  1.020(12)[0.49](8.48) &  1.033(13)[0.26](9.87) &  0.93(3)[0.43](0.72) \\
 $m_\Sigma$ &  1.216(10)[0.62](2.15) &  1.226(11)[0.98](3.03) &  1.16(2)[0.5](2.74) \\
 $m_\Xi$ &  1.319(9)[0.8](0.08) &  1.309(11)[0.63](0.69) &  1.334(20)[0.62](1.22) \\
 $m_\Lambda$ &  1.166(10)[1.18](4.51) &  1.167(12)[1.57](4.56) &  1.13(2)[0.53](1.27) \\
 $m_\Delta$ &  1.325(12)[0.97](7.57) &  1.367(15)[0.36](10.93) &  1.16(2)[4.81](5.78) \\
 $m_{\Sigma^*}$ &  1.461(9)[1.07](5.52) &  1.491(9)[0.4](7.67) &  1.34(2)[2.47](3.06) \\
 $m_{\Xi^*}$ &  1.566(6)[1.](2.17) &  1.582(5)[0.84](3.16) &  1.506(18)[1.81](1.79) \\
\hline\hline
\end{tabular}
\end{center}
\caption{\label{tab:baryon-final}Hadron masses (in GeV) obtained from $(m_H/m_\Omega)^n$ ($n=2$ for pseudoscalar mesons and 1 for the other hadrons) extrapolations using different sea-strange ensembles. The square brackets indicate the $\chi^2/{\rm dof}$ for each fit (with ${\rm dof}=6$), and the second parentheses indicate the percent deviation of the central value from experimental values.}
\end{table}

\subsection{Sommer Scale at the Physical Quark Masses}
   \label{sec:sommer_scale}

Using the static-potential data in Table~\ref{tab:r0}, we can extrapolate
$r_0 m_\Omega$ to the physical limit using $\{l,s\}_\Omega$ coordinates and the
simplest functional form:
\begin{eqnarray}\label{eq:r0-extrap}
r_0 m_\Omega = f_0 + f_l l_\Omega + f_s s_\Omega.
\end{eqnarray}
Once this parameterization is known, $m_\Omega$ serves as experimental input and
$r_0$ becomes a physical prediction.
Using this technique, we determine the Sommer scale in the physical limit, $r_0^{\rm phys}$. We compose a dimensionless ratio $\frac{r_0}{a_s}(a_t m_\Omega)$ and extrapolate using Eq.~\ref{eq:f-xyvar}. After a chiral extrapolation including all three strange ensembles, we find the dimensionless ratio $r_0m_\Omega/\xi_R=1.100(11)$. The ``sliced'' fits projected on to a single parameter $l_\Omega$ (left) and $s_\Omega$ (right) at each point are shown in Figure~\ref{fig:r0mO-extrap}. Then we substitute in the physical Omega mass to find
\beq
\label{eq:r0-phys}
r_0^{\rm phys}=0.454(5)\mbox{ fm},
\eeq
with $\chi^2/{\rm dof}=1.5(0.7)$ and ${\rm dof}=6$. The biggest $\chi^2$
contribution comes from the $a_tm_s=-0.0618$ ensemble. If we drop it, we improve
$\chi^2/{\rm dof}$ to $0.92(0.60)$ but find only a small change to $r_0^{\rm
  phys}=0.451(5)$~fm. Extrapolating using a ratio with $m_\phi$ rather than
$m_\Omega$ gives $r_0^{\rm phys}=0.446(4)$~fm. These data are consistent with
the MILC result in Ref.~\cite{Aubin:2004wf}, which gave a
continuum-extrapolated value of 0.462(12)~fm.

\begin{figure}
\includegraphics[width=0.45\textwidth]{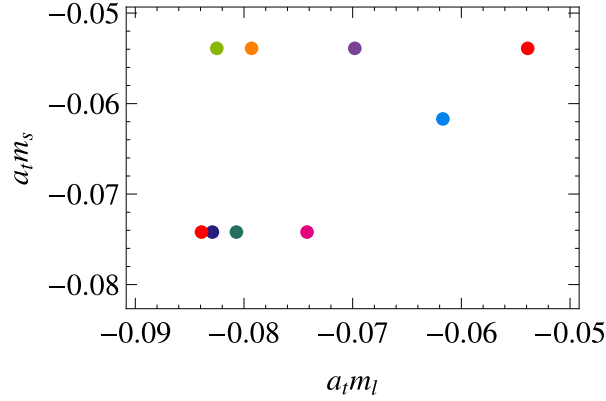}
\caption{The assignment of colors from different ensembles in coordinates $a_t\{m_l,m_s\}$. The convention will remain consistent when used again for later extrapolations.
}\label{fig:color}
\end{figure}

\begin{figure}
\includegraphics[angle=0,width=0.68\textwidth]{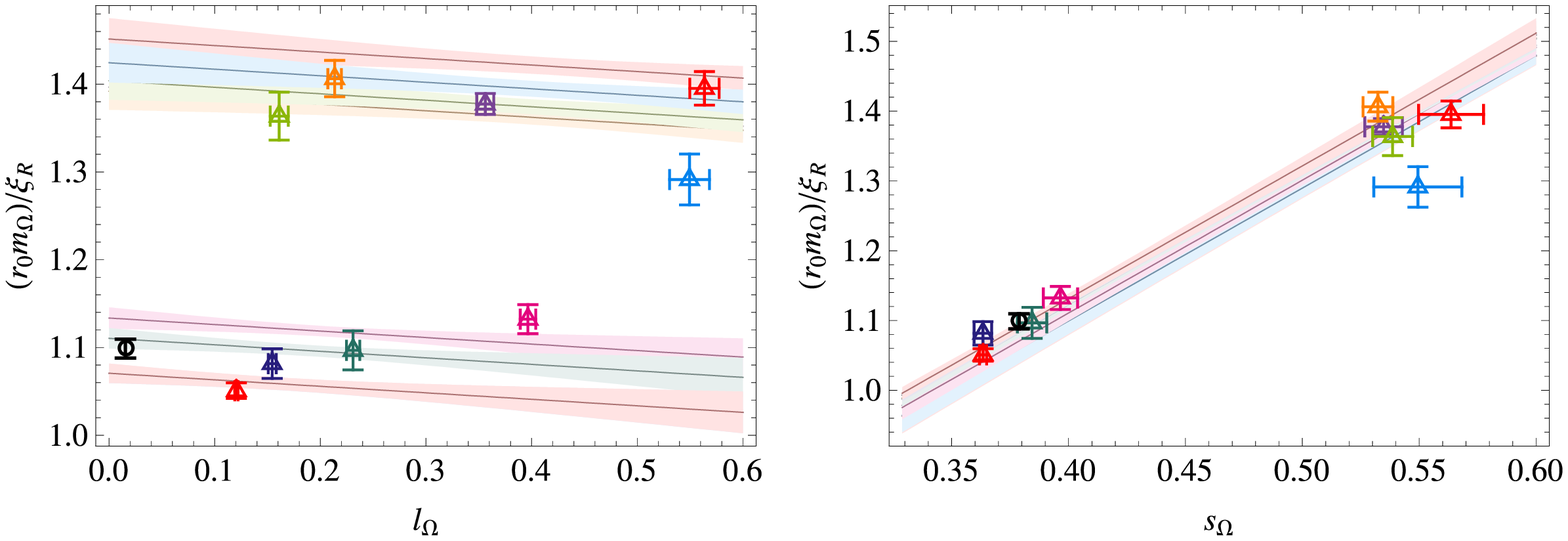}
\caption{``Slices'' of $r_0m_\Omega/\xi_R$ as functions of $l_\Omega$ (left) and $s_\Omega$ (middle). The straight lines are the fitted functions according to Eq.~\ref{eq:f-xyvar}. The assignment of colors is shown in the Figure~\ref{fig:color}.
}\label{fig:r0mO-extrap}
\end{figure}

\subsection{Scale Setting}\label{sec:scale_setting}

We determined the lattice cut-off (in physical units) at the physical point.
Again, we follow the strategy of using the Omega-baryon to set the scale in
physical units. A best fit of all simulation data to the model
\beq
  a_t m_\Omega = d_0 + d_l l_\Omega + d_s s_\Omega
\eeq
has  $\chi^2/N_{\rm dof} =3.10$.
Inputting the physical coordinates $\{l_\Omega,s_\Omega\}=\{0.0153,0.379\}$
yields $a_t=0.03506(23)$~fm and $a_s = 0.1227(8)$~fm.
The low quality of the fit provides us with
further incentive to avoid expressing the lattice cut-off scale in physical
units except in an extrapolation to the continuum. No continuum extrapolation is
possible with our current data set, since all our ensembles have a common
value of the gauge coupling $\beta$.

\section{Conclusion and Outlook}\label{Sec:Conclusion}
This paper presents our first investigation of a number of states in the light
hadron spectrum of QCD with $N_f=2+1$ dynamical flavors. Simulations were
performed on anisotropic lattices with the ratio of spatial and temporal scales
fixed non-perturbatively to $a_s/a_t=3.5$.

The focus of this work has been to
test a simple method for determining the bare strange-quark mass, to
allow us to approach the physical theory.
Conventionally, this has been difficult to achieve, as there is delicate coupling
between lattice action parameters and the cut-off scale. We found it
extremely useful to introduce a pair of coordinates, $s_\Omega$ and $l_\Omega$
to parameterize the two-dimensional space of quark mass values.
The degenerate three-flavor theory corresponds to the line $l_\Omega=s_\Omega$.
To leading order
in the chiral effective theory, these two
coordinates are proportional to the strange and light-quark masses.
These coordinates have been shown to be useful for our simulations since
$s_\Omega$ shows mild dependence on changes to the lattice light-quark mass.
This shows that a good approximate value of the strange quark mass can be found
by following the three-flavor degenerate line to the point where $s_\Omega$
takes its physical value before changing the light-quark masses, a strategy
adopted in this calculation.

With a lattice strange-quark mass close to the physical value, a number of the
simplest light hadrons were investigated. The finite-volume effects for the
mesons were checked on our data sets and found to be mild. Of the states we
investigated, the $a_0, a_1$ and $b_1$ mesons could have decayed on our
lattices. However, checking the overlap factors of the interpolating operators
with the ground states on different volumes suggests that the states we measured are
predominantly resonances. For the octet and decuplet baryons, a similar
analysis predicts that the $\Delta$, $\Sigma^*$ and $\Xi^*$ are stable in our study. At
the heavy strange-quark masses where calculations on $12^3$ lattices were
performed, some finite-volume effects were seen, but they were negligible on the
larger lattice volumes.

Physical predictions have been made by extrapolating simulation data as a
function of these coordinates to the physical point, $\{l_\Omega,s_\Omega\}=
\{0.0153,0.379\}$. These extrapolations have been seen to be robust and have the
advantage of making no reference to the lattice cut-off. This should enable
reliable contact with chiral effective theories to be made. In this analysis,
only the most naive extrapolations have been performed, and some discrepancy
between extrapolated hadron masses and experimental data remains. It is
encouraging to note that at worst, this discrepancy is less than 5\%
for mesons. The largest mismatch occurs in the nucleon-mass determination,
which disagrees with experiment by 8\%. It is very likely that the use of a naive
extrapolation is responsible. No extrapolation to the continuum limit has been
carried out; at present, calculations at a single value of the gauge coupling
$\beta$ have been performed, so no such analysis is possible.

The collaboration has begun to explore more challenging measurements on the
ensembles described in this work. The anisotropic lattice should allow us to
resolve heavier excited states and those states which have traditionally been
statistically rather imprecise with better accuracy. These more difficult
calculations include the hybrid and isoscalar mesons, including the glueballs.
We are confident that a detailed picture of a broad range of light-hadron physics
will emerge soon from these analyses.

\section*{Acknowledgements}
This work was done using the Chroma software suite~\cite{Edwards:2004sx} on clusters at Jefferson Laboratory using time awarded under the USQCD Initiative.
We thank Andreas Stathopoulos and Kostas Orginos for implementing the EigCG inverter\cite{Stathopoulos:2007zi} in the chroma library, which greatly sped up our calculations.
This research used resources of the National Center for Computational Sciences at Oak Ridge National Laboratory, which is supported by the Office of Science of the Department of Energy under Contract DE-AC05-00OR22725.
In particular, we made use of the Jaguar Cray XT facility, using time allocated through the US DOE INCITE program.
This research was supported in part by the National Science Foundation (NSF-PHY-0653315 and NSF-PHY-0510020) through TeraGrid resources provided by Pittsburgh Supercomputing Center (PSC), San Diego Supercomputing Center (SDSC) and the Texas Advanced Computing Center (TACC).
In particular we made use of the BigBen Cray XT3 system at PSC, the BlueGene/L system at SDSC, and the Ranger Infiniband
Constellation Cluster at TACC.
JJ, JF and CM were supported by grants NSF-PHY-0653315
and NSF-PHY-0510020;
EE and SW were supported by DOE grant DE-FG02-93ER-40762;
NM was supported under grant No. DST-SR/S2/RJN-19/2007.
MP and SR were supported by Science Foundation Ireland under research grants
04/BRG/P0275, 04/BRG/P0266, 06/RFP/PHY061 and 07/RFP/PHYF168.
MP and SR are extremely grateful for the generous hospitality of the theory
center at TJNAF while this research was carried out.
Authored by Jefferson Science Associates, LLC under U.S. DOE Contract No. DE-AC05-06OR23177. The U.S. Government retains a non-exclusive, paid-up, irrevocable, world-wide license to publish or reproduce this manuscript for U.S. Government purposes.


\appendix

\section{Alternative Possibilities for $\{l,s\}_X$}

In this work, we have been using the dimensionless parameters $\{l,s\}_\Omega$ (defined in Eq.~\ref{eq:ls-Omega}) to set the strange-quark mass and to extrapolate hadron mass ratios. In this section, we discuss alternatives to the $\Omega$: 1. the strange vector meson $\phi$, which like the $\Omega$ contains no valence up or down quarks; 2. the octet $\Xi$ which is statistically cleaner to measure than the decuplet $\Omega$; 3. a linear combination of octet baryons, $2\Xi-\Sigma$, where the linear combination is selected to hopefully cancel out the leading-order up/down-quark dependence.

\begin{figure}[h]
\includegraphics[width=0.32\textwidth]{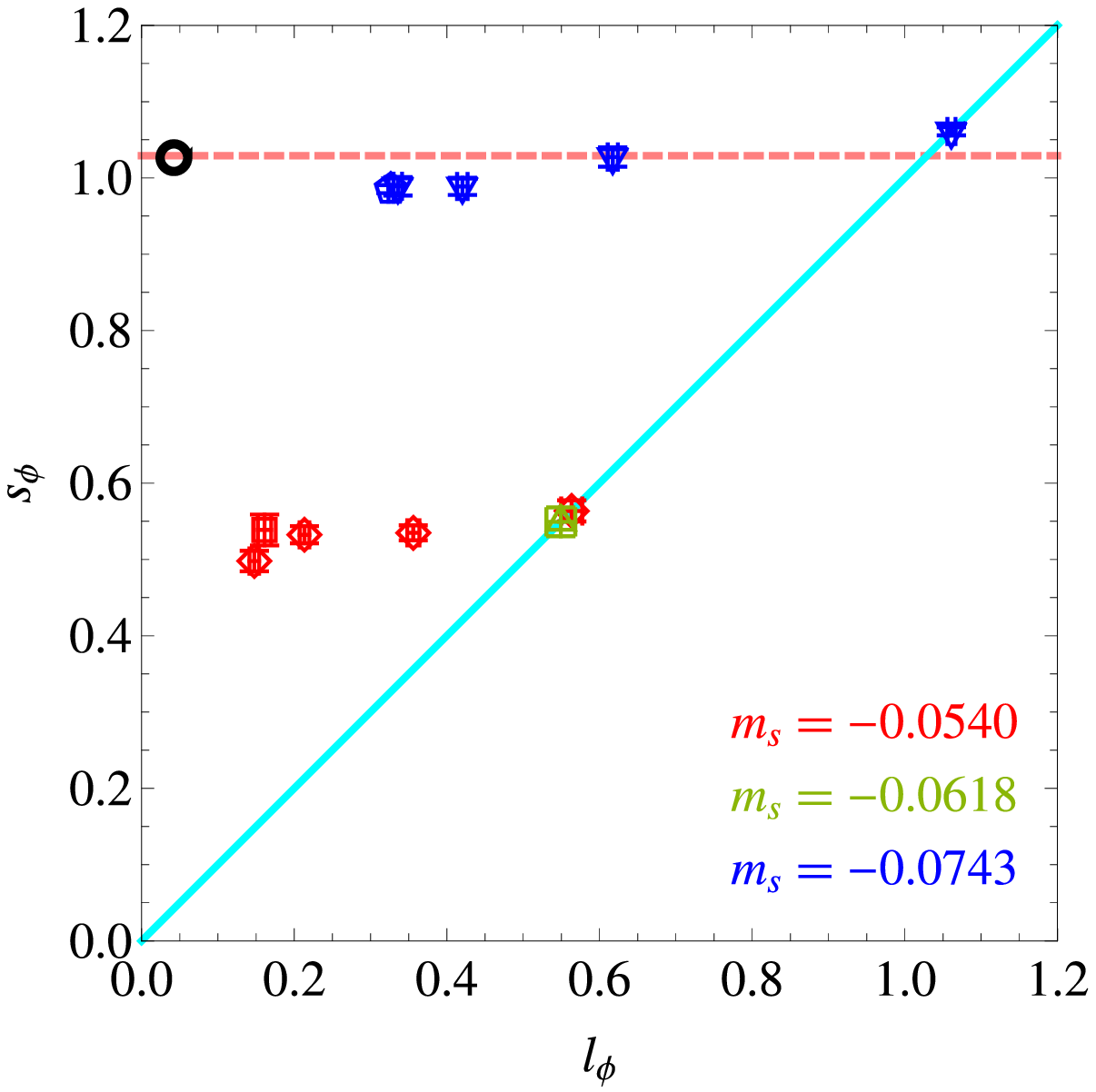}
\includegraphics[width=0.32\textwidth]{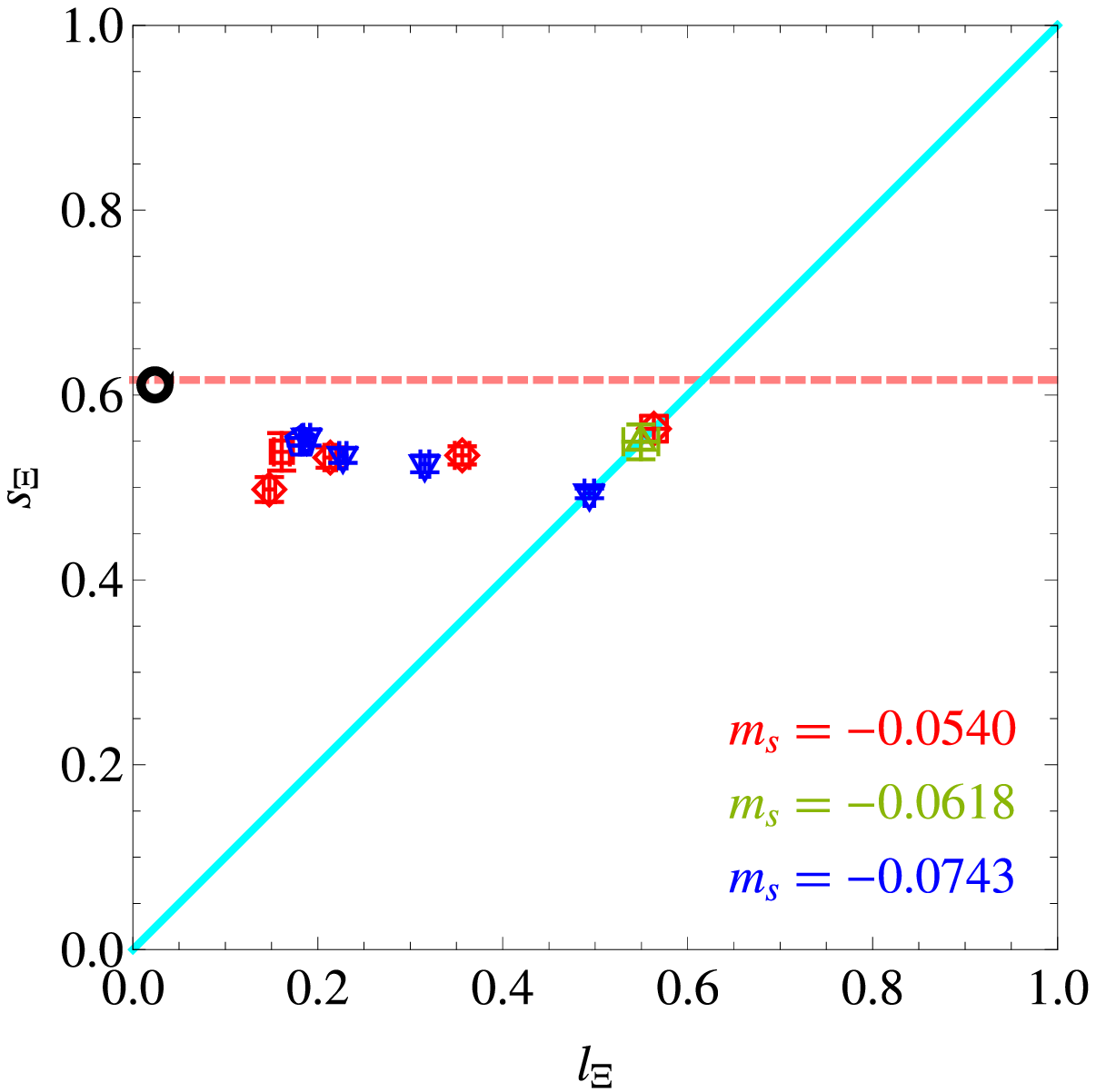}
\includegraphics[width=0.32\textwidth]{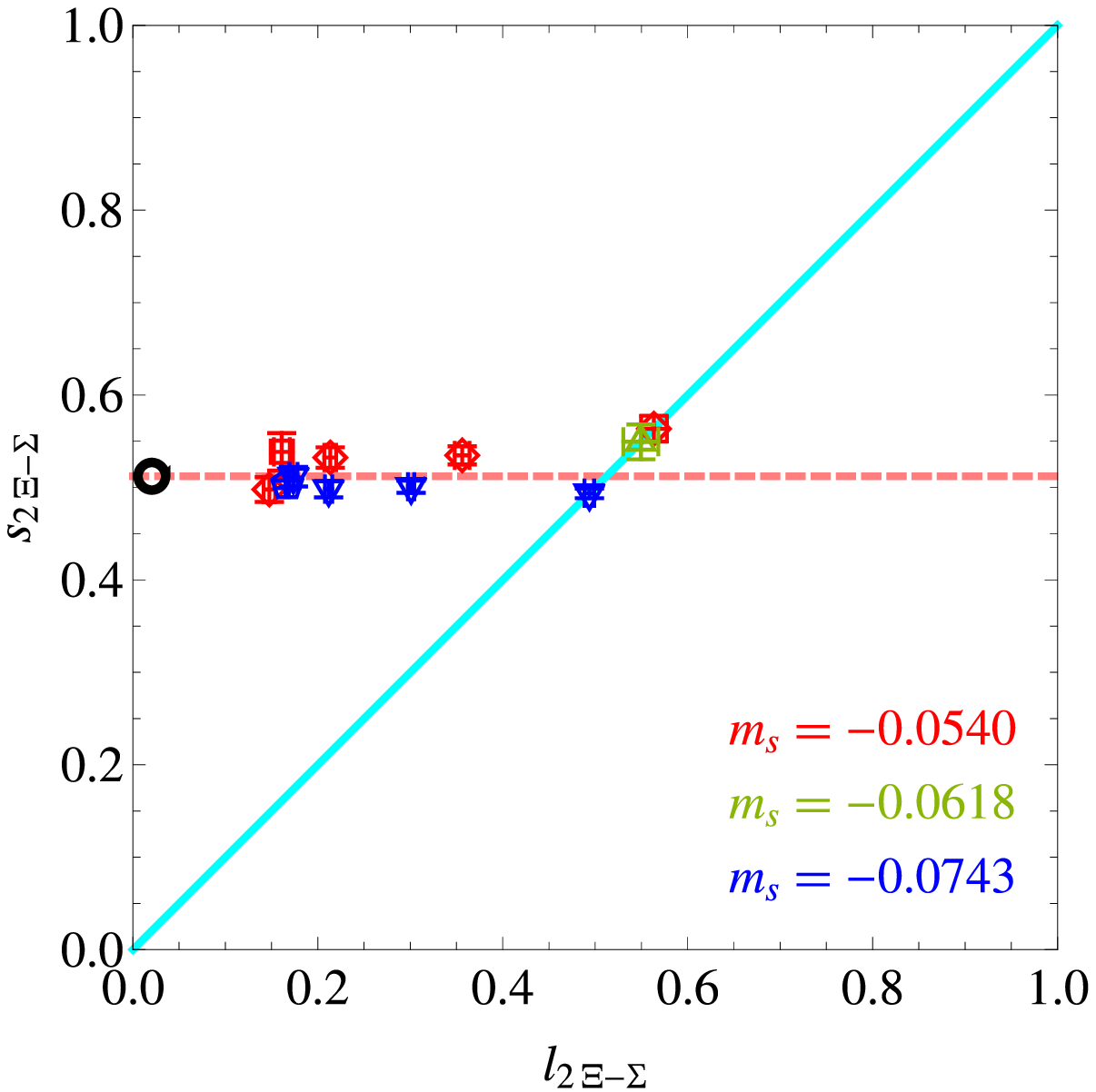}
\caption{The $s_X$-$l_X$ plot with $X=\phi$ (left) $\Xi$ (middle) and the linear
  combination $2\Xi-\Sigma$ for $N_f=3$ and $N_f=2+1$ at $\beta=1.5$. The circle (black) indicates
the physical point \{$l_X^{\rm phys}$, $s_X^{\rm phys}$\}. The (red)
    diamonds and squares are $12^3\times 96$ and $16^3\times96$ from
    $a_tm_s=-0.0540$ ensembles; (green) upper triangles are from $12^3\times 96$
    from $a_tm_s=-0.0618$ and (blue) upside-down triangles and pentagons are
    from $16^3\times128$ and $24^3\times128$ from $a_tm_s=-0.0743$ ensembles. Detailed parameters can be found in Table~\ref{tab:l_s}. The horizontal dashed (pink) line indicates physical $s_X$, and the straight diagonal lines indicate the SU(3) limit.}
\label{fig:jlab-alter}
\end{figure}

We first look at the $s_X$ dependence on the sea strange mass, a similar strategy as described in Sec.~\ref{Sec:Strange}.
Figure~\ref{fig:jlab-alter} is a similar to Figure~\ref{fig:jlab2}, which we
used to tune the sea strange quarks. The blue up-side-down triangles ($V=16^3$)
  and pentagon ($24^3$) are points from the $a_tm_s=-0.0743$ ensembles, the red
  diamonds  ($V=12^3$) and  squares ($16^3$) are from $a_tm_s=-0.0540$ and the
  green triangles are from $a_tm_s=-0.0618$. The leftmost plot corresponds to $X=\phi$, and it shows the strong dependence of
$s_X$ on $a_tm_s$ that we are looking for to tune the strange-quark mass. The $X=\phi$ could be an alternative for setting the strange quark mass. This
result also agrees with our choice of sea strange at $a_tm_s=-0.0743$. The results from $s_\Xi$ (middle plot) show negligible sea strange dependence on
$N_f=2+1$;  this makes it a poor index for tuning the bare strange quark, since
we cannot distinguish when sea quarks are not degenerate anymore. Somehow the
remaining freedom of the light quark in the $\Xi$ baryon dominates the chiral
behavior, since the $a_tm_s=-0.0743$ set is running up toward the physical line. This running is greatly improved when we consider $X=2\Xi-\Sigma$, which lies on the physical line for all $l_X$. However, it is a poor candidate for tuning, since it shows little dependence on $s_X$.

Let us move on to how the extrapolation behavior depends on the choice of $X$.
Table~\ref{tab:baryon-final-X} summarizes the results for all choices of $X$.
The extrapolation (performed according to the minimization process described in
    Eq.~\ref{eq:f-xyvar}) uses all three $a_tm_s$ ensembles without the smaller
volume on the lightest ensemble of $a_tm_s=-0.0540$ and $a_tm_s=-0.0743$ sets.
The $X=2\Xi-\Sigma$ has the poorest $\chi^2/{\rm dof}$ among them all; we will throw it away for reliable extrapolation comparison. The $X=\Xi$ fits have similar $\chi^2/{\rm dof}$ to the $\phi$ but slightly worse. This is possibly due to its insensitivity to $s_\Xi$ during the extrapolation. The $X=\phi$ should be quantitatively comparable to $X=\Omega$ coordinates. However, due to the lightness of the $\phi$ mass, it is not difficult to see that the next-leading-order contributions to the extrapolation form would be larger than the $X=\Omega$, causing it to be a slightly poorer fit at leading order. Even though the fit using $X=\phi$ has smaller statistical error, we expect the systematic error to be higher than $X=\Omega$. We will leave estimation of the systematics to future precision calculations where statistical error will be more reasonable. Still, we see good consistency between $X=\phi$ and $X=\Omega$ results, which reinforces our belief in the stability of extrapolations using the dimensionless coordinates $\{l,s\}_\Omega$.

\begin{table}
\caption{\label{tab:baryon-final-X}Hadron masses (in GeV) obtained from $(m_H/m_X)^n$ ($n=2$ for pseudoscalar mesons and 1 for the other hadrons) extrapolations in terms of $\{l,s\}_X$ with $X\in \{\Omega,\phi,\Xi,2\Xi-\Sigma\}$ using all sea-strange ensembles. The square brackets indicate the $\chi^2/{\rm dof}$ on the fit and the second parentheses denote the central value deviations from experimental values in percent.}
\begin{center}
\begin{tabular}{c|cccccccccc}
\hline\hline
 & $X=\Omega$ & $X=\phi$ & $X=\Xi$ & $X=2\Xi-\Sigma$ \\
\hline
 $a_tm_\eta$ &  0.570(5)[0.04](4.29) &  0.569(2)[0.14](4.04) &  0.570(3)[0.1](4.2) &  0.569(3)[0.11](4.1) \\
 $a_tm_\rho$ &  0.780(8)[0.72](1.27) &  0.795(5)[2.08](3.21) &  0.785(7)[2.44](1.93) &  0.819(7)[3.95](6.31) \\
 $a_tm_{K^*}$ &  0.896(7)[0.49](0.5) &  0.907(3)[1.77](1.69) &  0.905(5)[2.3](1.42) &  0.933(5)[3.73](4.55) \\
 $a_tm_\phi$ &  1.011(6)[0.42](0.84) &  n/a &  1.023(5)[1.92](0.32) &  1.044(5)[3.](2.32) \\
 $a_tm_{a_0}$ &  0.98(6)[0.3](0.13) &  0.99(6)[0.29](0.35) &  0.97(7)[0.33](1.98) &  1.00(7)[0.31](1.6) \\
 $a_tm_{a_1}$ &  1.19(3)[0.7](3.39) &  1.21(3)[0.76](2.02) &  1.18(3)[1.42](3.88) &  1.23(3)[1.73](0.32) \\
 $a_tm_{b_1}$ &  1.26(3)[1.39](2.28) &  1.28(3)[1.85](4.06) &  1.27(3)[1.9](2.89) &  1.32(3)[2.58](7.03) \\
\hline
 $a_tm_p$ &  1.020(12)[0.49](8.48) &  1.029(10)[0.89](9.45) &  1.007(6)[0.06](7.09) &  1.048(9)[0.63](11.5) \\
 $a_tm_\Sigma$ &  1.216(10)[0.62](2.15) &  1.226(8)[1.21](3.06) &  1.211(4)[2.7](1.75) &  1.243(7)[3.81](4.43) \\
 $a_tm_\Xi$ &  1.319(9)[0.8](0.08) &  1.323(7)[1.42](0.39) &  n/a &  1.345(4)[3.47](2.02) \\
 $a_tm_\Lambda$ &  1.166(10)[1.18](4.51) &  1.176(8)[1.86](5.38) &  1.161(3)[0.54](4.06) &  1.194(6)[1.6](6.97) \\
 $a_tm_\Delta$ &  1.325(12)[0.97](7.57) &  1.335(17)[1.38](8.36) &  1.312(18)[1.68](6.51) &  1.356(19)[2.66](10.07) \\
 $a_tm_{\Sigma^*}$ &  1.461(9)[1.07](5.52) &  1.464(15)[1.41](5.72) &  1.450(15)[1.41](4.66) &  1.490(17)[2.49](7.58) \\
 $a_tm_{\Xi^*}$ &  1.566(6)[1.](2.17) &  1.568(12)[1.03](2.3) &  1.561(13)[1.15](1.8) &  1.593(15)[2.22](3.94) \\
 $a_tm_\Omega$ &  n/a &  1.685(10)[0.37](0.8) &  1.683(12)[0.89](0.64) &  1.708(12)[1.82](2.17) \\
\hline\hline
\end{tabular}
\end{center}
\end{table}

\section{Strange-Setting Comparisons}
\begin{figure}
\includegraphics[width=0.45\textwidth]{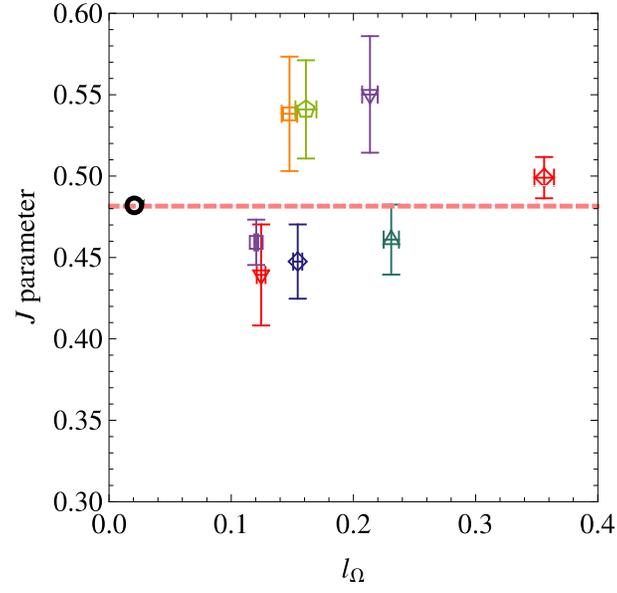}
\caption{$J$-parameter plot for $N_f=2+1$ at $\beta=1.5$. The upper diamond, upside-down triangle, pentagon ($V=12^3\times96$) and square ($V=16^3\times96$) points are from $m_s=-0.0540$ ensembles and the triangle, lower diamond, upside-down triangle ($V=16^3\times128$) and square ($V=24^3\times128$) points are from $m_s=-0.0743$; the circle (black) indicates
the physical point \{$l_\Omega^{\rm phys}$, $J^{\rm phys}$\}; the dashed line indicates the physical $J$ value.
}
\label{fig:J-param}
\end{figure}

The $J$ parameter\cite{Lacock:1995tq} is one common way to set the strange-quark mass; it is defined as
\begin{eqnarray}\label{eq:J-param}
J=\frac{dm_V}{dm_P}=\frac{m_{K^*}(m_\phi-m_\rho)}{2(m_K^2-m_\pi^2)}.
\end{eqnarray}
Here we examine how the parameter works for setting the strange-quark mass
in our calculation. The upper four points in Figure~\ref{fig:J-param} are from $N_f=2+1$ at fixed $m_s=-0.0540$ and $m_l\in\{-0.0699,-0.0794,-0.0826\}$ from right to left with two volumes on $-0.0826$ ($12^3$ and $16^3$). We hit the experimental value $J^{\rm exp}$ with the first $m_l=-0.0699$, and the remaining points are within $1\sigma$ of $J^{\rm exp}$. However, when we tried to extrapolate the kaon mass (see Sec.~\ref{Sec:Spec}), we found that it missed the experimental value by about 17\%. Such a mismatch resulting from tuning the strange mass using the $J$ parameter has previously been reported in the literature. For example, MILC also found their lattice $J$ parameter on their coarse and fine lattices agreed with $J^{\rm exp}$, but after extrapolation they found the sea strange-quark mass to be off by 25\% and 14\% on the coarse and fine lattices respectively\cite{Aubin:2004wf}. Although the discrepancy seems to become smaller for finer lattices, the $J$ parameter does not seem to be an ideal quantity for strange-quark tuning. Similar conclusions can also be reached by observing the 4 lower points in Figure~\ref{fig:J-param} which correspond to a $m_s=-0.0743$, $N_f=2+1$ simulation; these are only two $\sigma$ away from the points for $m_s=-0.0540$. The $J$ parameter is not sensitive enough to changes in the strange sea-quark mass.

\end{document}